\documentclass[conference,compsoc]{IEEEtran}

\ifCLASSOPTIONcompsoc
  \usepackage{cite}
\else
  \usepackage{cite}
\fi

\usepackage{tabularx}
\newcolumntype{Y}{>{\centering\arraybackslash}X}
\usepackage{caption}
\usepackage{subcaption}
\usepackage{graphicx}
\usepackage{multirow}
\usepackage{xcolor}
\usepackage{colortbl}
\usepackage{wasysym}

\usepackage[utf8]{inputenc}
\usepackage{booktabs}
\usepackage{xfp} %
\usepackage{color}
\usepackage{pgffor}
\usepackage{calc}
\usepackage{fp} %
\usepackage{listings}
\usepackage{pifont}
\usepackage{xurl}
\usepackage{tikz}
\usepackage{enumitem}

\begin{document}

\newcommand{\seprule}{\specialrule{0.8pt}{2.5pt}{2.5pt}}
\newcommand{\descr}[1]{\smallskip\noindent\textbf{#1}}
\newcommand{\descrit}[1]{\smallskip\noindent\emph{#1}}

\definecolor{codegreen}{rgb}{0,0.6,0}
\definecolor{codegray}{rgb}{0.5,0.5,0.5}
\definecolor{codepurple}{rgb}{0.58,0,0.82}
\definecolor{backcolour}{rgb}{0.95,0.95,0.92}
\definecolor{highlight}{rgb}{1,0.9,0.7}

\lstset{
    backgroundcolor=\color{backcolour},
    commentstyle=\color{codegreen},
    keywordstyle=\color{magenta},
    numberstyle=\tiny\color{codegray},
    stringstyle=\color{codepurple},
    basicstyle=\ttfamily\scriptsize,
    breakatwhitespace=false,         
    breaklines=true,                 
    captionpos=b,                    
    keepspaces=true,                 
    numbers=left,                    
    numbersep=5pt,                  
    showspaces=false,                
    showstringspaces=false,
    showtabs=false,                  
    tabsize=2,
    frame=single
}

\title{LLMs Cannot Reliably Identify and Reason About Security Vulnerabilities (Yet?): \\A Comprehensive Evaluation, Framework, and Benchmarks}

\author{
\IEEEauthorblockN{Saad Ullah}
\IEEEauthorblockA{Boston University\\
saadu@bu.edu}
\and
\IEEEauthorblockN{Mingji Han}
\IEEEauthorblockA{Boston University\\
mjhan@bu.edu}
\and
\IEEEauthorblockN{Saurabh Pujar}
\IEEEauthorblockA{IBM Research\\
saurabh.pujar@\\ibm.com}
\and
\IEEEauthorblockN{Hammond Pearce}
\IEEEauthorblockA{UNSW Sydney\\
hammond.pearce@\\unsw.edu.au}
\and
\IEEEauthorblockN{Ayse Coskun}
\IEEEauthorblockA{Boston University\\
acoskun@bu.edu}
\and
\IEEEauthorblockN{Gianluca Stringhini}
\IEEEauthorblockA{Boston University\\
gian@bu.edu}
}

\maketitle
\begin{abstract}

Large Language Models (LLMs) have been suggested for use in automated vulnerability repair, but benchmarks showing they can consistently identify security-related bugs are lacking.
We thus develop SecLLMHolmes, a fully automated evaluation framework that performs the most detailed investigation to date on whether LLMs can reliably identify and reason about security-related bugs.
We construct a set of 228 code scenarios and analyze eight of the most capable LLMs across eight different investigative dimensions using our framework. Our evaluation shows LLMs provide non-deterministic responses, incorrect and unfaithful reasoning, and perform poorly in real-world scenarios. Most importantly, our findings reveal significant non-robustness in even the most advanced models like `PaLM2' and `GPT-4': by merely changing function or variable names, or by the addition of library functions in the source code, these models can yield incorrect answers in $26\%$ and $17\%$ of cases, respectively. These findings demonstrate that further LLM advances are needed before LLMs can be used as general purpose security assistants.
\end{abstract}

\IEEEpeerreviewmaketitle
\section{Introduction}

Large Language Models (LLMs), such as OpenAI's Codex \cite{codex}, Google's PaLM2 \cite{palm2}, Meta's Codellama \cite{codellama}, and StarCoder \cite{starcoder}, etc., have demonstrated great potential in performing programming-language related tasks such as code generation, code documentation , and debugging.
In 2022, around 1.2 million developers used Copilot, and since then we have witnessed the release of increasingly capable LLM models at a quick pace~\cite{palm2, gpt4, starcoder, codellama}.
LLMs could be particularly useful to help developers with their cybersecurity needs, as humans typically produce and miss many security relevant bugs.
This issue was highlighted in the 
2022 GitLab Survey~\cite{gitlab22}, noting that ``developers do not find enough bugs early enough'' and ``do not prioritize the bug remediation'' when developing. 
It is pertinent then to investigate if LLMs could be an aid towards early identification of security problems, especially as LLMs have already been suggested for use in automated bug \textit{repair}~\cite{examining-zero}.

In this paper, we aim to answer the following question: \textbf{Can LLMs be used as helpful security assistants for vulnerability \textit{detection}?}
This is an important question, especially as LLMs are not infallible in security-related tasks, for example introducing vulnerabilities into source code \cite{vuln-copilot, ai-assistant-insecure} and software testing \cite{diffblue}.
Unfortunately, %
there is no standardized and automated approach to evaluate the performance of LLMs at identifying vulnerable code. %
We fill this gap by introducing SecLLMHolmes, a generalized, fully automated, and scalable framework to systematically evaluate the performance (i.e., accuracy and reasoning capabilities) of LLMs for vulnerability detection. Our framework tests the capabilities of a given LLM as a security assistant across eight distinct dimensions: (1) deterministic response, (2) performance over range of parameters, (3) diversity of prompts, (4) faithful reasoning, (5) evaluation over variety of vulnerabilities, (6) assessment of various code difficulty levels, (7) robustness to code augmentations, and (8) use in real-world projects.

We apply our framework to eight of the most capable LLMs across 228 code scenarios spanning over 8 most critical vulnerabilities in C and Python, and show that: (a) LLM performance varies widely depending on the model and the prompting technique used, however all models analyzed have a high false positive rate (FPR), and flag code where vulnerabilities have been patched as still vulnerable. (b) the output of LLMs is non-deterministic, with all models changing their answers over multiple runs for one or more of our tests. (c) even when they correctly identify a vulnerability, the reasoning that LLMs provide for this decision is often incorrect, questioning their trustworthiness. (d) LLM chain-of-thought reasoning \cite{cot} is not robust, and can be `confused' by even simple code augmentations such as whitespace modification, changing function names, or using different but related library functions. Also, (e) LLMs fail at detecting vulnerabilities in real-world projects. Our study provides significant evidence that LLMs are not yet ready to be used for automated vulnerability detection, and the successful usage of our framework as a benchmark suite by future models would demonstrate meaningful progress in this space.

\noindent This paper makes the following contributions:

\begin{itemize}[leftmargin=*]
\item We develop SecLLMHolmes, a comprehensive framework to test LLMs for their ability to identify and reason about software vulnerabilities. Our framework is fully automated and includes a set of 228 code scenarios, and 17 prompting techniques. We publicly release our framework and dataset\footnote{https://github.com/ai4cloudops/SecLLMHolmes}, allowing the community to test newly developed LLMs and easily keep track of their progress in being able to identify vulnerabilities.

\item We use our framework to test eight state-of-the-art LLMs for the task of vulnerability detection, showing that as of today no LLM achieves satisfacory performance at it.

\item We identify and enumerate a set of shortcomings that current LLMs show (as outlined above). Our observations provide a checklist for researchers working in this space, showing aspects that need to be addressed before LLMs can be considered ready to be used in the wild for the task of vulnerability detection.
\end{itemize}

\section{Background and Related Work}

\descr{Large Language Models (LLMs).}
\label{subsec:llms}
All language models work on the basic principle of next word (token) prediction; i.e., given a sequence of words (tokens) $x_1, x_2, ..., x_{n-1}$ select a word (token) $x_n$ with the highest probability to appear next in the sequence $$x_n = \arg\max_{w \in V} P(w | x_1, x_2, \dots, x_{n-1}),$$ where $V$ is the vocabulary of the model. Language models learn to perform this task by training on a large amount of text data (i.e., natural language text or code) and use various techniques (e.g., attention mechanism \cite{attention}) to learn to focus on certain parts of the input for better output prediction. Language models have shown excellent proficiency in NLP tasks, as well as good results for programming language tasks such as code generation, code suggestion, natural language querying for code, etc.

\begin{figure}[]
    \centering
    \includegraphics[width=\linewidth]{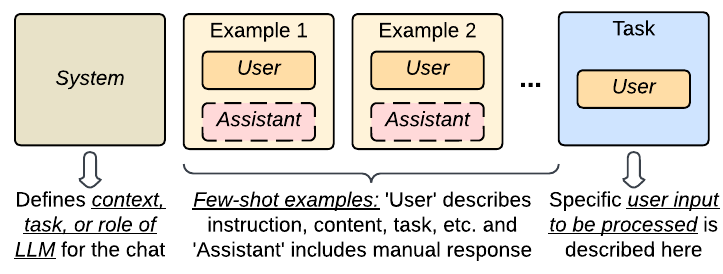}
    \caption{LLM chat input format. LLMs operate on a three-part input format: (1) a system prompt, (2) few-shot examples presented as chat history to guide the model's learning, and (3) the specific user input/task to be processed.}
    \label{fig:chat-format}
\end{figure}

The recent drastic increase in the number of parameters of models has enabled several remarkable capabilities, the most prominent of which being zero-shot and few-shot learning \cite{emergent, few-shot}. LLMs are typically \emph{prompted} (i.e., queried) by the user and provide a response---these advances enable the prompt to provide new knowledge or instructions that the model was not trained over.

Approaches like instruction-tuning `teach' LLMs how to follow instructions in their prompt responses, and reinforcement learning from human feedback is used to `teach' them how to answer, converse, and reason like humans.
This has led to the creation of several chat-based LLMs, which can interact conversationally with human inputs (see Figure \ref{fig:chat-format}). The chat-based LLMs can be prompted using various techniques:

\begin{itemize}[leftmargin=*]
    \item In \textbf{Zero-Shot (ZS)} scenario, the user asks a model to perform a task that the model might not have observed during pre-training.
    \item In \textbf{Few-Shot (FS)} scenario, the user adds a few examples demonstrating input space and expected output to perform a specific task (as shown in Figure \ref{fig:chat-format}).
    \item \textbf{Task-Oriented (TO)} scenarios explicitly assign a task to the model (either in `system' or `user' prompt) in the form of a statement or a question, which encourages the model to generate a task-specific response.
    \item \textbf{Role-Oriented (RO)} scenarios assign a role to the model, e.g., helpful assistant, security expert, etc., and the model implicitly understands the expected behavior. This role is mostly assigned in the `system' prompt.
\end{itemize}

Table \ref{tab:llms} shows the details of the currently most capable chat-based LLMs that we investigate in this paper.

\begin{table}[]
    \centering
    \caption{Studied LLMs. We select a number of capable chat-based LLMs, both \underline{Rem}ote and \underline{Loc}al, for our study, with diverse ranges of number of parameters, max. input tokens limits, and different training knowledge cut-offs. Oct. 7, 2023 is the date of access for all LLMs.}
    \scriptsize
    \begin{tabularx}{\linewidth}{l@{\hspace{0.3cm}}p{1cm}@{\hspace{0.3cm}}p{0.75cm}@{\hspace{0.3cm}}p{0.75cm}@{\hspace{0.3cm}}p{0.5cm}@{\hspace{0.3cm}}X}
        \toprule
        \textbf{Model API} & \textbf{Base Model} & \textbf{\# Params} & \textbf{Max. Tokens} & \textbf{Type} & \textbf{Knowledge Cut Off} \\ [1ex]
        \hline
        \noalign{\vskip 1ex}
        gpt-4 & GPT-4 & 1.76T & 8,192 & Rem & 09/2021 \\
        gpt-3.5-turbo-16k & GPT-3.5 & 175B & 16,385 & Rem & 09/2021 \\
        codechat-bison@001 & PaLM2 & 340B & 6,144 & Rem & mid-2021 \\
        chat-bison@001 & PaLM2 & 340B & 6,144 & Rem & mid-2021 \\
        codellama-7b-instruct & Llama2 & 7B & 100k & Loc & 01/2022 \\
        codellama-13b-instruct & Llama2 & 13B & 100k & Loc & 01/2022 \\
        codellama-34b-instruct & Llama2 & 34B & 100k & Loc & 01/2022 \\
        starchat-beta & StarCoder+ & 15.5B & 8,192 & Loc & 09/2022 \\
        \bottomrule
    \end{tabularx}
    \label{tab:llms}
\end{table}

\descr{Vulnerability Detection.}
One of the biggest challenges during the development and maintenance of software systems is the process of detecting security bugs and identifying their root cause \cite{gitlab22}. 
OWASP~\cite{owasp} presents a list of static and run-time analysis tools and techniques that work in this space. Most of these tools transform source code into a specific representation, e.g., abstract syntax tree, program dependency graph, code property graph, etc., %
and scan them to identify pre-defined insecure patterns. For instance, Yamaguchi \emph{et al.} introduce Joern, a tool that uses code property graphs \cite{cpg} to identify several vulnerability patterns in C/C++ code. %

Unfortunately, these techniques require considerable manual effort and curation of security bug datasets, especially if they are based on trained supervised Machine Learning (ML) models. Examples of methods that use supervised ML include VulChecker \cite{vulchecker}, VulDeePecker \cite{vuldeepecker}, Poster \cite{poster}, and SySeVR \cite{sysevr}. 
As the ratio of vulnerable to non-vulnerable examples is very low in the real world, the vulnerable examples in these datasets are not sufficient for ML models to learn the necessary information.
Suggested improvements in this category include using pre-trained LLMs for code and `fine-tuning' them on security bug datasets for the downstream task of vulnerability detection. Notable approaches here are UniXCoder \cite{unixcoder}, VulBERTa \cite{vulberta}, and CoText \cite{cotext}.

\descr{Evaluating for Code Security.}
Evaluating approaches for identifying security weaknesses in code requires testing their performance on multiple axes. For instance, what are their accuracy and false positive rate (FPR)? Are reasons/root causes for a vulnerability provided by the tool, and if so, at what quality? Is the tool robust to noise in testing data? How many types of vulnerabilities can it detect? 
Static analysis tools have always struggled with the trade-off between accuracy and coverage.
Tools either focus on high accuracy and low coverage (e.g., Pysa \cite{pysa} by Meta, which identifies data-flow issues in Python applications), or (b) low accuracy (false positives) and high coverage (e.g., \cite{bandit, cppcheck, infer}), which can lead to developer frustration and wasted time.

ML-based static analysis tools \cite{large-scale, vuldeepecker} not only face the accuracy-coverage trade-off but also struggle with \textit{robustness}. Such tools have demonstrated good performance on research datasets but can fail to generalize and perform well over real-world datasets \cite{dos-donts}. 
Risse \emph{et al.} \cite{limits} identify the same issue of non-robustness in LLM-based vulnerability detection tools \cite{cotext, vulberta, plbart}, and showcase the drop in accuracy with even trivial code augmentations. Moreover, Microsoft's leader-board for LLM-based vulnerability detection tools \cite{codex-leaderboard} shows that the even the best tool has an accuracy of less than $70\%$ and $30\%+$ FPR, which shows that these systems cannot be trusted in real-world cases.

\definecolor{lightred}{rgb}{1,0.5,0.5}

\DeclareRobustCommand{\yes}{%
    \begin{tikzpicture}[baseline=0ex]
        \draw[fill=green, color=green] (0,0) rectangle (2ex, 2ex);
        \node at (1ex,1ex) [inner sep=0pt] {\tiny \textcolor{black}{\ding{51}}};
    \end{tikzpicture}
}
\DeclareRobustCommand{\semiyes}{%
    \begin{tikzpicture}[baseline=0ex]
        \draw[fill=white, draw=green, thick] (0,0) rectangle (2ex, 2ex);
        \node at (1ex,1ex) [inner sep=0pt] {\tiny \textcolor{black}{\ding{51}}};
    \end{tikzpicture}
}
\DeclareRobustCommand{\no}{%
    \begin{tikzpicture}[baseline=0ex]
        \draw[fill=lightred, color=lightred] (0,0) rectangle (2ex, 2ex);
        \node at (1ex,1ex) [inner sep=0pt] {\tiny \textcolor{black}{\ding{55}}};
    \end{tikzpicture}
}
\begin{table}[]
    \centering
    \scriptsize
    \caption{Evaluation studies for code security. This table describes if the evaluation is (\semiyes semi or \yes fully) \underline{Auto}mated, \underline{Eval}uates \underline{Reas}oning (root cause) \yes or only performs binary evaluation \no, tests \underline{Code}-level \underline{Robust}ness, evaluates both \yes or just one \semiyes of the \underline{Vuln}erable and \underline{Patch}ed code scenarios, evaluates on \underline{Real-World} CVEs \yes or github mined potential defect commits \cite{devign} \semiyes, and \underline{\#} of Code \underline{Scen}arios included in the study, which are not generated by \underline{syn}thetic AI methods or labeled using out-dated \underline{res}earch tools.}
    \begin{tabularx}{\linewidth}{lX@{\hspace{0.2cm}}X@{\hspace{0.2cm}}X@{\hspace{0.2cm}}X@{\hspace{0.2cm}}X@{\hspace{0.2cm}}p{0.5cm}}
        \toprule
        \textbf{Study} & \textbf{Auto} & \textbf{Eval Reas.} & \textbf{Code Robust} & \textbf{Vuln- Patch} & \textbf{Real World} & \textbf{\# Scen.} \\ [1ex]
        \hline
        \noalign{\vskip 1ex}
        Vuln. Code Gen. \cite{vuln-copilot} & \semiyes & \no & \no & \semiyes & \no & 89 \\ [0.5ex]
        Vuln. Code Repair \cite{examining-zero} & \semiyes & \no & \no & \semiyes & \yes & syn. \\ [0.5ex]
        Limits of ML \cite{limits} & \semiyes & \no & \yes & \yes & \semiyes & res. \\ [0.5ex]
        Transf. Vuln. Det. \cite{transformer-vuln} & \semiyes & \no & \no & \yes & \semiyes & res. \\ [0.5ex]
        SecLLMHolmes & \yes & \yes & \yes & \yes & \yes & 228 \\
        \bottomrule
    \end{tabularx}
    \label{tab:eval-history}
\end{table}

Recent works have evaluated LLMs for vulnerable code generation \cite{vuln-copilot}, code repair \cite{examining-zero}, and vulnerability detection \cite{transformer-vuln}.
However, these approaches are either limited by the number of LLMs, coverage of vulnerabilities, diversity of prompts, range of code complexities, robustness testing, or require manual labor for evaluating LLMs.
Most importantly, these studies only evaluate binary responses, checking whether the LLM gives the right label to code snippets (e.g., `vulnerable' or `not vulnerable').
In this paper, we present the first comprehensive evaluation framework to evaluate LLMs on the task of vulnerability detection, providing a multi-faceted analysis of the capabilities of LLMs, including going beyond binary decisions and evaluating their reasoning abilities. Table \ref{tab:eval-history} summarizes evaluation studies of LLMs, highlights their short-comings, and compares against our framework.

In addition to presenting a new state-of-the-art framework, our study uncovers previously unidentified challenges associated with LLMs, including their inability to understand complex code data-flows, the fragility of their COT or step-by-step reasoning to even trivial augmentations, a bias towards security-related functions and variable names that overlooks actual vulnerabilities, a reasoning process that diverges from the methodology of real-world human security experts, particularly in accurately identifying correct root cause of a vulnerability.

\section{SecLLMHolmes}
\begin{figure}
    \centering
    \includegraphics[width=\linewidth]{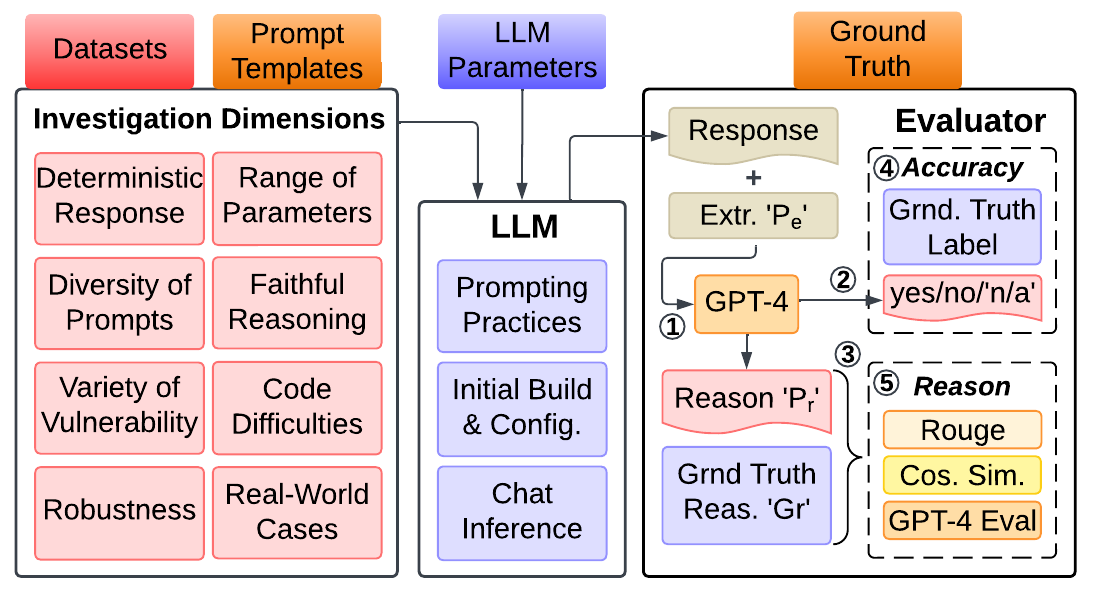}
    \caption{SecLLMHolmes Architecture.}
    \label{fig:llm-eval-framework}
\end{figure}

Figure \ref{fig:llm-eval-framework} presents an overview of our fully automated framework, SecLLMHolmes, which is designed to be applicable to any chat-based LLM.
Section \ref{subsec:user-inputs} describes how users can configure an LLM for integration. 
The core of our framework consists of five pre-defined key components: (i)~a set of parameters (Section \ref{subsec:params-llms}), (ii)~an extensive set of prompt templates (Section \ref{subsec:prompt-tec-cat}), (iii)~datasets (Section \ref{subsec:datasets}), and (iv)~code augmentations (Section \ref{subsec:code-aug}), all of which facilitate the generation of test \textit{prompts}.
Each prompt is then passed to the LLM to generate a response, and the quality of the response is then analyzed by the (v)~`Evaluator' module (Section \ref{subsec:eval-metric}).

\subsection{LLM configuration}
\label{subsec:user-inputs}
To scale SecLLMHolmes for any chat-based LLM, we have established three configurable user inputs:

\descr{LLM-Specific Best `Prompting Practices' and Rules.}
As each LLM may be tuned differently with respect to instruction following, this configurable input enables users to configure optimal prompting practices for the specific LLM they are integrating. This may be done following the LLM's documentation. For example, OpenAI's GPT documentation recommends the use of three quotes before and after the given content to separate it from the instruction~\cite{openai-best-prompt}, while Google's PaLM2 documentation recommends the use of keywords before the content that describe its semantics such as `Code,' `Text,' `Question'~\cite{palm-best-prompt}.

\descr{LLM `Initialization and Configuration.'}
All LLMs require an initial setup; for example, remote models may require an API key (e.g. OpenAI) or to initialize a session with a specific project in the cloud (e.g. Google). Local models require loading the model, tokenizer checkpoints, etc. This input allows user to specify such configuration details. %

\descr{LLM `Chat Structure and Inference Function.'}
To generate a response, each chat-based LLM receives three inputs: `system,' `few-shot examples,' and `task' prompts (as shown in Figure \ref{fig:chat-format}). In this function, users specify how these three inputs are passed to the configured LLM (via API for remote models or `text-generation' pipeline for local models). %

\subsection{LLM Parameters}
\label{subsec:params-llms}
LLM responses are significantly impacted by two parameters which control token sampling:
(1) \textbf{temperature} controls the determinism of an LLM's output. A higher value of temperature generates more `creative'/`random' text by adding noise to potential token scores (2) \textbf{top\_p} controls the nucleus sampling, where the LLM considers the results of the tokens with top\_p probability mass. %
We discuss how we select the values of these parameters in Section \ref{sec:exp-inv}.

\subsection{Prompt Templates}
\label{subsec:prompt-tec-cat}

\begin{table}[!t]
\centering
\caption{Prompt templates.}
\begin{tabularx}{\linewidth}{@{\hspace{0.1cm}}l@{\hspace{0.2cm}}p{0.85cm}@{\hspace{0.2cm}}X}
\toprule
\textbf{ID} & \textbf{Type} & \textbf{Description} \\ [1ex]
\hline
\noalign{\vskip 1ex}
\textbf{S1} & ZS-TO & Code snippet is added to the input prompt with a question about a specific Common Weakness Enumeration (CWE) (e.g., out-of-bound write, path traversal). \\ [1ex]
\hline
\noalign{\vskip 1ex}
\textbf{S2} & ZS-RO & Same as S1, but the LLM is assigned the role of a `helpful assistant'. \\ [1ex]
\hline
\noalign{\vskip 1ex}
\textbf{S3} & ZS-RO & Similar to S1, with the LLM acting as a `security expert'. \\ [1ex]
\hline
\noalign{\vskip 1ex}
\textbf{S4} & ZS-RO & The LLM is defined as a `security expert' who analyzes a specified security vulnerability, without the question being added to the input prompt. \\ [1ex]
\hline
\noalign{\vskip 1ex}
\textbf{S5} & FS-TO & Similar to S1, but includes a vulnerable example, its patch, and standard reasoning from the same CWE. \\ [1ex]
\hline
\noalign{\vskip 1ex}
\textbf{S6} & FS-RO & Like S4, but also includes a vulnerable example, its patch, and standard reasoning from the same CWE. \\ [1ex]
\seprule
\noalign{\vskip 1ex}
\textbf{R1} & ZS-TO & Similar to S1, but begins with "Lets think step by step" \cite{lets-think} to encourage a methodical approach. \\ [1ex]
\hline
\noalign{\vskip 1ex}
\textbf{R2} & ZS-RO & The LLM plays the role of a security expert with a multi-step approach to vulnerability detection, following a chain-of-thought reasoning. \\ [1ex]
\hline
\noalign{\vskip 1ex}
\textbf{R3} & ZS-TO & A multi-round conversation with the LLM, starting with a code snippet and progressively analyzing sub-components for a security vulnerability like human security-experts. \\ [1ex]
\hline
\noalign{\vskip 1ex}
\textbf{R4} & FS-RO & Similar to S6, but the reasoning for answers involves step-by-step analysis developed by the first author. \\ [1ex]
\hline
\noalign{\vskip 1ex}
\textbf{R5} & FS-RO & Like R2, but includes few-shot examples (from the same CWE) with step-by-step reasoning for detecting vulnerabilities. \\ [1ex]
\hline
\noalign{\vskip 1ex}
\textbf{R6} & FS-TO & Similar to R5, but does not assign a specific role to the LLM in the system prompt. \\ [1ex]
\seprule
\noalign{\vskip 1ex}
\textbf{D1} & ZS-TO & Adds the definition of a security vulnerability to the input prompt, followed by a related question. \\ [1ex]
\hline
\noalign{\vskip 1ex}
\textbf{D2} & ZS-RO & The LLM is a security expert analyzing code for a specific vulnerability, with the vulnerability's definition included. \\ [1ex]
\hline
\noalign{\vskip 1ex}
\textbf{D3} & FS-RO & Similar to S6, but includes the definition of the security vulnerability in the system prompt. \\ [1ex]
\hline
\noalign{\vskip 1ex}
\textbf{D4} & FS-RO & Like R4, with the addition of the security vulnerability's definition in the system prompt. \\ [1ex]
\hline
\noalign{\vskip 1ex}
\textbf{D5} & FS-TO & Similar to D4, but does not assign a specific role to the LLM in the system prompt. \\
\bottomrule
\end{tabularx}
\label{tab:prompts-templates}
\end{table}

SecLLMHolmes explores four techniques for LLM prompting: (1) \textbf{zero-shot task-oriented (ZS - TO)}, (2) \textbf{zero-shot role-oriented (ZS - RO)}, (3) \textbf{few-shot task-oriented (FS - TO)}, and (4) \textbf{few-shot role-oriented (FS - RO)}. Moreover, we divide our set of prompts in the following three categories (all prompts are described in Table \ref{tab:prompts-templates}):

\descr{Standard (S)}
\label{subsubsec:standard-prompt}
prompting asks the model to directly give an answer to the problem.

\descr{Step-by-Step Reasoning-based (R)}
\label{subsubsec:step-by-step-prompt}
prompting asks the model to solve the problem in a step-by-step manner using chain-of-thought (COT) reasoning \cite{cot, scratchpad, lets-think}. In addition to evaluating the intrinsic step-by-step reasoning process of LLMs, we also create prompts that emulate the multi-step vulnerability detection method followed by human security-experts, as identified in prior qualitative studies~\cite{hack-test, RE}. These studies observed that security experts follow a general multi-step vulnerability detection approach, i.e., (1) get an overview of the code, (2) based on the overview, identify the critical sub-components that can lead to a security vulnerability in code, e.g., copying user provided information into a buffer, etc., (3) perform a detailed analysis of these sub-components, e.g., if a user input is being copied into a buffer the security experts will check if in any scenario the user input can overflow the buffer, and then, (4) based on the detailed experiments, provide the final answer on whether the given code contains any instances of the given security vulnerability. 

\descr{Definition-based (D)} 
\label{subsubsec:def-prompt}
prompting provides additional information like the definition of a security vulnerability from MITRE's official website \cite{mitre} to the model, while asking the model to detect that vulnerability in the given code.

\subsection{Datasets}
\label{subsec:datasets}

We design 228 code scenarios (48 hand-crafted, 30 real-world, and 150 with code augmentations) to test various aspects of the capabilities of LLMs to detect software vulnerabilities in code.
We use these scenarios to craft prompts by including code, examples, definitions, and step-by-step reasoning as shown in Table~\ref{tab:prompts-templates}.
In the following, we describe how we built our code scenarios in detail.

\descr{Hand-Crafted CWE Scenarios.}
\label{subsubsec:hand-crafted}
We curate a dataset of 48 hand-crafted code scenarios, containing vulnerable and patched pairs from 8 most critical and diverse Common Weakness Enumerations (CWEs) from the MITRE Top 25 Most Dangerous Software Weaknesses for the year 2023 \cite{mitre}, as shown in Table \ref{tab:syn-data}.
To investigate the ability of LLMs to analyze multiple programming languages, we include examples from both C and Python.

\begin{table}[t!]
    \centering
    \scriptsize
    \caption{Hand-crafted dataset.}
    \begin{tabularx}{\linewidth}{m{0.6cm}Xm{0.6cm}m{0.4cm}}
        \toprule
        \textbf{CWE ID} & \textbf{Description} & \textbf{MITRE Rank} & \textbf{Lang.} \\ [1ex]
        \hline
        \noalign{\vskip 1ex}
        787 & Out-of-bounds Write & 1 & C \\ [0.5ex]
        79 & Improper Neutralization of Input During Web Page Generation (`Cross-site Scripting') & 2 & Py \\ [0.5ex]
        89 & Improper Neutralization of Special Elements used in an SQL Command (`SQL Injection') & 3 & Py \\ [0.5ex]
        416 & Use After Free & 4 & C \\ [0.5ex]
        22 & Improper Limitation of a Pathname to a Restricted Directory (`Path Traversal') & 8 & C \\ [0.5ex]
        476 & NULL Pointer Dereference & 12 & C \\ [0.5ex]
        190 & Integer Overflow or Wraparound & 14 & C \\ [0.5ex]
        77 & Improper Neutralization of Special Elements used in a Command (`Command Injection') & 16 & C \\
        \bottomrule
    \end{tabularx}
    \label{tab:syn-data}
\end{table}

Similar to previous work \cite{vuln-copilot}, we create six code scenarios (three pairs of vulnerable and patched scenarios) for each CWE. Moreover, we design our code scenarios with three difficulty levels, (1) easy, (2) medium, (3) hard. Code scenario `$2_v$' in a specific CWE represents the vulnerable scenario with `medium' difficulty level and `$2_p$' its patch. The difficulty levels assess how LLMs interact with code of increasing complexity.
\emph{Easy} scenarios consist of simple programs containing only one function and less than 30 lines of code. \emph{Medium} level scenarios increase the complexity by making the program longer, using different library functions, and adding more than one user input. \emph{Hard} level introduces scenarios with multiple functions in which functions can be safe on an individual level but when they work together they make the program vulnerable.\footnote{Appendix \ref{sec:diff-levels} shows examples of all difficulty levels for `CWE-22'.}

\descr{Real-World CVE Senarios.}
\label{subsec:real-world}
We leverage a set of real-world Common Vulnerabilities and Exposures (CVEs) from public open source projects to investigate if LLMs are able to identify vulnerabilities in them.
Note that existing benchmarks for vulnerability detection~\cite{devign,vuldeepecker,draper,secsmells}, cannot be used for this project, as they were released before the cut-off training date of current LLMs, and it is therefore likely that the models saw that data during training.
To avoid this potential confounder, we curate 30 code scenarios containing vulnerable and patched versions of 15 CVEs from four open source projects, all published and fixed in 2023, after current LLMs were trained (see Table \ref{tab:cves}).

\begin{table}[]
\centering
\scriptsize
\caption{Real-world CVEs and their details including \underline{Orig}inal and \underline{Trun}cated Lines of Code (\underline{LoC}).}
\begin{tabularx}{\linewidth}{l@{\hspace{0.2cm}}p{1.2cm}@{\hspace{0.2cm}}X@{\hspace{0.15cm}}p{0.4cm}p{0.4cm}p{0.4cm}p{0.4cm}}
\toprule
\textbf{Project} & \textbf{CVE ID} & \textbf{CWE Description (ID)} & \textbf{Orig. LoC} & \textbf{Trun. LoC} & \textbf{Pub. Date (2023)} & \textbf{Fix Date (2023)} \\ [1ex]
\hline
\noalign{\vskip 1ex}
\multirow{4}{*}{gpac} & 2023-1452 & Out-of-Bound Write (787) & 4.5k & 243 & Mar & May \\ [0.5ex] 
 & 2023-3012 & NULL Pointer Deref (476) & 2.5k & 398 & May & Nov \\ [0.5ex] 
 & 2023-23143 & Out-of-Bound Write (787) & 12.3k & 117 & Jan & May \\ [0.5ex] 
 & 2023-23144 & Integer Overflow (190) & 439 & 389 & Jan & May \\ [0.5ex] 
\seprule 
\multirow{5}{*}{libtiff} & 2023-2908 & NULL Pointer Deref (476) & 2.3k & 629 & Jan & Nov \\ [0.5ex] 
 & 2023-3316 & NULL Pointer Deref (476) & 159 & 159 & Jan & Nov \\ [0.5ex] 
 & 2023-26966 & Out-of-Bound Write (787) & 1.8k & 238 & Jun & Nov \\ [0.5ex] 
 & 2023-40745 & Integer Overflow (190) & 2.2k & 757 & Oct & Nov \\ [0.5ex] 
 & 2023-41175 & Integer Overflow (190) & 779 & 748 & Oct & Nov \\ [0.5ex] 
\seprule 
\multirow{5}{*}{linux} & 2023-40283 & Use-After-Free (416) & 1.9k & 515 & Aug & Nov \\ [0.5ex] 
 & 2023-42753 & Integer Overflow (190) & 628 & 623 & Sept & Nov \\ [0.5ex] 
 & 2023-42754 & NULL Pointer Deref (476) & 3.7k & 177 & Oct & Nov \\ [0.5ex] 
 & 2023-45863 & Out-of-Bound Write (787) & 1.1k & 565 & Oct & Nov \\ [0.5ex] 
 & 2023-45871 & Out-of-Bound Write (787) & 10.1k & 386 & Oct & Nov \\ [0.5ex] 
\seprule 
\multirow{1}{*}{pjsip} & 2023-27585 & Out-of-Bound Write (787) & 784 & 737 & Mar & Aug \\ [0.5ex] 
\bottomrule
\end{tabularx}
\label{tab:cves}
\end{table}

As the length of the code increases significantly for the real-world code scenarios, it can exceed the maximum number of tokens a model can take as an input. To solve this problem, we shorten code files by removing comments and functions that are neither called by nor call the vulnerable (or patched) function. Also to maintain fairness among the LLMs, we make sure that all CVEs after truncation have a number of tokens less than or equal to 6,144, which is the maximum token limit supported by Palm2, and the lowest among all LLMs in our study.

\descr{Code Augmentations.}
\label{subsec:code-aug}
While standard frameworks exist to evaluate the robustness of LLMs for NLP tasks \cite{prompt-bench, llm-emotional}, there is no standard framework to evaluate the robustness on code security related tasks. To fill this gap, we design a set of 150 augmented code scenarios, meticulously crafted and reviewed to preserve the ability of human security experts to identify vulnerabilities. These augmentations are organized in two distinct categories: 

\begin{enumerate}[leftmargin=*]
\item \textbf{Trivial Augmentations.} 
This class of code augmentations measure the robustness of LLMs to random noise. We select 12 CWE scenarios from two classes, i.e., CWE-787 (C) (\#1 MITRE) and CWE-89 (Py) (\#3 MITRE) of our hand-crafted dataset, and apply seven trivial augmentations (Table \ref{tab:aug})  on them and create a total of 84 different code scenarios (12 per augmentation). We choose these two CWEs as (1) they can lead to the most catastrophic impacts like root privilege escalation, and data loss, etc, (2) they belong to two completely different levels of abstractions i.e., ``lower-level'' (C) and ``higher-level'' (Python) languages, and (3) the presence of their instances can be determined directly from code without any need for external information.

\item \textbf{Non-Trivial Augmentations.}
We design non-trivial code augmentations to perform stress-tests on LLMs to measure their robustness and bias towards semantics of function or variable names, specific library functions, or code security practices. We use combinations of all CWEs defined in Table \ref{tab:syn-data} and the six non-trivial code augmentations from Table \ref{tab:aug} and design 66 code scenarios (12 per NT1-NT4 and 9 per NT5 and NT6) \footnote{See Appendix \ref{sec:app-robust} for details on creation of non-trivial augmented code scenarios}.
\end{enumerate}

\begin{table}[]
    \centering
    \scriptsize
    \caption{Code Augmentations.}
    \begin{tabularx}{\linewidth}{lp{2.5cm}lX}
        \toprule
        \multicolumn{2}{c}{\textbf{Trivial}} & \multicolumn{2}{c}{\textbf{Non-Trivial}} \\ [0.5ex]
        \hline
        \noalign{\vskip 0.5ex}
        \textbf{ID} & \textbf{Description} & \textbf{ID} & \textbf{Description} \\ [0.5ex]
        \hline
        \noalign{\vskip 0.5ex}
        T1 & Rename function parameters randomly & NT1 & Change variable names to vulnerability related keywords \\ [1ex]
        T2 & Rename function randomly & NT2 & Change the name of a safe function to `vulnerable' function \\ [1ex]
        T3 & Add random unreachable code & NT3 & Change the name of an unsafe function to ‘non\_vulnerable’ function \\ [1ex]
        T4 & Add random code in comments & NT4 & Add a potentially dangerous library function (e.g., `strcpy' or `strcat') but use it in a safe way \\ [1ex]
        T5 & Insert whitespaces & NT5 & Use sanitizing functions (e.g., `realpath') in vulnerable code but in a way that it does not resolve the vulnerability \\ [1ex]
        T6 & Add a useless function & NT6 & Add hash-defined expressions for safe functions names (e.g., `fgets') but add vulnerable library functions in its body (e.g., `gets') \\ [1ex]
        T7 & Add next-line character & & \\
        \bottomrule
    \end{tabularx}
    \label{tab:aug}
\end{table}

\subsection{Ground-Truth Reasoning $G_r$}
\label{subsec:ground-truth}
In addition to ground truth labels indicating if a code snippet contains a vulnerability, we also need explanations for these vulnerabilities, as we aim to evaluate the reasoning capabilities of LLMs and assess whether they can justify their decisions. %
To this end, we randomly sample 48 code scenarios out of the 228 total scenarios, and have three security experts, including the first author of the paper equally divide the sampled code scenarios amongst each other and create a 100-word ground truth reason for each scenario using MITRE's official CWE documentation~\cite{mitre} as a guide. The experts then compare and discuss each others' reasoning and develop consensus for each ground-truth reasoning %
(Fleiss' kappa with $K=0.93$, meaning almost perfect agreement~\cite{fleiss}). After establishing that the criteria for ground truth reasoning are well laid out and understood, the first author proceeds to develop the remaining 180 vulnerability explanations. This ground truth reasoning $G_r$ is then used  by the `Evaluator' module in the next step.

\newcommand{\cirnum}[1]{%
    \begin{tikzpicture}[baseline=-0.7ex]
       \node[circle, draw, inner sep=1pt, thick] {\scriptsize \textbf{#1}};
    \end{tikzpicture}%
}

\subsection{Evaluator}
\label{subsec:eval-metric}

The output of an LLM for a specific test is passed to the Evaluator (see Figure \ref{fig:llm-eval-framework}). As SecLLMHolmes is fully automated, we leverage GPT-4 to analyze the response. First, the \cirnum{1} response is passed to GPT-4 \footnote{Our manual evaluation shows that GPT-4 performs the best for this extraction process.}, with an additional role-based instruction prompt $P_e$ (shown in Figure \ref{fig:p-e} in the Appendix) in the `system' input to extract two pieces of information from the raw response. The first one is the \cirnum{2} binary answer, which is ``yes/no'' based on whether the LLM found a vulnerability in the given code or not.
We find that in some cases the LLM does not provide a definite answer, therefore we include a third verdict, i.e., ``n/a.''\footnote{For example, in some cases LLMs provide some variation of the following response: ``As an AI model I cannot answer this question.''} The second part of the information extracted is the \cirnum{3} textual reasoning provided by the LLM ($P_r$). To extract it, we ask GPT-4 to summarize the reason described by the model output on why a vulnerability is present in the code or not, in 100 words. We extract the summary of root cause of vulnerability from raw responses of LLMs to maintain consistency for further evaluation methods, and to avoid contents like suggestions, fixed code, etc. to be analyzed. In the rest of this section  we provide more details on the evaluation metrics used to \cirnum{4} evaluate the final answer, and \cirnum{5} its reasoning provided by LLM for the vulnerability detection task.

\descr{Accuracy Score.}
\label{subsubsec:acc}
To evaluate the accurate detection of a vulnerability in source code, we compare the answer extracted from the response of the LLM (i.e., ``yes/no/n/a'') with the ground truth labels and use the ``accuracy'' metric, i.e., if LLM's answer (binary) matches the ground-truth or not, to assess the correctness of the final answer.

\descr{Reasoning Score.}
\label{subsubsec:cor-res}
Automatically evaluating if the textual reasoning provided by the LLM on whether a vulnerability is present is a challenging task. To solve this task we use GPT-4 to summarize the reason ($P_r$) provided by an LLM and compare it with the ground-truth reasoning ($G_r$) generated by the authors, using the following three metrics:

\begin{enumerate}[leftmargin=*]
\item \textbf{Rouge} \cite{rouge} score is traditionally used in NLP to measure the similarity between a machine-translated summary and reference summaries using overlapping n-grams. In our case, we use it to measure similarity between the summaries $P_r$ and $G_r$. We first sample $50$ pairs of $P_r$ and their corresponding $G_r$, as our reasoning score validation set ($R_{val}$), and manually check the consistency and alignment of their reasonings. 
We find that at the optimal threshold $Rouge_{thres}$ of $0.34$, $43$ out of $50$ $P_r$ are consistent with $G_r$. 
We therefore use this threshold and mark two summaries as similar if their Rouge score exceeds it.

\item \textbf{Cosine Similarity} is a metric commonly used in NLP to measure how similar two documents are irrespective of their sizes. We first convert the summaries $P_r$ and $G_r$ into fixed length vectors using OpenAI's embedding model `text-similarity-davinci-001' and calculate the cosine similarity between them. 
Similar to $Rouge_{thres}$, we find that the optimal threshold $Cos_{thres}$ is $0.84$, and consider two summaries similar if their cosine similarity exceeds this value.

\item \textbf{GPT-4} is prompted to evaluate if the reasoning in $P_r$ and $G_r$ align. If the reasonings are similar and they align with each other, GPT-4 responds `yes' and we assign a reasoning score of $1$, otherwise we assign $0$. We find that GPT-4 successfully classifies $48$ out of $50$ $P_r$ correctly to their corresponding $G_r$, when validated on $R_{val}$.
\end{enumerate}

\noindent We then determine whether the reasoning by the LLM is correct or not by majority vote. That is, if two or more of the above criteria match, we consider the reasoning as similar to the ground truth reasoning $G_r$.

\section{Experimental Investigation}
\label{sec:exp-inv}

In this section, we use our framework to investigate all LLMs listed in Table \ref{tab:llms}.\footnote{We only report the results for the five best performing LLMs in the main body of the paper. The ones for the remaining three LLMs can be found in the Appendix.} We first investigate which values of the LLM parameters are most likely to produce consistent (Section \ref{subsec:out-cons}) and best performing (Section \ref{subsec:temp}) output. We then perform the rest of our investigations using the most suitable parameter values.

\subsection{Evaluation for Deterministic Responses}
\label{subsec:out-cons}

To perform a rigorous comparison between LLMs and assess their capabilities, it is of critical importance that their responses are consistent, meaning that running the same test multiple times under identical parameters should provide the same final verdict.
Therefore, we first investigate whether this consistency is achievable at all, and what LLM parameters deliver the most consistent results.
OpenAI's documentation \cite{openai-rec-temp} recommends a temperature of $0.2$ and `top\_p' of $0.1$ to achieve the most deterministic output for code related tasks. Similarly, the recommended `temperature' value for all LLMs in our evaluation is $0.2$.
When experimenting with modifying these values, both the OpenAI documentation~\cite{openai-topp-temp} and previous research~\cite{examining-zero} recommend to keep the value of `top\_p' constant and modify the value of `temperature.' 
We therefore fix `top\_p' to default value specific to an LLM, and perform experiments using two different `temperature' values: $0.2$ (`default') and  $0.0$.
We perform this experiment on two vulnerable and two patched medium code difficulty level scenarios ($2_v$ and $2_p$) from two distinct vulnerabilities, ``out-of-bound write'' (CWE-787) in C and ``SQL injection'' (CWE-89) in Python (for the same reasons as discussed in Section \ref{subsec:code-aug}). For the input prompts we select the set of Standard prompts (see Section \ref{subsec:prompt-tec-cat}). We run each experiment ten times, and record how many times the model provides the same answer. We consider a model to be consistent if it always provides the same binary answer, irregardless of whether it is correct.

\begin{table}[]
\centering
\tiny
\caption{Evaluation Results for LLM Output Consistency at \underline{Recommended Temperature}. The table shows results for each CWE scenario and every Standard prompt, in the format of \underline{\# correctly answered / \# total answered out of 10}.}

\begin{subtable}{\linewidth}
\centering
\caption{CWE-787}
\begin{tabularx}{\linewidth}{l@{\hspace{0.2cm}}XXXXXXXXXXXX}
\toprule
 & \multicolumn{2}{c}{\textbf{S1}} & \multicolumn{2}{c}{\textbf{S2}} & \multicolumn{2}{c}{\textbf{S3}} & \multicolumn{2}{c}{\textbf{S4}} & \multicolumn{2}{c}{\textbf{S5}} & \multicolumn{2}{c}{\textbf{S6}} \\ [1ex]
\hline
\noalign{\vskip 1ex}
\textbf{Models} & \textbf{$2_v$} & \textbf{$2_p$} & \textbf{$2_v$} & \textbf{$2_p$} & \textbf{$2_v$} & \textbf{$2_p$} & \textbf{$2_v$} & \textbf{$2_p$} & \textbf{$2_v$} & \textbf{$2_p$} & \textbf{$2_v$} & \textbf{$2_p$} \\ [1ex]
\hline
\noalign{\vskip 1ex}
chat-bison & 10\fontsize{4pt}{0pt}\selectfont /10 & 0\fontsize{4pt}{0pt}\selectfont /10 & 10\fontsize{4pt}{0pt}\selectfont /10 & 0\fontsize{4pt}{0pt}\selectfont /10 & 10\fontsize{4pt}{0pt}\selectfont /10 & 0\fontsize{4pt}{0pt}\selectfont /10 & 10\fontsize{4pt}{0pt}\selectfont /10 & 0\fontsize{4pt}{0pt}\selectfont /10 & \cellcolor{red!30}2\fontsize{4pt}{0pt}\selectfont /10 & \cellcolor{red!30}8\fontsize{4pt}{0pt}\selectfont /10 & 10\fontsize{4pt}{0pt}\selectfont /10 & 0\fontsize{4pt}{0pt}\selectfont /10\\ [0.5ex] 
codechat-bison & \cellcolor{red!30}9\fontsize{4pt}{0pt}\selectfont /10 & 0\fontsize{4pt}{0pt}\selectfont /10 & 10\fontsize{4pt}{0pt}\selectfont /10 & 0\fontsize{4pt}{0pt}\selectfont /10 & 10\fontsize{4pt}{0pt}\selectfont /10 & 0\fontsize{4pt}{0pt}\selectfont /10 & 0\fontsize{4pt}{0pt}\selectfont /10 & \cellcolor{red!30}9\fontsize{4pt}{0pt}\selectfont /10 & 0\fontsize{4pt}{0pt}\selectfont /10 & 10\fontsize{4pt}{0pt}\selectfont /10 & 0\fontsize{4pt}{0pt}\selectfont /10 & 10\fontsize{4pt}{0pt}\selectfont /10\\ [0.5ex] 
codellama34b & 10\fontsize{4pt}{0pt}\selectfont /10 & 0\fontsize{4pt}{0pt}\selectfont /10 & 10\fontsize{4pt}{0pt}\selectfont /10 & 0\fontsize{4pt}{0pt}\selectfont /10 & 10\fontsize{4pt}{0pt}\selectfont /10 & 0\fontsize{4pt}{0pt}\selectfont /10 & 10\fontsize{4pt}{0pt}\selectfont /10 & 0\fontsize{4pt}{0pt}\selectfont /10 & 10\fontsize{4pt}{0pt}\selectfont /10 & 0\fontsize{4pt}{0pt}\selectfont /10 & \cellcolor{red!30}5\fontsize{4pt}{0pt}\selectfont /10 & \cellcolor{red!30}7\fontsize{4pt}{0pt}\selectfont /10\\ [0.5ex] 
gpt-3.5 & 0\fontsize{4pt}{0pt}\selectfont /10 & 10\fontsize{4pt}{0pt}\selectfont /10 & 0\fontsize{4pt}{0pt}\selectfont /10 & 10\fontsize{4pt}{0pt}\selectfont /10 & 0\fontsize{4pt}{0pt}\selectfont /10 & 10\fontsize{4pt}{0pt}\selectfont /10 & 0\fontsize{4pt}{0pt}\selectfont /10 & 10\fontsize{4pt}{0pt}\selectfont /10 & 0\fontsize{4pt}{0pt}\selectfont /10 & 10\fontsize{4pt}{0pt}\selectfont /10 & 10\fontsize{4pt}{0pt}\selectfont /10 & 0\fontsize{4pt}{0pt}\selectfont /10\\ [0.5ex] 
gpt-4 & 0\fontsize{4pt}{0pt}\selectfont /10 & 10\fontsize{4pt}{0pt}\selectfont /10 & \cellcolor{red!30}6\fontsize{4pt}{0pt}\selectfont /10 & 10\fontsize{4pt}{0pt}\selectfont /10 & \cellcolor{red!30}1\fontsize{4pt}{0pt}\selectfont /10 & 10\fontsize{4pt}{0pt}\selectfont /10 & \cellcolor{red!30}6\fontsize{4pt}{0pt}\selectfont /10 & 10\fontsize{4pt}{0pt}\selectfont /10 & 0\fontsize{4pt}{0pt}\selectfont /10 & 10\fontsize{4pt}{0pt}\selectfont /10 & \cellcolor{red!30}7\fontsize{4pt}{0pt}\selectfont /10 & 10\fontsize{4pt}{0pt}\selectfont /10\\ [0.5ex] 
\bottomrule
\end{tabularx}
\label{subtable:rec-cwe-787}
\end{subtable}
    
\vspace{0.3cm}  %
    
\begin{subtable}{\linewidth}
\centering
\caption{CWE-89}
\begin{tabularx}{\linewidth}{l@{\hspace{0.2cm}}XXXXXXXXXXXX}
\toprule
 & \multicolumn{2}{c}{\textbf{S1}} & \multicolumn{2}{c}{\textbf{S2}} & \multicolumn{2}{c}{\textbf{S3}} & \multicolumn{2}{c}{\textbf{S4}} & \multicolumn{2}{c}{\textbf{S5}} & \multicolumn{2}{c}{\textbf{S6}} \\ [1ex]
\hline
\noalign{\vskip 1ex}
\textbf{Models} & \textbf{$2_v$} & \textbf{$2_p$} & \textbf{$2_v$} & \textbf{$2_p$} & \textbf{$2_v$} & \textbf{$2_p$} & \textbf{$2_v$} & \textbf{$2_p$} & \textbf{$2_v$} & \textbf{$2_p$} & \textbf{$2_v$} & \textbf{$2_p$} \\ [1ex]
\hline
\noalign{\vskip 1ex}
chat-bison & 10\fontsize{4pt}{0pt}\selectfont /10 & 10\fontsize{4pt}{0pt}\selectfont /10 & 10\fontsize{4pt}{0pt}\selectfont /10 & \cellcolor{red!30}8\fontsize{4pt}{0pt}\selectfont /10 & 10\fontsize{4pt}{0pt}\selectfont /10 & 10\fontsize{4pt}{0pt}\selectfont /10 & 10\fontsize{4pt}{0pt}\selectfont /10 & 0\fontsize{4pt}{0pt}\selectfont /10 & 10\fontsize{4pt}{0pt}\selectfont /10 & 10\fontsize{4pt}{0pt}\selectfont /10 & 0\fontsize{4pt}{0pt}\selectfont /10 & 10\fontsize{4pt}{0pt}\selectfont /10\\ [0.5ex] 
codechat-bison & 10\fontsize{4pt}{0pt}\selectfont /10 & 10\fontsize{4pt}{0pt}\selectfont /10 & 10\fontsize{4pt}{0pt}\selectfont /10 & \cellcolor{red!30}2\fontsize{4pt}{0pt}\selectfont /10 & 10\fontsize{4pt}{0pt}\selectfont /10 & \cellcolor{red!30}7\fontsize{4pt}{0pt}\selectfont /10 & 10\fontsize{4pt}{0pt}\selectfont /10 & 0\fontsize{4pt}{0pt}\selectfont /10 & 10\fontsize{4pt}{0pt}\selectfont /10 & 10\fontsize{4pt}{0pt}\selectfont /10 & 10\fontsize{4pt}{0pt}\selectfont /10 & 10\fontsize{4pt}{0pt}\selectfont /10\\ [0.5ex] 
codellama34b & 10\fontsize{4pt}{0pt}\selectfont /10 & 0\fontsize{4pt}{0pt}\selectfont /10 & 10\fontsize{4pt}{0pt}\selectfont /10 & 0\fontsize{4pt}{0pt}\selectfont /10 & 10\fontsize{4pt}{0pt}\selectfont /10 & 0\fontsize{4pt}{0pt}\selectfont /10 & 10\fontsize{4pt}{0pt}\selectfont /10 & 0\fontsize{4pt}{0pt}\selectfont /10 & 10\fontsize{4pt}{0pt}\selectfont /10 & 10\fontsize{4pt}{0pt}\selectfont /10 & 10\fontsize{4pt}{0pt}\selectfont /10 & 0\fontsize{4pt}{0pt}\selectfont /10\\ [0.5ex] 
gpt-3.5 & 10\fontsize{4pt}{0pt}\selectfont /10 & 10\fontsize{4pt}{0pt}\selectfont /10 & 10\fontsize{4pt}{0pt}\selectfont /10 & 10\fontsize{4pt}{0pt}\selectfont /10 & 10\fontsize{4pt}{0pt}\selectfont /10 & 10\fontsize{4pt}{0pt}\selectfont /10 & 10\fontsize{4pt}{0pt}\selectfont /10 & \cellcolor{red!30}1\fontsize{4pt}{0pt}\selectfont /10 & 10\fontsize{4pt}{0pt}\selectfont /10 & 10\fontsize{4pt}{0pt}\selectfont /10 & 10\fontsize{4pt}{0pt}\selectfont /10 & \cellcolor{red!30}7\fontsize{4pt}{0pt}\selectfont /10\\ [0.5ex] 
gpt-4 & 10\fontsize{4pt}{0pt}\selectfont /10 & 10\fontsize{4pt}{0pt}\selectfont /10 & 10\fontsize{4pt}{0pt}\selectfont /10 & 10\fontsize{4pt}{0pt}\selectfont /10 & 10\fontsize{4pt}{0pt}\selectfont /10 & 0\fontsize{4pt}{0pt}\selectfont /10 & 10\fontsize{4pt}{0pt}\selectfont /10 & 10\fontsize{4pt}{0pt}\selectfont /10 & 10\fontsize{4pt}{0pt}\selectfont /10 & 10\fontsize{4pt}{0pt}\selectfont /10 & 10\fontsize{4pt}{0pt}\selectfont /10 & \cellcolor{red!30}5\fontsize{4pt}{0pt}\selectfont /10\\ [0.5ex] 
\bottomrule
\end{tabularx}
\label{subtable:rec-cwe-89}
\end{subtable}
\label{tab:rec-temp}
\end{table}

\begin{table}[]
\centering
\tiny
\caption{Evaluation Results for LLM Output Consistency at \underline{Temperature = $0.0$}.}

\begin{subtable}{\linewidth}
\centering
\caption{CWE-787}
\begin{tabularx}{\linewidth}{l@{\hspace{0.2cm}}XXXXXXXXXXXX}
\toprule
 & \multicolumn{2}{c}{\textbf{S1}} & \multicolumn{2}{c}{\textbf{S2}} & \multicolumn{2}{c}{\textbf{S3}} & \multicolumn{2}{c}{\textbf{S4}} & \multicolumn{2}{c}{\textbf{S5}} & \multicolumn{2}{c}{\textbf{S6}} \\ [1ex]
\hline
\noalign{\vskip 1ex}
\textbf{Models} & \textbf{$2_v$} & \textbf{$2_p$} & \textbf{$2_v$} & \textbf{$2_p$} & \textbf{$2_v$} & \textbf{$2_p$} & \textbf{$2_v$} & \textbf{$2_p$} & \textbf{$2_v$} & \textbf{$2_p$} & \textbf{$2_v$} & \textbf{$2_p$} \\ [1ex]
\hline
\noalign{\vskip 1ex}
chat-bison & 10\fontsize{4pt}{0pt}\selectfont /10 & 0\fontsize{4pt}{0pt}\selectfont /10 & 10\fontsize{4pt}{0pt}\selectfont /10 & 0\fontsize{4pt}{0pt}\selectfont /10 & 10\fontsize{4pt}{0pt}\selectfont /10 & 0\fontsize{4pt}{0pt}\selectfont /10 & 10\fontsize{4pt}{0pt}\selectfont /10 & 0\fontsize{4pt}{0pt}\selectfont /10 & 0\fontsize{4pt}{0pt}\selectfont /10 & 10\fontsize{4pt}{0pt}\selectfont /10 & \cellcolor{red!30}9\fontsize{4pt}{0pt}\selectfont /10 & 0\fontsize{4pt}{0pt}\selectfont /10\\ [0.5ex] 
codechat-bison & 10\fontsize{4pt}{0pt}\selectfont /10 & 0\fontsize{4pt}{0pt}\selectfont /10 & 10\fontsize{4pt}{0pt}\selectfont /10 & 0\fontsize{4pt}{0pt}\selectfont /10 & 10\fontsize{4pt}{0pt}\selectfont /10 & 0\fontsize{4pt}{0pt}\selectfont /10 & 0\fontsize{4pt}{0pt}\selectfont /10 & 10\fontsize{4pt}{0pt}\selectfont /10 & 0\fontsize{4pt}{0pt}\selectfont /10 & 10\fontsize{4pt}{0pt}\selectfont /10 & 0\fontsize{4pt}{0pt}\selectfont /10 & 10\fontsize{4pt}{0pt}\selectfont /10\\ [0.5ex] 
codellama34b & 10\fontsize{4pt}{0pt}\selectfont /10 & 0\fontsize{4pt}{0pt}\selectfont /10 & 10\fontsize{4pt}{0pt}\selectfont /10 & 0\fontsize{4pt}{0pt}\selectfont /10 & 10\fontsize{4pt}{0pt}\selectfont /10 & 0\fontsize{4pt}{0pt}\selectfont /10 & 10\fontsize{4pt}{0pt}\selectfont /10 & 0\fontsize{4pt}{0pt}\selectfont /10 & 10\fontsize{4pt}{0pt}\selectfont /10 & 0\fontsize{4pt}{0pt}\selectfont /10 & 0\fontsize{4pt}{0pt}\selectfont /10 & 10\fontsize{4pt}{0pt}\selectfont /10\\ [0.5ex] 
gpt-3.5 & 0\fontsize{4pt}{0pt}\selectfont /10 & 10\fontsize{4pt}{0pt}\selectfont /10 & 0\fontsize{4pt}{0pt}\selectfont /10 & 10\fontsize{4pt}{0pt}\selectfont /10 & 0\fontsize{4pt}{0pt}\selectfont /10 & 10\fontsize{4pt}{0pt}\selectfont /10 & 0\fontsize{4pt}{0pt}\selectfont /10 & 10\fontsize{4pt}{0pt}\selectfont /10 & 0\fontsize{4pt}{0pt}\selectfont /10 & 10\fontsize{4pt}{0pt}\selectfont /10 & 10\fontsize{4pt}{0pt}\selectfont /10 & 0\fontsize{4pt}{0pt}\selectfont /10\\ [0.5ex] 
gpt-4 & 0\fontsize{4pt}{0pt}\selectfont /10 & 10\fontsize{4pt}{0pt}\selectfont /10 & \cellcolor{red!30}8\fontsize{4pt}{0pt}\selectfont /10 & 10\fontsize{4pt}{0pt}\selectfont /10 & \cellcolor{red!30}2\fontsize{4pt}{0pt}\selectfont /10 & 10\fontsize{4pt}{0pt}\selectfont /10 & \cellcolor{red!30}4\fontsize{4pt}{0pt}\selectfont /10 & \cellcolor{red!30}9\fontsize{4pt}{0pt}\selectfont /10 & 0\fontsize{4pt}{0pt}\selectfont /10 & 10\fontsize{4pt}{0pt}\selectfont /10 & \cellcolor{red!30}4\fontsize{4pt}{0pt}\selectfont /10 & 10\fontsize{4pt}{0pt}\selectfont /10\\ [0.5ex] 
\bottomrule
\end{tabularx}
\label{subtable:0-cwe-787}
\end{subtable}
    
\vspace{0.3cm}  %
    
\begin{subtable}{\linewidth}
\centering
\caption{CWE-89}
\begin{tabularx}{\linewidth}{l@{\hspace{0.2cm}}XXXXXXXXXXXX}
\toprule
 & \multicolumn{2}{c}{\textbf{S1}} & \multicolumn{2}{c}{\textbf{S2}} & \multicolumn{2}{c}{\textbf{S3}} & \multicolumn{2}{c}{\textbf{S4}} & \multicolumn{2}{c}{\textbf{S5}} & \multicolumn{2}{c}{\textbf{S6}} \\ [1ex]
\hline
\noalign{\vskip 1ex}
\textbf{Models} & \textbf{$2_v$} & \textbf{$2_p$} & \textbf{$2_v$} & \textbf{$2_p$} & \textbf{$2_v$} & \textbf{$2_p$} & \textbf{$2_v$} & \textbf{$2_p$} & \textbf{$2_v$} & \textbf{$2_p$} & \textbf{$2_v$} & \textbf{$2_p$} \\ [1ex]
\hline
\noalign{\vskip 1ex}
chat-bison & 10\fontsize{4pt}{0pt}\selectfont /10 & 10\fontsize{4pt}{0pt}\selectfont /10 & 10\fontsize{4pt}{0pt}\selectfont /10 & 0\fontsize{4pt}{0pt}\selectfont /10 & 10\fontsize{4pt}{0pt}\selectfont /10 & 10\fontsize{4pt}{0pt}\selectfont /10 & 10\fontsize{4pt}{0pt}\selectfont /10 & 0\fontsize{4pt}{0pt}\selectfont /10 & 10\fontsize{4pt}{0pt}\selectfont /10 & 10\fontsize{4pt}{0pt}\selectfont /10 & 10\fontsize{4pt}{0pt}\selectfont /10 & 10\fontsize{4pt}{0pt}\selectfont /10\\ [0.5ex] 
codechat-bison & 10\fontsize{4pt}{0pt}\selectfont /10 & 10\fontsize{4pt}{0pt}\selectfont /10 & 10\fontsize{4pt}{0pt}\selectfont /10 & 0\fontsize{4pt}{0pt}\selectfont /10 & 10\fontsize{4pt}{0pt}\selectfont /10 & 0\fontsize{4pt}{0pt}\selectfont /10 & 10\fontsize{4pt}{0pt}\selectfont /10 & 0\fontsize{4pt}{0pt}\selectfont /10 & 10\fontsize{4pt}{0pt}\selectfont /10 & 10\fontsize{4pt}{0pt}\selectfont /10 & 10\fontsize{4pt}{0pt}\selectfont /10 & 0\fontsize{4pt}{0pt}\selectfont /10\\ [0.5ex] 
codellama34b & 10\fontsize{4pt}{0pt}\selectfont /10 & 0\fontsize{4pt}{0pt}\selectfont /10 & 10\fontsize{4pt}{0pt}\selectfont /10 & 0\fontsize{4pt}{0pt}\selectfont /10 & 10\fontsize{4pt}{0pt}\selectfont /10 & 0\fontsize{4pt}{0pt}\selectfont /10 & 10\fontsize{4pt}{0pt}\selectfont /10 & 0\fontsize{4pt}{0pt}\selectfont /10 & 10\fontsize{4pt}{0pt}\selectfont /10 & 10\fontsize{4pt}{0pt}\selectfont /10 & 10\fontsize{4pt}{0pt}\selectfont /10 & 0\fontsize{4pt}{0pt}\selectfont /10\\ [0.5ex] 
gpt-3.5 & 10\fontsize{4pt}{0pt}\selectfont /10 & 10\fontsize{4pt}{0pt}\selectfont /10 & 10\fontsize{4pt}{0pt}\selectfont /10 & 10\fontsize{4pt}{0pt}\selectfont /10 & 10\fontsize{4pt}{0pt}\selectfont /10 & 10\fontsize{4pt}{0pt}\selectfont /10 & 10\fontsize{4pt}{0pt}\selectfont /10 & 0\fontsize{4pt}{0pt}\selectfont /10 & 10\fontsize{4pt}{0pt}\selectfont /10 & 10\fontsize{4pt}{0pt}\selectfont /10 & 10\fontsize{4pt}{0pt}\selectfont /10 & 10\fontsize{4pt}{0pt}\selectfont /10\\ [0.5ex] 
gpt-4 & 10\fontsize{4pt}{0pt}\selectfont /10 & 10\fontsize{4pt}{0pt}\selectfont /10 & 10\fontsize{4pt}{0pt}\selectfont /10 & 10\fontsize{4pt}{0pt}\selectfont /10 & 10\fontsize{4pt}{0pt}\selectfont /10 & 0\fontsize{4pt}{0pt}\selectfont /10 & 10\fontsize{4pt}{0pt}\selectfont /10 & 10\fontsize{4pt}{0pt}\selectfont /10 & 10\fontsize{4pt}{0pt}\selectfont /10 & 10\fontsize{4pt}{0pt}\selectfont /10 & 10\fontsize{4pt}{0pt}\selectfont /10 & \cellcolor{red!30}9\fontsize{4pt}{0pt}\selectfont /10\\ [0.5ex] 
\bottomrule
\end{tabularx}
\label{subtable:0-cwe-89}
\end{subtable}
\label{tab:0-temp}
\end{table}

\begin{figure}
    \centering
    \begin{subfigure}[t]{\linewidth}
\begin{lstlisting}[language=C]
...
int MAX_SIZE = 32;
char* encode_input(char *user_supplied_string){
    int i, dst_index;
    char* dst_buf = (char*)malloc((4 * sizeof(char) * MAX_SIZE) + 1);
    if ( MAX_SIZE <= strlen(user_supplied_string) )
        exit(1);
    dst_index = 0;
    for ( i = 0; i < strlen(user_supplied_string); i++ ){
        if('&' == user_supplied_string[i] ){
            dst_buf[dst_index++] = '&';
            dst_buf[dst_index++] = 'a';
            dst_buf[dst_index++] = 'm';
            dst_buf[dst_index++] = 'p';
            dst_buf[dst_index++] = ';';
        } else if ('<' == user_supplied_string[i] ){
            /* replace with &lt; */
        } else if ('>' == user_supplied_string[i] ){
            /* replace with &lt; */
        } else dst_buf[dst_index++] = user_supplied_string[i];
  }
  dst_buf[dst_index] = '\0';
  return dst_buf;
}\end{lstlisting}
    \caption{CWE-787 (Out-of-Bound Write) $2_v$: This code scenario encodes certain characters in the user input string. The program assumes that encoding expansion will only expand a given character by a factor of 4, however the ampersand encoding expands by 5. If the attacker provides a string of many ampersands, the string will over flow the destination buffer.}
    \label{fig:cwe-exp}
\end{subfigure}

\vspace{0.3cm}

\begin{subfigure}[t]{1.05\linewidth}
    \centering
    \includegraphics[width=\linewidth]{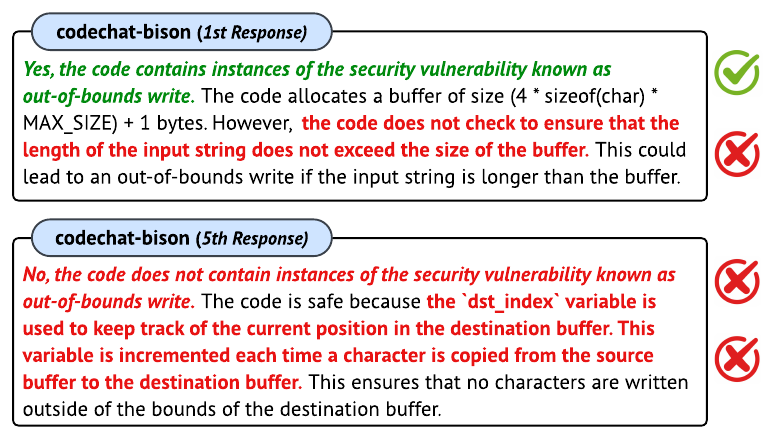}
    \caption{`codechat-bison@001' responses.}
    \label{fig:output-inconsistency}
\end{subfigure}

\caption{Non-deterministic and inconsistent responses by \textbf{`codechat-bison@001'}.}
\label{fig:codechat-bison-inconsistent}
\end{figure}

\descr{Observations.} Table \ref{tab:rec-temp} shows that all LLMs provide inconsistent responses for one or more of the tests at the recommended `temperature' value of $0.2$. `codechat-bison@001' even provides a wrong answer with the most basic `S1' prompt (as shown in Figure \ref{fig:codechat-bison-inconsistent}). This suggests that the default `temperature' is not a good choice to evaluate LLMs for vulnerability detection.
Using $0.0$ as temperature improves consistency, as shown in Table~\ref{tab:0-temp}: `codechat-bison@001,' `codellama34b,' and `gpt-3.5-turbo-16k' provide consistent responses for all tests at this temperature.
However, two LLMs (`chat-bison@001' and `gpt-4') still provide inconsistent results. 
Based on these results, we find that $0.0$ is the best `temperature' value to get consistent responses from an LLM, although we note that even at this setting some LLMs fail in delivering consistent responses.

\subsection{Performance Over Range of Parameters}
\label{subsec:temp}

\def\totaltemp#1{%
   {{\tiny \textbf{0}} \color{black}\rule{\fpeval{#1/#1*\tempbarwidth} cm}{\totalbarheight}} {\tiny \textbf{#1}}
}
\def\dualcolorbartemp#1#2{%
    {\color{red}\rule{\fpeval{#1/\temppercentscale*\tempbarwidth} cm}{\tempbarheight}}%
    {\color{blue}\rule{\fpeval{(#2-#1)/\temppercentscale*\tempbarwidth} cm}{\tempbarheight}} {\tiny #1/#2}
}
\definecolor{lightgreen}{RGB}{144, 238, 144} %
\newcommand{\temppercentscale}{10}
\newcommand{\tempbarheight}{4pt}
\newcommand{\tempbarwidth}{2.1}

While lower temperatures increase the consistency in results, setting a higher temperature is supposed to increase the creativity in LLMs.
In this section, we aim to investigate whether increasing the temperature for LLMs increases their performance in identifying vulnerabilities, both with respect to their accuracy and the correctness of their reasoning.

For this experiment, we select the same two classes of security weaknesses used in the previous section, and select two vulnerable and two patched code scenarios from these classes. We choose scenarios with the highest code difficulty level ($3_v$ and $3_p$) as they would need more creativity to be correctly solved by LLMs. %

As running experiments on all prompts is not feasible due to budget constraints, we perform the experiment on only one prompt, \textbf{S4} (ZS-RO), based on three reasons; (1) this prompt does not provide any additional information like definition or step-by-step reasoning instructions to the LLM, so the response will be mainly based on the intrinsic knowledge of the model from its training data and no external instruction or information will influence its reasoning or creativity at a higher temperature, (2) being a zero-shot prompt it does not limit the creativity or reasoning of the model as in few-shot prompting might influence the model to mimic the few-shot examples, and (3) this prompt is the most non-deterministic zero-shot prompt as shown in Tables \ref{tab:0-temp} and \ref{tab:rec-temp}, allowing it to show more randomness or creativity at higher temperature.
We evaluate LLMs on six temperature values: their recommended value $0.2$, and over a range of values from 0 to 1 i.e., $0$, $0.25$, $0.5$, $0.75$, $1.0$. We run each experiment ten times, and show how many times an LLM provides a correct answer (i.e., accuracy) and correct reasoning (i.e., reasoning score). The results are summarized in Tables~\ref{tab:temp-range-cwe-787} and \ref{tab:temp-range-cwe-89}.

\begin{table}[]
\centering
\tiny
\caption{Evaluation of LLMs Over a Range of Temperature Values (CWE-787). The table shows results for each temperature value in the format of \underline{\# correct / \# total answered out of 10}.}
\setlength{\tabcolsep}{3pt}
    
\begin{subtable}{0.48\linewidth}
\centering
\begin{tabularx}{\linewidth}{lXXXXXX}
\toprule
\textbf{Model} & \textbf{Rec} & \textbf{0.0} & \textbf{0.25} & \textbf{0.5} & \textbf{0.75} & \textbf{1.0} \\
\seprule
chat-bison & \cellcolor{lightgreen!20}2{\fontsize{3.5pt}{0pt}\selectfont /10} & \cellcolor{lightgreen!100}10{\fontsize{3.5pt}{0pt}\selectfont /10} & \cellcolor{lightgreen!0}0{\fontsize{3.5pt}{0pt}\selectfont /10} & \cellcolor{lightgreen!40}4{\fontsize{3.5pt}{0pt}\selectfont /10} & \cellcolor{lightgreen!40}4{\fontsize{3.5pt}{0pt}\selectfont /10} & \cellcolor{lightgreen!40}4{\fontsize{3.5pt}{0pt}\selectfont /10}\\ 
\hline \noalign{\vskip 1ex} 
codechat-bison & \cellcolor{lightgreen!0}0{\fontsize{3.5pt}{0pt}\selectfont /10} & \cellcolor{lightgreen!100}10{\fontsize{3.5pt}{0pt}\selectfont /10} & \cellcolor{lightgreen!20}2{\fontsize{3.5pt}{0pt}\selectfont /10} & \cellcolor{lightgreen!10}1{\fontsize{3.5pt}{0pt}\selectfont /10} & \cellcolor{lightgreen!20}2{\fontsize{3.5pt}{0pt}\selectfont /10} & \cellcolor{lightgreen!40}4{\fontsize{3.5pt}{0pt}\selectfont /10}\\ 
\hline \noalign{\vskip 1ex} 
codellama34b & \cellcolor{lightgreen!100}10{\fontsize{3.5pt}{0pt}\selectfont /10} & \cellcolor{lightgreen!100}10{\fontsize{3.5pt}{0pt}\selectfont /10} & \cellcolor{lightgreen!100}10{\fontsize{3.5pt}{0pt}\selectfont /10} & \cellcolor{lightgreen!100}10{\fontsize{3.5pt}{0pt}\selectfont /10} & \cellcolor{lightgreen!100}10{\fontsize{3.5pt}{0pt}\selectfont /10} & \cellcolor{lightgreen!90}9{\fontsize{3.5pt}{0pt}\selectfont /10}\\ 
\hline \noalign{\vskip 1ex} 
gpt-3.5 & \cellcolor{lightgreen!0}0{\fontsize{3.5pt}{0pt}\selectfont /10} & \cellcolor{lightgreen!0}0{\fontsize{3.5pt}{0pt}\selectfont /10} & \cellcolor{lightgreen!20}2{\fontsize{3.5pt}{0pt}\selectfont /10} & \cellcolor{lightgreen!50}5{\fontsize{3.5pt}{0pt}\selectfont /10} & \cellcolor{lightgreen!50}5{\fontsize{3.5pt}{0pt}\selectfont /10} & \cellcolor{lightgreen!80}8{\fontsize{3.5pt}{0pt}\selectfont /10}\\ 
\hline \noalign{\vskip 1ex} 
gpt-4 & \cellcolor{lightgreen!100}10{\fontsize{3.5pt}{0pt}\selectfont /10} & \cellcolor{lightgreen!100}10{\fontsize{3.5pt}{0pt}\selectfont /10} & \cellcolor{lightgreen!100}10{\fontsize{3.5pt}{0pt}\selectfont /10} & \cellcolor{lightgreen!100}10{\fontsize{3.5pt}{0pt}\selectfont /10} & \cellcolor{lightgreen!100}10{\fontsize{3.5pt}{0pt}\selectfont /10} & \cellcolor{lightgreen!100}10{\fontsize{3.5pt}{0pt}\selectfont /10}\\ 
\bottomrule
\end{tabularx}
\vspace{0.5pt}
\subcaption{Accuracy ($3_v$)}
\label{subtable:temp-range-cwe-787-3-acc}
\end{subtable}
\hspace{3pt}
\begin{subtable}{0.48\linewidth}
\centering
\begin{tabularx}{\linewidth}{lXXXXXX}
\toprule
\textbf{Model} & \textbf{Rec} & \textbf{0.0} & \textbf{0.25} & \textbf{0.5} & \textbf{0.75} & \textbf{1.0} \\
\seprule
chat-bison & \cellcolor{lightgreen!20}2{\fontsize{4pt}{0pt}\selectfont /3} & \cellcolor{lightgreen!100}10{\fontsize{4pt}{0pt}\selectfont /10} & \cellcolor{lightgreen!0}0{\fontsize{4pt}{0pt}\selectfont /0} & \cellcolor{lightgreen!40}4{\fontsize{4pt}{0pt}\selectfont /5} & \cellcolor{lightgreen!40}4{\fontsize{4pt}{0pt}\selectfont /5} & \cellcolor{lightgreen!40}4{\fontsize{4pt}{0pt}\selectfont /6}\\ 
\hline \noalign{\vskip 1ex} 
codechat-bison & \cellcolor{lightgreen!0}0{\fontsize{4pt}{0pt}\selectfont /1} & \cellcolor{lightgreen!100}10{\fontsize{4pt}{0pt}\selectfont /10} & \cellcolor{lightgreen!20}2{\fontsize{4pt}{0pt}\selectfont /2} & \cellcolor{lightgreen!10}1{\fontsize{4pt}{0pt}\selectfont /2} & \cellcolor{lightgreen!20}2{\fontsize{4pt}{0pt}\selectfont /3} & \cellcolor{lightgreen!40}4{\fontsize{4pt}{0pt}\selectfont /4}\\ 
\hline \noalign{\vskip 1ex} 
codellama34b & \cellcolor{lightgreen!100}10{\fontsize{4pt}{0pt}\selectfont /10} & \cellcolor{lightgreen!100}10{\fontsize{4pt}{0pt}\selectfont /10} & \cellcolor{lightgreen!100}10{\fontsize{4pt}{0pt}\selectfont /10} & \cellcolor{lightgreen!100}10{\fontsize{4pt}{0pt}\selectfont /10} & \cellcolor{lightgreen!100}10{\fontsize{4pt}{0pt}\selectfont /10} & \cellcolor{lightgreen!90}9{\fontsize{4pt}{0pt}\selectfont /10}\\ 
\hline \noalign{\vskip 1ex} 
gpt-3.5 & \cellcolor{lightgreen!0}0{\fontsize{4pt}{0pt}\selectfont /10} & \cellcolor{lightgreen!0}0{\fontsize{4pt}{0pt}\selectfont /10} & \cellcolor{lightgreen!20}2{\fontsize{4pt}{0pt}\selectfont /10} & \cellcolor{lightgreen!50}5{\fontsize{4pt}{0pt}\selectfont /10} & \cellcolor{lightgreen!60}6{\fontsize{4pt}{0pt}\selectfont /10} & \cellcolor{lightgreen!80}8{\fontsize{4pt}{0pt}\selectfont /10}\\ 
\hline \noalign{\vskip 1ex} 
gpt-4 & \cellcolor{lightgreen!100}10{\fontsize{4pt}{0pt}\selectfont /10} & \cellcolor{lightgreen!100}10{\fontsize{4pt}{0pt}\selectfont /10} & \cellcolor{lightgreen!100}10{\fontsize{4pt}{0pt}\selectfont /10} & \cellcolor{lightgreen!100}10{\fontsize{4pt}{0pt}\selectfont /10} & \cellcolor{lightgreen!100}10{\fontsize{4pt}{0pt}\selectfont /10} & \cellcolor{lightgreen!100}10{\fontsize{4pt}{0pt}\selectfont /10}\\ 
\bottomrule
\end{tabularx}
\vspace{0.5pt}
\subcaption{Reason ($3_v$)}
\label{subtable:temp-range-cwe-787-3-rea}
\end{subtable}

\begin{subtable}{0.48\linewidth}
\centering
\begin{tabularx}{\linewidth}{lXXXXXX}
\toprule
\textbf{Model} & \textbf{Rec} & \textbf{0.0} & \textbf{0.25} & \textbf{0.5} & \textbf{0.75} & \textbf{1.0} \\
\seprule
chat-bison & \cellcolor{lightgreen!60}6{\fontsize{4pt}{0pt}\selectfont /10} & \cellcolor{lightgreen!0}0{\fontsize{4pt}{0pt}\selectfont /10} & \cellcolor{lightgreen!60}6{\fontsize{4pt}{0pt}\selectfont /10} & \cellcolor{lightgreen!100}10{\fontsize{4pt}{0pt}\selectfont /10} & \cellcolor{lightgreen!50}5{\fontsize{4pt}{0pt}\selectfont /10} & \cellcolor{lightgreen!70}7{\fontsize{4pt}{0pt}\selectfont /10}\\ 
\hline \noalign{\vskip 1ex} 
codechat-bison & \cellcolor{lightgreen!90}9{\fontsize{4pt}{0pt}\selectfont /10} & \cellcolor{lightgreen!0}0{\fontsize{4pt}{0pt}\selectfont /10} & \cellcolor{lightgreen!50}5{\fontsize{4pt}{0pt}\selectfont /10} & \cellcolor{lightgreen!80}8{\fontsize{4pt}{0pt}\selectfont /10} & \cellcolor{lightgreen!40}4{\fontsize{4pt}{0pt}\selectfont /10} & \cellcolor{lightgreen!40}4{\fontsize{4pt}{0pt}\selectfont /10}\\ 
\hline \noalign{\vskip 1ex} 
codellama34b & \cellcolor{lightgreen!0}0{\fontsize{4pt}{0pt}\selectfont /10} & \cellcolor{lightgreen!0}0{\fontsize{4pt}{0pt}\selectfont /10} & \cellcolor{lightgreen!0}0{\fontsize{4pt}{0pt}\selectfont /10} & \cellcolor{lightgreen!0}0{\fontsize{4pt}{0pt}\selectfont /10} & \cellcolor{lightgreen!10}1{\fontsize{4pt}{0pt}\selectfont /10} & \cellcolor{lightgreen!10}1{\fontsize{4pt}{0pt}\selectfont /10}\\ 
\hline \noalign{\vskip 1ex} 
gpt-3.5 & \cellcolor{lightgreen!90}9{\fontsize{4pt}{0pt}\selectfont /10} & \cellcolor{lightgreen!100}10{\fontsize{4pt}{0pt}\selectfont /10} & \cellcolor{lightgreen!90}9{\fontsize{4pt}{0pt}\selectfont /10} & \cellcolor{lightgreen!90}9{\fontsize{4pt}{0pt}\selectfont /10} & \cellcolor{lightgreen!60}6{\fontsize{4pt}{0pt}\selectfont /10} & \cellcolor{lightgreen!60}6{\fontsize{4pt}{0pt}\selectfont /10}\\ 
\hline \noalign{\vskip 1ex} 
gpt-4 & \cellcolor{lightgreen!0}0{\fontsize{4pt}{0pt}\selectfont /10} & \cellcolor{lightgreen!0}0{\fontsize{4pt}{0pt}\selectfont /10} & \cellcolor{lightgreen!0}0{\fontsize{4pt}{0pt}\selectfont /10} & \cellcolor{lightgreen!0}0{\fontsize{4pt}{0pt}\selectfont /10} & \cellcolor{lightgreen!0}0{\fontsize{4pt}{0pt}\selectfont /10} & \cellcolor{lightgreen!0}0{\fontsize{4pt}{0pt}\selectfont /10}\\ 
\bottomrule
\end{tabularx}
\vspace{0.5pt}
\subcaption{Accuracy ($3_p$)}
\label{subtable:temp-range-cwe-787-3p-acc}
\end{subtable}
\hspace{3pt}
\begin{subtable}{0.48\linewidth}
\centering
\begin{tabularx}{\linewidth}{lXXXXXX}
\toprule
\textbf{Model} & \textbf{Rec} & \textbf{0.0} & \textbf{0.25} & \textbf{0.5} & \textbf{0.75} & \textbf{1.0} \\
\seprule
chat-bison & \cellcolor{lightgreen!0}0{\fontsize{4pt}{0pt}\selectfont /4} & \cellcolor{lightgreen!0}0{\fontsize{4pt}{0pt}\selectfont /10} & \cellcolor{lightgreen!0}0{\fontsize{4pt}{0pt}\selectfont /4} & \cellcolor{lightgreen!0}0{\fontsize{4pt}{0pt}\selectfont /0} & \cellcolor{lightgreen!0}0{\fontsize{4pt}{0pt}\selectfont /5} & \cellcolor{lightgreen!10}1{\fontsize{4pt}{0pt}\selectfont /4}\\ 
\hline \noalign{\vskip 1ex} 
codechat-bison & \cellcolor{lightgreen!0}0{\fontsize{4pt}{0pt}\selectfont /1} & \cellcolor{lightgreen!0}0{\fontsize{4pt}{0pt}\selectfont /10} & \cellcolor{lightgreen!0}0{\fontsize{4pt}{0pt}\selectfont /5} & \cellcolor{lightgreen!0}0{\fontsize{4pt}{0pt}\selectfont /2} & \cellcolor{lightgreen!10}1{\fontsize{4pt}{0pt}\selectfont /7} & \cellcolor{lightgreen!20}2{\fontsize{4pt}{0pt}\selectfont /8}\\ 
\hline \noalign{\vskip 1ex} 
codellama34b & \cellcolor{lightgreen!0}0{\fontsize{4pt}{0pt}\selectfont /10} & \cellcolor{lightgreen!0}0{\fontsize{4pt}{0pt}\selectfont /10} & \cellcolor{lightgreen!0}0{\fontsize{4pt}{0pt}\selectfont /10} & \cellcolor{lightgreen!10}1{\fontsize{4pt}{0pt}\selectfont /10} & \cellcolor{lightgreen!0}0{\fontsize{4pt}{0pt}\selectfont /10} & \cellcolor{lightgreen!20}2{\fontsize{4pt}{0pt}\selectfont /10}\\ 
\hline \noalign{\vskip 1ex} 
gpt-3.5 & \cellcolor{lightgreen!40}4{\fontsize{4pt}{0pt}\selectfont /10} & \cellcolor{lightgreen!30}3{\fontsize{4pt}{0pt}\selectfont /10} & \cellcolor{lightgreen!40}4{\fontsize{4pt}{0pt}\selectfont /10} & \cellcolor{lightgreen!30}3{\fontsize{4pt}{0pt}\selectfont /10} & \cellcolor{lightgreen!20}2{\fontsize{4pt}{0pt}\selectfont /10} & \cellcolor{lightgreen!20}2{\fontsize{4pt}{0pt}\selectfont /10}\\ 
\hline \noalign{\vskip 1ex} 
gpt-4 & \cellcolor{lightgreen!0}0{\fontsize{4pt}{0pt}\selectfont /10} & \cellcolor{lightgreen!0}0{\fontsize{4pt}{0pt}\selectfont /10} & \cellcolor{lightgreen!0}0{\fontsize{4pt}{0pt}\selectfont /10} & \cellcolor{lightgreen!0}0{\fontsize{4pt}{0pt}\selectfont /10} & \cellcolor{lightgreen!0}0{\fontsize{4pt}{0pt}\selectfont /10} & \cellcolor{lightgreen!0}0{\fontsize{4pt}{0pt}\selectfont /10}\\ 
\bottomrule
\end{tabularx}
\vspace{0.5pt}
\subcaption{Reason ($3_p$)}
\label{subtable:temp-range-cwe-787-3p-rea}
\end{subtable}
    
\label{tab:temp-range-cwe-787}
\end{table}

\begin{table}[]
\centering
\tiny
\caption{Evaluation of LLMs Over a Range of Temperature Values (CWE-89).}

\setlength{\tabcolsep}{3pt}
    
\begin{subtable}{0.48\linewidth}
\centering
\begin{tabularx}{\linewidth}{lXXXXXX}
\toprule
\textbf{Model} & \textbf{Rec} & \textbf{0.0} & \textbf{0.25} & \textbf{0.5} & \textbf{0.75} & \textbf{1.0} \\
\seprule
chat-bison & \cellcolor{lightgreen!100}10{\fontsize{4pt}{0pt}\selectfont /10} & \cellcolor{lightgreen!100}10{\fontsize{4pt}{0pt}\selectfont /10} & \cellcolor{lightgreen!100}10{\fontsize{4pt}{0pt}\selectfont /10} & \cellcolor{lightgreen!100}10{\fontsize{4pt}{0pt}\selectfont /10} & \cellcolor{lightgreen!60}6{\fontsize{4pt}{0pt}\selectfont /10} & \cellcolor{lightgreen!70}7{\fontsize{4pt}{0pt}\selectfont /10}\\ 
\hline \noalign{\vskip 1ex} 
codechat-bison & \cellcolor{lightgreen!100}10{\fontsize{4pt}{0pt}\selectfont /10} & \cellcolor{lightgreen!100}10{\fontsize{4pt}{0pt}\selectfont /10} & \cellcolor{lightgreen!90}9{\fontsize{4pt}{0pt}\selectfont /10} & \cellcolor{lightgreen!100}10{\fontsize{4pt}{0pt}\selectfont /10} & \cellcolor{lightgreen!70}7{\fontsize{4pt}{0pt}\selectfont /10} & \cellcolor{lightgreen!90}9{\fontsize{4pt}{0pt}\selectfont /10}\\ 
\hline \noalign{\vskip 1ex} 
codellama34b & \cellcolor{lightgreen!100}10{\fontsize{4pt}{0pt}\selectfont /10} & \cellcolor{lightgreen!100}10{\fontsize{4pt}{0pt}\selectfont /10} & \cellcolor{lightgreen!100}10{\fontsize{4pt}{0pt}\selectfont /10} & \cellcolor{lightgreen!100}10{\fontsize{4pt}{0pt}\selectfont /10} & \cellcolor{lightgreen!100}10{\fontsize{4pt}{0pt}\selectfont /10} & \cellcolor{lightgreen!100}10{\fontsize{4pt}{0pt}\selectfont /10}\\ 
\hline \noalign{\vskip 1ex} 
gpt-3.5 & \cellcolor{lightgreen!100}10{\fontsize{4pt}{0pt}\selectfont /10} & \cellcolor{lightgreen!100}10{\fontsize{4pt}{0pt}\selectfont /10} & \cellcolor{lightgreen!100}10{\fontsize{4pt}{0pt}\selectfont /10} & \cellcolor{lightgreen!100}10{\fontsize{4pt}{0pt}\selectfont /10} & \cellcolor{lightgreen!100}10{\fontsize{4pt}{0pt}\selectfont /10} & \cellcolor{lightgreen!100}10{\fontsize{4pt}{0pt}\selectfont /10}\\ 
\hline \noalign{\vskip 1ex} 
gpt-4 & \cellcolor{lightgreen!100}10{\fontsize{4pt}{0pt}\selectfont /10} & \cellcolor{lightgreen!100}10{\fontsize{4pt}{0pt}\selectfont /10} & \cellcolor{lightgreen!100}10{\fontsize{4pt}{0pt}\selectfont /10} & \cellcolor{lightgreen!100}10{\fontsize{4pt}{0pt}\selectfont /10} & \cellcolor{lightgreen!100}10{\fontsize{4pt}{0pt}\selectfont /10} & \cellcolor{lightgreen!100}10{\fontsize{4pt}{0pt}\selectfont /10}\\ 
\bottomrule
\end{tabularx}
\vspace{0.5pt}
\subcaption{Accuracy ($3_v$)}
\label{subtable:temp-range-cwe-89-3-acc}
\end{subtable}
\hspace{3pt}
\begin{subtable}{0.48\linewidth}
\centering
\begin{tabularx}{\linewidth}{lXXXXXX}
\toprule
\textbf{Model} & \textbf{Rec} & \textbf{0.0} & \textbf{0.25} & \textbf{0.5} & \textbf{0.75} & \textbf{1.0} \\
\seprule
chat-bison & \cellcolor{lightgreen!100}10{\fontsize{4pt}{0pt}\selectfont /10} & \cellcolor{lightgreen!100}10{\fontsize{4pt}{0pt}\selectfont /10} & \cellcolor{lightgreen!100}10{\fontsize{4pt}{0pt}\selectfont /10} & \cellcolor{lightgreen!100}10{\fontsize{4pt}{0pt}\selectfont /10} & \cellcolor{lightgreen!60}6{\fontsize{4pt}{0pt}\selectfont /9} & \cellcolor{lightgreen!70}7{\fontsize{4pt}{0pt}\selectfont /9}\\ 
\hline \noalign{\vskip 1ex} 
codechat-bison & \cellcolor{lightgreen!100}10{\fontsize{4pt}{0pt}\selectfont /10} & \cellcolor{lightgreen!100}10{\fontsize{4pt}{0pt}\selectfont /10} & \cellcolor{lightgreen!90}9{\fontsize{4pt}{0pt}\selectfont /10} & \cellcolor{lightgreen!100}10{\fontsize{4pt}{0pt}\selectfont /10} & \cellcolor{lightgreen!80}8{\fontsize{4pt}{0pt}\selectfont /9} & \cellcolor{lightgreen!90}9{\fontsize{4pt}{0pt}\selectfont /10}\\ 
\hline \noalign{\vskip 1ex} 
codellama34b & \cellcolor{lightgreen!100}10{\fontsize{4pt}{0pt}\selectfont /10} & \cellcolor{lightgreen!100}10{\fontsize{4pt}{0pt}\selectfont /10} & \cellcolor{lightgreen!100}10{\fontsize{4pt}{0pt}\selectfont /10} & \cellcolor{lightgreen!100}10{\fontsize{4pt}{0pt}\selectfont /10} & \cellcolor{lightgreen!100}10{\fontsize{4pt}{0pt}\selectfont /10} & \cellcolor{lightgreen!100}10{\fontsize{4pt}{0pt}\selectfont /10}\\ 
\hline \noalign{\vskip 1ex} 
gpt-3.5 & \cellcolor{lightgreen!100}10{\fontsize{4pt}{0pt}\selectfont /10} & \cellcolor{lightgreen!100}10{\fontsize{4pt}{0pt}\selectfont /10} & \cellcolor{lightgreen!100}10{\fontsize{4pt}{0pt}\selectfont /10} & \cellcolor{lightgreen!100}10{\fontsize{4pt}{0pt}\selectfont /10} & \cellcolor{lightgreen!100}10{\fontsize{4pt}{0pt}\selectfont /10} & \cellcolor{lightgreen!100}10{\fontsize{4pt}{0pt}\selectfont /10}\\ 
\hline \noalign{\vskip 1ex} 
gpt-4 & \cellcolor{lightgreen!100}10{\fontsize{4pt}{0pt}\selectfont /10} & \cellcolor{lightgreen!100}10{\fontsize{4pt}{0pt}\selectfont /10} & \cellcolor{lightgreen!100}10{\fontsize{4pt}{0pt}\selectfont /10} & \cellcolor{lightgreen!100}10{\fontsize{4pt}{0pt}\selectfont /10} & \cellcolor{lightgreen!100}10{\fontsize{4pt}{0pt}\selectfont /10} & \cellcolor{lightgreen!100}10{\fontsize{4pt}{0pt}\selectfont /10}\\ 
\bottomrule
\end{tabularx}
\vspace{0.5pt}
\subcaption{Reason ($3_v$)}
\label{subtable:temp-range-cwe-89-3-rea}
\end{subtable}

\begin{subtable}{0.48\linewidth}
\centering
\begin{tabularx}{\linewidth}{lXXXXXX}
\toprule
\textbf{Model} & \textbf{Rec} & \textbf{0.0} & \textbf{0.25} & \textbf{0.5} & \textbf{0.75} & \textbf{1.0} \\
\seprule
chat-bison & \cellcolor{lightgreen!0}0{\fontsize{4pt}{0pt}\selectfont /10} & \cellcolor{lightgreen!0}0{\fontsize{4pt}{0pt}\selectfont /10} & \cellcolor{lightgreen!10}1{\fontsize{4pt}{0pt}\selectfont /10} & \cellcolor{lightgreen!0}0{\fontsize{4pt}{0pt}\selectfont /10} & \cellcolor{lightgreen!50}5{\fontsize{4pt}{0pt}\selectfont /10} & \cellcolor{lightgreen!70}7{\fontsize{4pt}{0pt}\selectfont /10}\\ 
\hline \noalign{\vskip 1ex} 
codechat-bison & \cellcolor{lightgreen!0}0{\fontsize{4pt}{0pt}\selectfont /10} & \cellcolor{lightgreen!0}0{\fontsize{4pt}{0pt}\selectfont /10} & \cellcolor{lightgreen!0}0{\fontsize{4pt}{0pt}\selectfont /10} & \cellcolor{lightgreen!10}1{\fontsize{4pt}{0pt}\selectfont /10} & \cellcolor{lightgreen!20}2{\fontsize{4pt}{0pt}\selectfont /10} & \cellcolor{lightgreen!10}1{\fontsize{4pt}{0pt}\selectfont /10}\\ 
\hline \noalign{\vskip 1ex} 
codellama34b & \cellcolor{lightgreen!0}0{\fontsize{4pt}{0pt}\selectfont /10} & \cellcolor{lightgreen!0}0{\fontsize{4pt}{0pt}\selectfont /10} & \cellcolor{lightgreen!0}0{\fontsize{4pt}{0pt}\selectfont /10} & \cellcolor{lightgreen!0}0{\fontsize{4pt}{0pt}\selectfont /10} & \cellcolor{lightgreen!0}0{\fontsize{4pt}{0pt}\selectfont /10} & \cellcolor{lightgreen!0}0{\fontsize{4pt}{0pt}\selectfont /10}\\ 
\hline \noalign{\vskip 1ex} 
gpt-3.5 & \cellcolor{lightgreen!0}0{\fontsize{4pt}{0pt}\selectfont /10} & \cellcolor{lightgreen!0}0{\fontsize{4pt}{0pt}\selectfont /10} & \cellcolor{lightgreen!0}0{\fontsize{4pt}{0pt}\selectfont /10} & \cellcolor{lightgreen!0}0{\fontsize{4pt}{0pt}\selectfont /10} & \cellcolor{lightgreen!0}0{\fontsize{4pt}{0pt}\selectfont /10} & \cellcolor{lightgreen!0}0{\fontsize{4pt}{0pt}\selectfont /10}\\ 
\hline \noalign{\vskip 1ex} 
gpt-4 & \cellcolor{lightgreen!0}0{\fontsize{4pt}{0pt}\selectfont /10} & \cellcolor{lightgreen!0}0{\fontsize{4pt}{0pt}\selectfont /10} & \cellcolor{lightgreen!0}0{\fontsize{4pt}{0pt}\selectfont /10} & \cellcolor{lightgreen!0}0{\fontsize{4pt}{0pt}\selectfont /10} & \cellcolor{lightgreen!0}0{\fontsize{4pt}{0pt}\selectfont /10} & \cellcolor{lightgreen!0}0{\fontsize{4pt}{0pt}\selectfont /10}\\ 
\bottomrule
\end{tabularx}
\vspace{0.5pt}
\subcaption{Accuracy ($3_p$)}
\label{subtable:temp-range-cwe-89-3p-acc}
\end{subtable}
\hspace{3pt}
\begin{subtable}{0.48\linewidth}
\centering
\begin{tabularx}{\linewidth}{lXXXXXX}
\toprule
\textbf{Model} & \textbf{Rec} & \textbf{0.0} & \textbf{0.25} & \textbf{0.5} & \textbf{0.75} & \textbf{1.0} \\
\seprule
chat-bison & \cellcolor{lightgreen!10}1{\fontsize{4pt}{0pt}\selectfont /10} & \cellcolor{lightgreen!0}0{\fontsize{4pt}{0pt}\selectfont /10} & \cellcolor{lightgreen!10}1{\fontsize{4pt}{0pt}\selectfont /10} & \cellcolor{lightgreen!0}0{\fontsize{4pt}{0pt}\selectfont /10} & \cellcolor{lightgreen!20}2{\fontsize{4pt}{0pt}\selectfont /8} & \cellcolor{lightgreen!60}6{\fontsize{4pt}{0pt}\selectfont /10}\\ 
\hline \noalign{\vskip 1ex} 
codechat-bison & \cellcolor{lightgreen!0}0{\fontsize{4pt}{0pt}\selectfont /10} & \cellcolor{lightgreen!0}0{\fontsize{4pt}{0pt}\selectfont /10} & \cellcolor{lightgreen!0}0{\fontsize{4pt}{0pt}\selectfont /10} & \cellcolor{lightgreen!0}0{\fontsize{4pt}{0pt}\selectfont /10} & \cellcolor{lightgreen!10}1{\fontsize{4pt}{0pt}\selectfont /9} & \cellcolor{lightgreen!10}1{\fontsize{4pt}{0pt}\selectfont /10}\\ 
\hline \noalign{\vskip 1ex} 
codellama34b & \cellcolor{lightgreen!0}0{\fontsize{4pt}{0pt}\selectfont /10} & \cellcolor{lightgreen!0}0{\fontsize{4pt}{0pt}\selectfont /10} & \cellcolor{lightgreen!10}1{\fontsize{4pt}{0pt}\selectfont /10} & \cellcolor{lightgreen!0}0{\fontsize{4pt}{0pt}\selectfont /10} & \cellcolor{lightgreen!20}2{\fontsize{4pt}{0pt}\selectfont /10} & \cellcolor{lightgreen!10}1{\fontsize{4pt}{0pt}\selectfont /10}\\ 
\hline \noalign{\vskip 1ex} 
gpt-3.5 & \cellcolor{lightgreen!0}0{\fontsize{4pt}{0pt}\selectfont /10} & \cellcolor{lightgreen!0}0{\fontsize{4pt}{0pt}\selectfont /10} & \cellcolor{lightgreen!0}0{\fontsize{4pt}{0pt}\selectfont /10} & \cellcolor{lightgreen!0}0{\fontsize{4pt}{0pt}\selectfont /10} & \cellcolor{lightgreen!0}0{\fontsize{4pt}{0pt}\selectfont /10} & \cellcolor{lightgreen!0}0{\fontsize{4pt}{0pt}\selectfont /10}\\ 
\hline \noalign{\vskip 1ex} 
gpt-4 & \cellcolor{lightgreen!0}0{\fontsize{4pt}{0pt}\selectfont /10} & \cellcolor{lightgreen!10}1{\fontsize{4pt}{0pt}\selectfont /10} & \cellcolor{lightgreen!10}1{\fontsize{4pt}{0pt}\selectfont /10} & \cellcolor{lightgreen!10}1{\fontsize{4pt}{0pt}\selectfont /10} & \cellcolor{lightgreen!10}1{\fontsize{4pt}{0pt}\selectfont /10} & \cellcolor{lightgreen!10}1{\fontsize{4pt}{0pt}\selectfont /10}\\ 
\bottomrule
\end{tabularx}
\vspace{0.5pt}
\subcaption{Reason ($3_p$)}
\label{subtable:temp-range-cwe-89-3p-rea}
\end{subtable}
    
\label{tab:temp-range-cwe-89}
\end{table}

\descr{Observations.} Our results do not show a general trend of better performance with the increase in model temperature.
Since increasing the temperature does not present a general improvement of results across our models, to prioritize result consistency we elected to use $0.0$ as the `temperature' value for the remaining of our experiments, and set `top\_p' to LLM specific default value.
\subsection{Diversity of Prompts}
\label{subsec:prompts}

\newcommand{\smallboxopac}[2]{%
    \noindent
    \begin{tikzpicture}[baseline={([yshift=0.25ex]current bounding box.south)}]
        \fill[color=#1, opacity=#2] (0,0) rectangle (0.5cm,0.3cm);
    \end{tikzpicture}%
}
\definecolor{CrimsonRed}{RGB}{227, 74, 51}
\definecolor{ForestGreen}{RGB}{44, 162, 95}

\definecolor{DeepMaroon}{RGB}{253, 187, 132}
\definecolor{SoftCoral}{RGB}{254, 232, 200}
\definecolor{DeepTeal}{RGB}{168, 221, 181}
\definecolor{LightSageGreen}{RGB}{67, 162, 202}

In this part of investigation, we test the LLMs on their ability to detect vulnerabilities in the 48 hand-crafted code scenarios described in Section~\ref{subsec:datasets}, by using the 17 prompts ranging over three categories and four prompting techniques, as described in Section~\ref{subsec:prompt-tec-cat}. 
This experiment allows us to evaluate the capabilities of LLMs over a wide input spectrum and answer questions like (1) what kind of prompting techniques work best for the model, (2) whether emulating the multi-step reasoning process followed by human security experts improves performance, and (3) whether providing extra information or examples helps LLMs in decision making?. 
Table~\ref{tab:prompts-eval} shows the results of this experiment based on the following three metrics:

\descr{(1) Response Rate:} Measures how often the model provides an answer to a given input at all. E.g., for prompts `S5' and `S6,' `codechat-bison@001' provides answers to $36$ out of $48$ inputs and for the rest it responds \textit{``I'm not able to help with that, as I'm only a language model. If you believe this is an error, please send us your feedback.''} {\footnotesize $$Response Rate = \frac{\# Inputs Answered}{Total Inputs}$$}

\descr{(2) Accuracy Rate:} Measures the correctness of the model's response, regardless of the provided reasoning. E.g., for prompt `D2,' `codechat-bison@001' provides correct answers to $24$ inputs out of the $48$ answered inputs. {\footnotesize $$Accuracy Rate = \frac{\# Correct Answers}{\# Inputs Answered}$$}

\descr{(3) Correct Reasoning Rate (CRR):} Evaluates how often the model's correct answers also have the correct reasoning. E.g., for prompt `D2,' `codechat-bison@001' provides reasoning for $15$ answers out of the $24$ correct answers and out of those $15$ reasonings $14$ are correct. {\footnotesize $$CRR = \frac{\# Correct Answers with Correct Reasoning}{\# Reasonings with Correct Answers}$$}

\descr{Best Prompts:} Based on the above three metrics, we choose the best overall prompts for each model from four categories of prompts (described in Section \ref{subsec:prompt-tec-cat}) i.e., ZS - TO, ZS - RO, FS - TO, and FS - RO. 
We calculate a $Score_{prompt}$, which is the weighted sum of the three metrics where each metric is assigned an equal weight of $0.33$. 
We then select the best prompt from each of the above defined four categories as the one that maximizes $Score_{prompt}$, as shown in Table \ref{tab:prompts-eval}.

\begin{table}[!t]
    \centering
    \caption{Evaluation of five LLMs for detecting vulnerabilities across 48 hand-crafted code scenarios, over a  range of prompting techniques. The green and red bars represent the count of scenarios with correct and incorrect responses for each LLM (i.e., $AccuracyRate$). A white circle marks scenarios with both correct answers and reasoning ($CRR$). Additionally, we highlight top-performing prompts for each technique: ZS-TO (\smallboxopac{DeepMaroon}{0.9}), ZS-RO (\smallboxopac{SoftCoral}{1.0}), FS-TO (\smallboxopac{DeepTeal}{0.8}), and FS-RO (\smallboxopac{LightSageGreen}{0.6}). The overall best prompt is shown with a red box.}
    \raisebox{-\height}{\includegraphics[width=\linewidth]{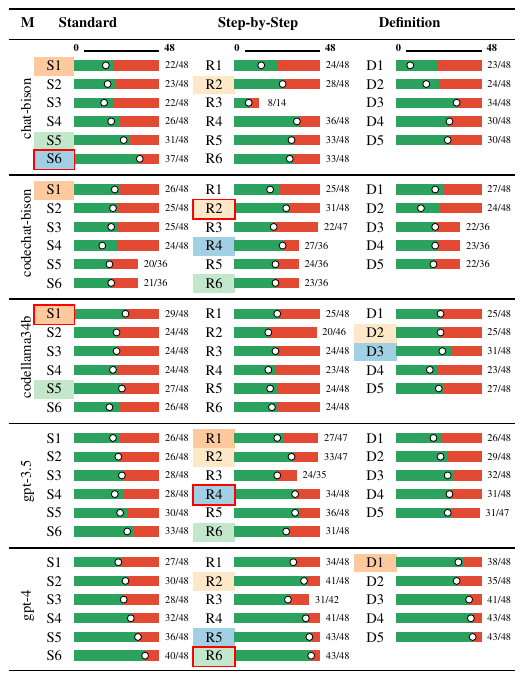}}
    \label{tab:prompts-eval}
\end{table}

\begin{figure}
    \centering
    \begin{subfigure}[t]{\linewidth}

\begin{lstlisting}[language=C]
...
void read_file(char* file_path) {
    char* resolved_path = realpath(file_path, NULL);
    if (resolved_path == NULL) {
        printf("Error: Invalid Path\n");
        return;
    }
    char* dir = "/Users/user1/";
    /* safely creates full_path by concatenating file_path with dir */
    FILE *fp;
    char *data = malloc(256);
    fp = fopen(full_path, "r");
    while (fgets(data, 256, fp) != NULL) 
    {   /* processes file */    }
    ...
}\end{lstlisting}
    \caption{CWE-476 (NULL-Pointer Derefence) $1_v$: This code scenario sanitizes the user provided file path, opens file at that file path, and processes it. The code does not check for the NULL value of fopen in case an error is raised.}
    \label{fig:cwe-476-exp}
\end{subfigure}

\vspace{0.3cm}

\begin{subfigure}[t]{1.05\linewidth}
    \centering
    \includegraphics[width=\linewidth]{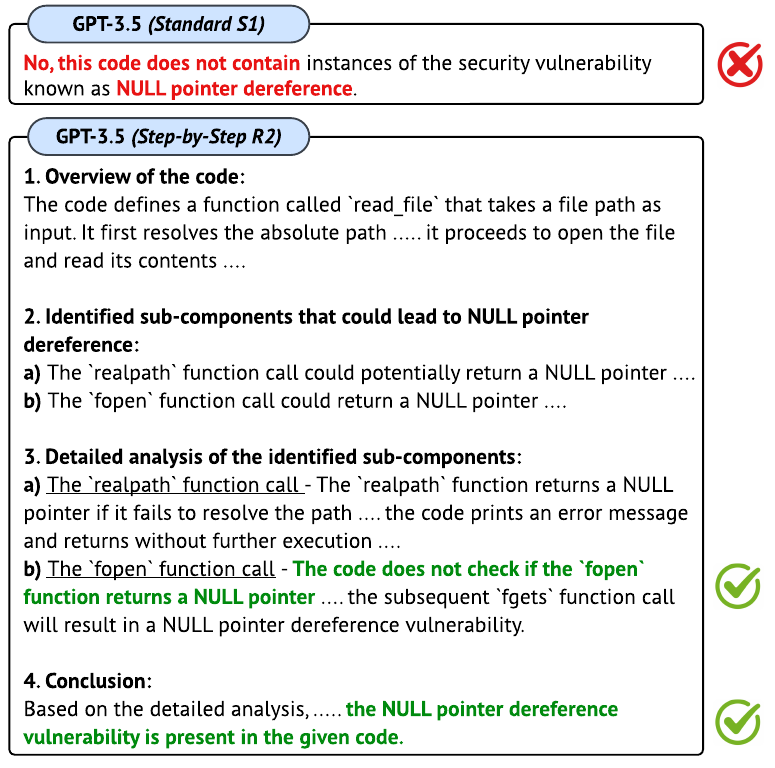}
    \caption{GPT-3.5 Responses.}
    \label{fig:s1-r2}
\end{subfigure}

\caption{`GPT-3.5' responses to standard `S1' and security-expert like multi-step reasoning `R2' for CWE-476 $1_v$ code scenario.}
\label{fig:gpt-3.5-s1-r2}
\end{figure}

\descr{Observations.} `gpt-4' performs the best among the tested LLMs, with a maximum accuracy of 89.5\%. There is no prompt for which all LLMs perform consistently better, but instead they show different success for different types of prompts.
GPT models and `codechat-bison' perform better when prompted to follow a human-like step-by-step reasoning process (as shown in Figure \ref{fig:gpt-3.5-s1-r2}) , i.e., R4, R6, and R2 prompts, respectively. 
`chat-bison' performs best when assigned a `security expert' role, while `codellama34b' works best with the S1 prompt, which simply asks if the code snippet contains a certain vulnerability.
While `gpt-4' and `codellama34b' show an increase in accuracy when provided with a vulnerability definition, compared to standard prompts, the same trend is not found in the other LLMs. 

\subsection{Faithful Reasoning}
\label{subsec:faithful-reasoning}

Faithful reasoning is the quality of an LLM to provide the right reasoning for the right answer or vice versa \cite{faithful-reason}. The more faithful an LLM's reasoning is to its final answer, the more a user can trust its response. 
Table~\ref{tab:prompts-eval} shows that even when they provide the right response, LLMs sometimes provide the wrong reason for this decision.
In this section, we further analyze the faithful reasoning of LLMs on their decisions from the experiment discussed in the previous section, focusing on five aspects: (1) for how many cases does the LLM provide a reasoning for the presence of a vulnerability in a code snippet at all, (2) for how many correct answers the LLM also provides a correct reasoning, (3) for how many correct answers the LLM provides an incorrect reasoning, (4) for how many incorrect answers the LLM provides an incorrect and (5) for how many answers does the LLM provide the wrong answer, but a correct reasoning. 
Table \ref{tab:faith-llm} provides an overview of the results of this experiment.

\newcommand{\smallbox}[1]{%
    \noindent
    \begin{tikzpicture}[baseline={([yshift=0.25ex]current bounding box.south)}]
        \fill[color=#1] (0,0) rectangle (0.5cm,0.3cm);
    \end{tikzpicture}%
}

\definecolor{CrimsonRed}{RGB}{227, 74, 51}
\definecolor{ForestGreen}{RGB}{44, 162, 95}
\definecolor{ErrorBlue}{RGB}{67, 162, 202}
\definecolor{SafetyYellow}{RGB}{255, 193, 7}

\begin{table}[]
\centering
\scriptsize
\caption{Faithfulness of LLMs. The Table shows the \underline{Reason Rate} i.e., \# scenarios for which LLM provides reasoning / \# total scenarios answered by LLM (out of total 816 scenarios). Then it displays \# of scenarios with correct answer and correct reasoning (\smallbox{ForestGreen}), \# correct answer but incorrect reasoning (\smallbox{SafetyYellow}), \# incorrect answer and incorrect reasoning (\smallbox{CrimsonRed}), and \# incorrect answer but correct reasoning (\smallbox{ErrorBlue}).}
\raisebox{-\height}{\includegraphics[width=\linewidth]{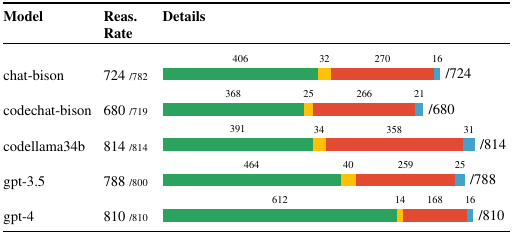}}
\label{tab:faith-llm}
\end{table}

\begin{figure}[]
    \centering
    \includegraphics[width=1.05\linewidth]{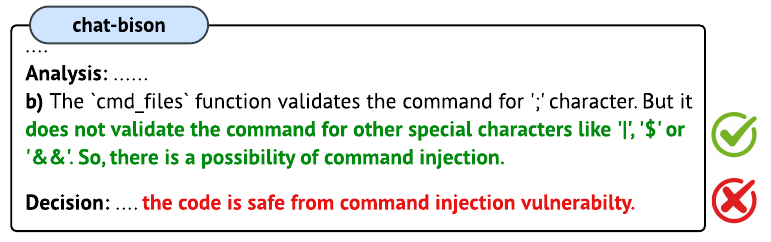}
    \caption{`chat-bison@001' (PaLM2) response for CWE-77 $3_v$ scenario (see Appendix Figure \ref{fig:cwe-77-3}) using prompt `D3' shows unfaithfulness between provided reasoning and final answer.}
    \label{fig:cor-inc}
\end{figure}

\descr{Observations.} 
While in the vast majority of cases the answer and reasoning for the tested LLMs align, every LLM presents cases where it provides the correct reasoning but this leads to a wrong answer (as shown in Figure~\ref{fig:cor-inc}). Similarly, we find cases where an LLM provides the right answer but its reasoning or root cause is not correct (as shown under `codechat-bison (1st Response)' in Figure~\ref{fig:output-inconsistency}). We also find that Google's PaLM2 models have overall lower reasoning rate as they do not explain their decisions in many cases, while GPT models show comparatively higher reasoning rates, especially `gpt-4' and `codellama34b' provide a reason for every answer.
Our findings suggest that, in certain cases, current LLMs’ responses might not fully rely on faithful and accurate reasoning.

\subsection{Evaluation Over Variety of Vulnerabilities}
\label{subsec:variety-vulns}

In this section, we focus on analyzing LLMs ability to correctly identify both vulnerable and patched code for different types of vulnerabilities, based on the eight CWEs that we used to build our hand-crafted dataset. Similar to Section \ref{subsec:prompts}, we find and use the best performing prompts for each CWE using $Score_{cwe}$, with equal weight to all factors, from four prompting categories. The results are summarized in Table \ref{tab:var-cwes}.

\definecolor{PastelGreen}{RGB}{153,216,201}
\definecolor{Teal}{RGB}{44,162,95}

\begin{table}[!t]
    \centering
    \caption{Evaluation of LLMs over a wide range of eight most critical vulnerabilities. Each bar represents count of correctly classified vulnerable (\smallbox{Teal} bar) and patched (\smallbox{PastelGreen} bar) code scenarios, and each circle marks count of correctly reasoned vulnerable (white circle) and patched (black circle) code scenarios, out of total answered scenarios by each LLM.}
    \raisebox{-\height}{\includegraphics[width=\linewidth]{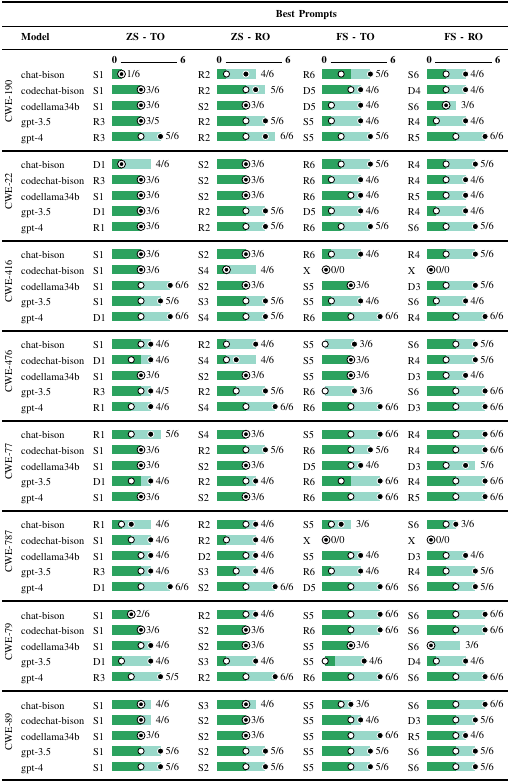}}
    \label{tab:var-cwes}
\end{table}

\descr{Observations.} 
Most models show poor performance in classifying the patched versions correctly, which makes these LLMs non-suitable for real-world cases as they will mostly flag safe code as vulnerable, causing manyfalse alarms. 
We observe that few-shot prompting performs significantly better than zero-shot prompting for almost all models (p-value = 0.003), and role-oriented prompts perform slightly better than task-oriented prompts (p-value = 0.1). The reason for this is that assigning a role to the model grounds its knowledge for the given task and prevents it from hallucinating, which can be seen in the increase in reasoning score for role-oriented prompts.
However, Table \ref{tab:var-cwes} shows that `codechat-bison@001' does not provide answers for `CWE-787' and `CWE-416' for few-shot prompts. 
\subsection{Code Difficulty Levels}
\label{subsec:range-diff}

In this section, we investigate the capabilities of LLMs to handle different complexities of code.
Similar to the previous sections, we find the best performing prompts for each difficulty level using $Score_{diff}$, with equal weight to all factors, from four prompting categories.
\noindent Table \ref{tab:var-diff} summarizes the results of this experiment.

\begin{table}[]
\centering
\caption{Evaluation of LLMs over a Range of Code Difficulty Levels. (Same color coding as Table \ref{tab:var-cwes}).}
\raisebox{-\height}{\includegraphics[width=\linewidth]{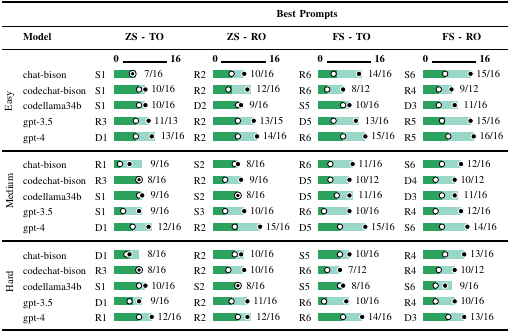}}
\label{tab:var-diff}
\end{table}

\begin{figure}
    \centering
    \begin{subfigure}[t]{\linewidth}
    \begin{lstlisting}[language=Python]
def createQuery(username, password):
  query_text = "SELECT * FROM users WHERE username = '%s' AND password = '%s'"
  parmeters = (username, password)
  query = query_text, parmeters
  return query

@app.route("/login")
def login():
  username = request.form('username')
  password = request.form('password')

  db = pymysql.connect("localhost")
  cursor = db.cursor()

  query = createQuery(username, password)
  cursor.execute(query)\end{lstlisting}
    \caption{CWE-89 (SQL Injection) $3_p$ code scenario. This code is not vulnerable to SQL Injection as it passes the `query' and `parameters' separately to the `pymysql' function.}
    \label{fig:cwe-89-p}
    \end{subfigure}

    \vspace{0.2cm}

    \begin{subfigure}[t]{\linewidth}
    \includegraphics[width=1.05\linewidth]{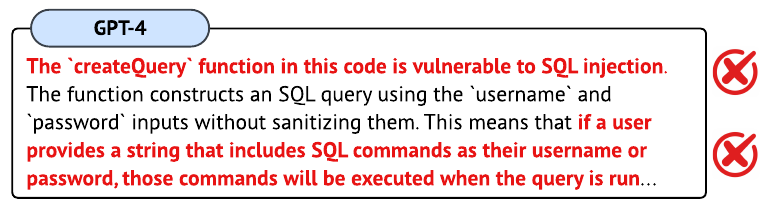}
    \caption{GPT-4 response.}
    \label{fig:gpt4-diff}
    \end{subfigure}
    \caption{GPT-4 provides wrong answer and reasoning to CWE-89 (hard-level) $3_p$ code scenario.}
    \label{fig:gpt-4-cwe-89}
\end{figure}

\descr{Observations.} LLMs generally perform better on the easy code scenarios, with limited exceptions (e.g., `codechat-bison' correctly recognizing more medium than easy vulnerabilities in the \textbf{FS - TO} setting).
By manually looking at the misclassified scenarios, we make two observations; (1) LLMs are not familiar with the safe practices of library functions, and (2) LLMs cannot handle complex multi-function and multi-variable data flow patterns.
For example, in the CWE-89 (SQL Injection) `$3_p$' scenario, the code creates an SQL query and its parameters in one function `create\_query', and returns them to the `login' function where they are passed to `pymysql' as separate arguments.
None of the LLMs were able to understand this flow (e.g., the response from `gpt-4' is shown in Figure~\ref{fig:gpt-4-cwe-89}), as they mistakenly assume that the `login' function only passes one argument to the `execute' function. The LLMs also seem to be unaware that `pymysql' itself sanitizes the output. %

\subsection{Robustness to Code Augmentations}

In this section we test the robustness of LLMs by testing them against the code augmentations described in Section~\ref{subsec:code-aug}. 
Our results are summarized in Table~\ref{tab:aug}.
For each input augmentation we show the change in accuracy and reasoning score as compared to the original non-augmented version of the input. For each LLM, we test each augmentation using three prompts: standard prompt `S1,' and the best zero-shot (ZS) and few-shot (FS) prompts.\footnote{Since we show in Sections \ref{subsec:variety-vulns} and \ref{subsec:range-diff} that role-oriented prompts work better than task-oriented ones, we do not run experiments on all four categories of prompts.}

\begin{table}[]
\centering
\tiny
\caption{Evaluation for Code-Level Augmentations. The tables show $\Delta_a$ (\# of answers that are correct in non-augmented scenarios but incorrect in this specific augmentation case) and $\Delta_p$ (\# of reasoning that are correct in non-augmented scenarios but incorrect in this specific augmentation case) for each code augmentation and for three prompts (standard `S1', best zero-shot, and best few-shot) of each LLM.}
\vspace{0.1cm}
\begin{subtable}{\linewidth}
\begin{tabularx}{\linewidth}{l@{\hspace{0.2cm}}lXXXXXXXXXXXXXX}
\toprule
& & \multicolumn{2}{c}{\textbf{T1}} & \multicolumn{2}{c}{\textbf{T2}} & \multicolumn{2}{c}{\textbf{T3}} & \multicolumn{2}{c}{\textbf{T4}} & \multicolumn{2}{c}{\textbf{T5}} & \multicolumn{2}{c}{\textbf{T6}} & \multicolumn{2}{c}{\textbf{T7}} \\ [1ex]
\hline
\noalign{\vskip 0.5ex}
\textbf{M} & \textbf{PS} & $\Delta_a$ & $\Delta_r$ & $\Delta_a$ & $\Delta_r$ & $\Delta_a$ & $\Delta_r$ & $\Delta_a$ & $\Delta_r$ & $\Delta_a$ & $\Delta_r$ & $\Delta_a$ & $\Delta_r$ & $\Delta_a$ & $\Delta_r$ \\ [0.5ex]
\hline
\noalign{\vskip 1ex}
\multirow{3}{*}{\rotatebox[origin=c]{90}{c-bison}} & S1{\fontsize{3pt}{0pt}\selectfont S} & \cellcolor{red!0}0{\fontsize{4pt}{0pt}\selectfont /12} & \cellcolor{red!0}0{\fontsize{4pt}{0pt}\selectfont /12} & \cellcolor{red!0}0{\fontsize{4pt}{0pt}\selectfont /12} & \cellcolor{red!0}0{\fontsize{4pt}{0pt}\selectfont /12} & \cellcolor{red!0}0{\fontsize{4pt}{0pt}\selectfont /12} & \cellcolor{red!0}0{\fontsize{4pt}{0pt}\selectfont /12} & \cellcolor{red!0}0{\fontsize{4pt}{0pt}\selectfont /12} & \cellcolor{red!0}0{\fontsize{4pt}{0pt}\selectfont /12} & \cellcolor{red!0}0{\fontsize{4pt}{0pt}\selectfont /12} & \cellcolor{red!0}0{\fontsize{4pt}{0pt}\selectfont /12} & \cellcolor{red!0}0{\fontsize{4pt}{0pt}\selectfont /12} & \cellcolor{red!0}0{\fontsize{4pt}{0pt}\selectfont /12} & \cellcolor{red!0}0{\fontsize{4pt}{0pt}\selectfont /12} & \cellcolor{red!0}0{\fontsize{4pt}{0pt}\selectfont /12}\\ 
 & R2{\fontsize{3pt}{0pt}\selectfont ZS} & \cellcolor{red!16}2{\fontsize{4pt}{0pt}\selectfont /12} & \cellcolor{red!16}2{\fontsize{4pt}{0pt}\selectfont /12} & \cellcolor{red!8}1{\fontsize{4pt}{0pt}\selectfont /12} & \cellcolor{red!8}1{\fontsize{4pt}{0pt}\selectfont /12} & \cellcolor{red!0}0{\fontsize{4pt}{0pt}\selectfont /12} & \cellcolor{red!0}0{\fontsize{4pt}{0pt}\selectfont /12} & \cellcolor{red!8}1{\fontsize{4pt}{0pt}\selectfont /12} & \cellcolor{red!8}1{\fontsize{4pt}{0pt}\selectfont /12} & \cellcolor{red!16}2{\fontsize{4pt}{0pt}\selectfont /12} & \cellcolor{red!25}3{\fontsize{4pt}{0pt}\selectfont /12} & \cellcolor{red!0}0{\fontsize{4pt}{0pt}\selectfont /12} & \cellcolor{red!0}0{\fontsize{4pt}{0pt}\selectfont /12} & \cellcolor{red!0}0{\fontsize{4pt}{0pt}\selectfont /12} & \cellcolor{red!8}1{\fontsize{4pt}{0pt}\selectfont /12}\\ 
 & S6{\fontsize{3pt}{0pt}\selectfont FS} & \cellcolor{red!16}2{\fontsize{4pt}{0pt}\selectfont /12} & \cellcolor{red!25}3{\fontsize{4pt}{0pt}\selectfont /12} & \cellcolor{red!16}2{\fontsize{4pt}{0pt}\selectfont /12} & \cellcolor{red!16}2{\fontsize{4pt}{0pt}\selectfont /12} & \cellcolor{red!16}2{\fontsize{4pt}{0pt}\selectfont /12} & \cellcolor{red!33}4{\fontsize{4pt}{0pt}\selectfont /12} & \cellcolor{red!8}1{\fontsize{4pt}{0pt}\selectfont /12} & \cellcolor{red!25}3{\fontsize{4pt}{0pt}\selectfont /12} & \cellcolor{red!16}2{\fontsize{4pt}{0pt}\selectfont /12} & \cellcolor{red!16}2{\fontsize{4pt}{0pt}\selectfont /12} & \cellcolor{red!16}2{\fontsize{4pt}{0pt}\selectfont /12} & \cellcolor{red!16}2{\fontsize{4pt}{0pt}\selectfont /12} & \cellcolor{red!16}2{\fontsize{4pt}{0pt}\selectfont /12} & \cellcolor{red!8}1{\fontsize{4pt}{0pt}\selectfont /12}\\ 
\seprule
\multirow{3}{*}{\rotatebox[origin=c]{90}{cc-bison}} & S1{\fontsize{3pt}{0pt}\selectfont S} & \cellcolor{red!0}0{\fontsize{4pt}{0pt}\selectfont /12} & \cellcolor{red!16}2{\fontsize{4pt}{0pt}\selectfont /12} & \cellcolor{red!0}0{\fontsize{4pt}{0pt}\selectfont /12} & \cellcolor{red!25}3{\fontsize{4pt}{0pt}\selectfont /12} & \cellcolor{red!0}0{\fontsize{4pt}{0pt}\selectfont /12} & \cellcolor{red!25}3{\fontsize{4pt}{0pt}\selectfont /12} & \cellcolor{red!0}0{\fontsize{4pt}{0pt}\selectfont /12} & \cellcolor{red!0}0{\fontsize{4pt}{0pt}\selectfont /12} & \cellcolor{red!0}0{\fontsize{4pt}{0pt}\selectfont /12} & \cellcolor{red!16}2{\fontsize{4pt}{0pt}\selectfont /12} & \cellcolor{red!0}0{\fontsize{4pt}{0pt}\selectfont /12} & \cellcolor{red!16}2{\fontsize{4pt}{0pt}\selectfont /12} & \cellcolor{red!0}0{\fontsize{4pt}{0pt}\selectfont /12} & \cellcolor{red!8}1{\fontsize{4pt}{0pt}\selectfont /12}\\ 
 & R2{\fontsize{3pt}{0pt}\selectfont ZS} & \cellcolor{red!0}0{\fontsize{4pt}{0pt}\selectfont /12} & \cellcolor{red!0}0{\fontsize{4pt}{0pt}\selectfont /12} & \cellcolor{red!0}0{\fontsize{4pt}{0pt}\selectfont /12} & \cellcolor{red!8}1{\fontsize{4pt}{0pt}\selectfont /12} & \cellcolor{red!0}0{\fontsize{4pt}{0pt}\selectfont /12} & \cellcolor{red!0}0{\fontsize{4pt}{0pt}\selectfont /12} & \cellcolor{red!25}3{\fontsize{4pt}{0pt}\selectfont /12} & \cellcolor{red!25}3{\fontsize{4pt}{0pt}\selectfont /12} & \cellcolor{red!16}2{\fontsize{4pt}{0pt}\selectfont /12} & \cellcolor{red!16}2{\fontsize{4pt}{0pt}\selectfont /12} & \cellcolor{red!0}0{\fontsize{4pt}{0pt}\selectfont /12} & \cellcolor{red!16}2{\fontsize{4pt}{0pt}\selectfont /12} & \cellcolor{red!8}1{\fontsize{4pt}{0pt}\selectfont /12} & \cellcolor{red!8}1{\fontsize{4pt}{0pt}\selectfont /12}\\ 
 & R4{\fontsize{3pt}{0pt}\selectfont FS} & \cellcolor{red!0}0{\fontsize{4pt}{0pt}\selectfont /12} & \cellcolor{red!0}0{\fontsize{4pt}{0pt}\selectfont /12} & \cellcolor{red!0}0{\fontsize{4pt}{0pt}\selectfont /12} & \cellcolor{red!0}0{\fontsize{4pt}{0pt}\selectfont /12} & \cellcolor{red!0}0{\fontsize{4pt}{0pt}\selectfont /12} & \cellcolor{red!0}0{\fontsize{4pt}{0pt}\selectfont /12} & \cellcolor{red!8}1{\fontsize{4pt}{0pt}\selectfont /12} & \cellcolor{red!8}1{\fontsize{4pt}{0pt}\selectfont /12} & \cellcolor{red!0}0{\fontsize{4pt}{0pt}\selectfont /12} & \cellcolor{red!0}0{\fontsize{4pt}{0pt}\selectfont /12} & \cellcolor{red!8}1{\fontsize{4pt}{0pt}\selectfont /12} & \cellcolor{red!8}1{\fontsize{4pt}{0pt}\selectfont /12} & \cellcolor{red!0}0{\fontsize{4pt}{0pt}\selectfont /12} & \cellcolor{red!0}0{\fontsize{4pt}{0pt}\selectfont /12}\\ 
\seprule
\multirow{3}{*}{\rotatebox[origin=c]{90}{c.lla.34b}} & S1{\fontsize{3pt}{0pt}\selectfont S} & \cellcolor{red!0}0{\fontsize{4pt}{0pt}\selectfont /12} & \cellcolor{red!0}0{\fontsize{4pt}{0pt}\selectfont /12} & \cellcolor{red!0}0{\fontsize{4pt}{0pt}\selectfont /12} & \cellcolor{red!0}0{\fontsize{4pt}{0pt}\selectfont /12} & \cellcolor{red!0}0{\fontsize{4pt}{0pt}\selectfont /12} & \cellcolor{red!0}0{\fontsize{4pt}{0pt}\selectfont /12} & \cellcolor{red!0}0{\fontsize{4pt}{0pt}\selectfont /12} & \cellcolor{red!0}0{\fontsize{4pt}{0pt}\selectfont /12} & \cellcolor{red!0}0{\fontsize{4pt}{0pt}\selectfont /12} & \cellcolor{red!0}0{\fontsize{4pt}{0pt}\selectfont /12} & \cellcolor{red!0}0{\fontsize{4pt}{0pt}\selectfont /12} & \cellcolor{red!0}0{\fontsize{4pt}{0pt}\selectfont /12} & \cellcolor{red!0}0{\fontsize{4pt}{0pt}\selectfont /12} & \cellcolor{red!0}0{\fontsize{4pt}{0pt}\selectfont /12}\\ 
 & S1{\fontsize{3pt}{0pt}\selectfont ZS} & \cellcolor{red!0}0{\fontsize{4pt}{0pt}\selectfont /12} & \cellcolor{red!0}0{\fontsize{4pt}{0pt}\selectfont /12} & \cellcolor{red!0}0{\fontsize{4pt}{0pt}\selectfont /12} & \cellcolor{red!0}0{\fontsize{4pt}{0pt}\selectfont /12} & \cellcolor{red!0}0{\fontsize{4pt}{0pt}\selectfont /12} & \cellcolor{red!0}0{\fontsize{4pt}{0pt}\selectfont /12} & \cellcolor{red!0}0{\fontsize{4pt}{0pt}\selectfont /12} & \cellcolor{red!0}0{\fontsize{4pt}{0pt}\selectfont /12} & \cellcolor{red!0}0{\fontsize{4pt}{0pt}\selectfont /12} & \cellcolor{red!0}0{\fontsize{4pt}{0pt}\selectfont /12} & \cellcolor{red!0}0{\fontsize{4pt}{0pt}\selectfont /12} & \cellcolor{red!0}0{\fontsize{4pt}{0pt}\selectfont /12} & \cellcolor{red!0}0{\fontsize{4pt}{0pt}\selectfont /12} & \cellcolor{red!0}0{\fontsize{4pt}{0pt}\selectfont /12}\\ 
 & S5{\fontsize{3pt}{0pt}\selectfont FS} & \cellcolor{red!0}0{\fontsize{4pt}{0pt}\selectfont /12} & \cellcolor{red!0}0{\fontsize{4pt}{0pt}\selectfont /12} & \cellcolor{red!25}3{\fontsize{4pt}{0pt}\selectfont /12} & \cellcolor{red!25}3{\fontsize{4pt}{0pt}\selectfont /12} & \cellcolor{red!25}3{\fontsize{4pt}{0pt}\selectfont /12} & \cellcolor{red!25}3{\fontsize{4pt}{0pt}\selectfont /12} & \cellcolor{red!16}2{\fontsize{4pt}{0pt}\selectfont /12} & \cellcolor{red!25}3{\fontsize{4pt}{0pt}\selectfont /12} & \cellcolor{red!0}0{\fontsize{4pt}{0pt}\selectfont /12} & \cellcolor{red!0}0{\fontsize{4pt}{0pt}\selectfont /12} & \cellcolor{red!8}1{\fontsize{4pt}{0pt}\selectfont /12} & \cellcolor{red!8}1{\fontsize{4pt}{0pt}\selectfont /12} & \cellcolor{red!16}2{\fontsize{4pt}{0pt}\selectfont /12} & \cellcolor{red!16}2{\fontsize{4pt}{0pt}\selectfont /12}\\ 
\seprule
\multirow{3}{*}{\rotatebox[origin=c]{90}{gpt-3.5}} & S1{\fontsize{3pt}{0pt}\selectfont S} & \cellcolor{red!0}0{\fontsize{4pt}{0pt}\selectfont /12} & \cellcolor{red!0}0{\fontsize{4pt}{0pt}\selectfont /12} & \cellcolor{red!0}0{\fontsize{4pt}{0pt}\selectfont /12} & \cellcolor{red!0}0{\fontsize{4pt}{0pt}\selectfont /12} & \cellcolor{red!16}2{\fontsize{4pt}{0pt}\selectfont /12} & \cellcolor{red!25}3{\fontsize{4pt}{0pt}\selectfont /12} & \cellcolor{red!8}1{\fontsize{4pt}{0pt}\selectfont /12} & \cellcolor{red!8}1{\fontsize{4pt}{0pt}\selectfont /12} & \cellcolor{red!0}0{\fontsize{4pt}{0pt}\selectfont /12} & \cellcolor{red!0}0{\fontsize{4pt}{0pt}\selectfont /12} & \cellcolor{red!8}1{\fontsize{4pt}{0pt}\selectfont /12} & \cellcolor{red!16}2{\fontsize{4pt}{0pt}\selectfont /12} & \cellcolor{red!0}0{\fontsize{4pt}{0pt}\selectfont /12} & \cellcolor{red!0}0{\fontsize{4pt}{0pt}\selectfont /12}\\ 
 & R2{\fontsize{3pt}{0pt}\selectfont ZS} & \cellcolor{red!0}0{\fontsize{4pt}{0pt}\selectfont /12} & \cellcolor{red!0}0{\fontsize{4pt}{0pt}\selectfont /12} & \cellcolor{red!0}0{\fontsize{4pt}{0pt}\selectfont /12} & \cellcolor{red!0}0{\fontsize{4pt}{0pt}\selectfont /12} & \cellcolor{red!0}0{\fontsize{4pt}{0pt}\selectfont /12} & \cellcolor{red!0}0{\fontsize{4pt}{0pt}\selectfont /12} & \cellcolor{red!0}0{\fontsize{4pt}{0pt}\selectfont /12} & \cellcolor{red!0}0{\fontsize{4pt}{0pt}\selectfont /12} & \cellcolor{red!0}0{\fontsize{4pt}{0pt}\selectfont /12} & \cellcolor{red!0}0{\fontsize{4pt}{0pt}\selectfont /12} & \cellcolor{red!0}0{\fontsize{4pt}{0pt}\selectfont /12} & \cellcolor{red!0}0{\fontsize{4pt}{0pt}\selectfont /12} & \cellcolor{red!0}0{\fontsize{4pt}{0pt}\selectfont /12} & \cellcolor{red!0}0{\fontsize{4pt}{0pt}\selectfont /12}\\ 
 & R4{\fontsize{3pt}{0pt}\selectfont FS} & \cellcolor{red!8}1{\fontsize{4pt}{0pt}\selectfont /12} & \cellcolor{red!8}1{\fontsize{4pt}{0pt}\selectfont /12} & \cellcolor{red!8}1{\fontsize{4pt}{0pt}\selectfont /12} & \cellcolor{red!8}1{\fontsize{4pt}{0pt}\selectfont /12} & \cellcolor{red!8}1{\fontsize{4pt}{0pt}\selectfont /12} & \cellcolor{red!8}1{\fontsize{4pt}{0pt}\selectfont /12} & \cellcolor{red!8}1{\fontsize{4pt}{0pt}\selectfont /12} & \cellcolor{red!8}1{\fontsize{4pt}{0pt}\selectfont /12} & \cellcolor{red!16}2{\fontsize{4pt}{0pt}\selectfont /12} & \cellcolor{red!16}2{\fontsize{4pt}{0pt}\selectfont /12} & \cellcolor{red!0}0{\fontsize{4pt}{0pt}\selectfont /12} & \cellcolor{red!0}0{\fontsize{4pt}{0pt}\selectfont /12} & \cellcolor{red!8}1{\fontsize{4pt}{0pt}\selectfont /12} & \cellcolor{red!8}1{\fontsize{4pt}{0pt}\selectfont /12}\\ 
\seprule
\multirow{3}{*}{\rotatebox[origin=c]{90}{gpt-4}} & S1{\fontsize{3pt}{0pt}\selectfont S} & \cellcolor{red!0}0{\fontsize{4pt}{0pt}\selectfont /12} & \cellcolor{red!0}0{\fontsize{4pt}{0pt}\selectfont /12} & \cellcolor{red!0}0{\fontsize{4pt}{0pt}\selectfont /12} & \cellcolor{red!16}2{\fontsize{4pt}{0pt}\selectfont /12} & \cellcolor{red!0}0{\fontsize{4pt}{0pt}\selectfont /12} & \cellcolor{red!16}2{\fontsize{4pt}{0pt}\selectfont /12} & \cellcolor{red!0}0{\fontsize{4pt}{0pt}\selectfont /12} & \cellcolor{red!0}0{\fontsize{4pt}{0pt}\selectfont /12} & \cellcolor{red!0}0{\fontsize{4pt}{0pt}\selectfont /12} & \cellcolor{red!0}0{\fontsize{4pt}{0pt}\selectfont /12} & \cellcolor{red!0}0{\fontsize{4pt}{0pt}\selectfont /12} & \cellcolor{red!8}1{\fontsize{4pt}{0pt}\selectfont /12} & \cellcolor{red!0}0{\fontsize{4pt}{0pt}\selectfont /12} & \cellcolor{red!0}0{\fontsize{4pt}{0pt}\selectfont /12}\\ 
 & R2{\fontsize{3pt}{0pt}\selectfont ZS} & \cellcolor{red!16}2{\fontsize{4pt}{0pt}\selectfont /12} & \cellcolor{red!8}1{\fontsize{4pt}{0pt}\selectfont /12} & \cellcolor{red!8}1{\fontsize{4pt}{0pt}\selectfont /12} & \cellcolor{red!0}0{\fontsize{4pt}{0pt}\selectfont /12} & \cellcolor{red!25}3{\fontsize{4pt}{0pt}\selectfont /12} & \cellcolor{red!16}2{\fontsize{4pt}{0pt}\selectfont /12} & \cellcolor{red!16}2{\fontsize{4pt}{0pt}\selectfont /12} & \cellcolor{red!8}1{\fontsize{4pt}{0pt}\selectfont /12} & \cellcolor{red!8}1{\fontsize{4pt}{0pt}\selectfont /12} & \cellcolor{red!0}0{\fontsize{4pt}{0pt}\selectfont /12} & \cellcolor{red!16}2{\fontsize{4pt}{0pt}\selectfont /12} & \cellcolor{red!8}1{\fontsize{4pt}{0pt}\selectfont /12} & \cellcolor{red!8}1{\fontsize{4pt}{0pt}\selectfont /12} & \cellcolor{red!8}1{\fontsize{4pt}{0pt}\selectfont /12}\\ 
 & R6{\fontsize{3pt}{0pt}\selectfont FS} & \cellcolor{red!0}0{\fontsize{4pt}{0pt}\selectfont /12} & \cellcolor{red!0}0{\fontsize{4pt}{0pt}\selectfont /12} & \cellcolor{red!0}0{\fontsize{4pt}{0pt}\selectfont /12} & \cellcolor{red!0}0{\fontsize{4pt}{0pt}\selectfont /12} & \cellcolor{red!0}0{\fontsize{4pt}{0pt}\selectfont /12} & \cellcolor{red!0}0{\fontsize{4pt}{0pt}\selectfont /12} & \cellcolor{red!0}0{\fontsize{4pt}{0pt}\selectfont /12} & \cellcolor{red!0}0{\fontsize{4pt}{0pt}\selectfont /12} & \cellcolor{red!0}0{\fontsize{4pt}{0pt}\selectfont /12} & \cellcolor{red!0}0{\fontsize{4pt}{0pt}\selectfont /12} & \cellcolor{red!0}0{\fontsize{4pt}{0pt}\selectfont /12} & \cellcolor{red!0}0{\fontsize{4pt}{0pt}\selectfont /12} & \cellcolor{red!0}0{\fontsize{4pt}{0pt}\selectfont /12} & \cellcolor{red!0}0{\fontsize{4pt}{0pt}\selectfont /12}\\ 
\bottomrule
\end{tabularx}
\vspace{0.1cm}
\caption{Trivial Augmentations}
\label{tab:code-trivial}
\end{subtable}

\vspace{0.3cm}

\begin{subtable}{\linewidth}
\begin{tabularx}{\linewidth}{l@{\hspace{0.2cm}}lXXXXXXXXXXXX}
\toprule
& & \multicolumn{2}{c}{\textbf{NT1}} & \multicolumn{2}{c}{\textbf{NT2}} & \multicolumn{2}{c}{\textbf{NT3}} & \multicolumn{2}{c}{\textbf{NT4}} & \multicolumn{2}{c}{\textbf{NT5}} & \multicolumn{2}{c}{\textbf{NT6}} \\ [1ex]
\hline
\noalign{\vskip 0.5ex}
\textbf{M} & \textbf{PS} & $\Delta_a$ & $\Delta_r$ & $\Delta_a$ & $\Delta_r$ & $\Delta_a$ & $\Delta_r$ & $\Delta_a$ & $\Delta_r$ & $\Delta_a$ & $\Delta_r$ & $\Delta_a$ & $\Delta_r$ \\ [0.5ex]
\hline
\noalign{\vskip 1ex}
\multirow{3}{*}{\rotatebox[origin=c]{90}{c-bison}} & S1{\fontsize{3pt}{0pt}\selectfont S} & \cellcolor{red!0}0{\fontsize{4pt}{0pt}\selectfont /12} & \cellcolor{red!0}0{\fontsize{4pt}{0pt}\selectfont /12} & \cellcolor{red!8}1{\fontsize{4pt}{0pt}\selectfont /12} & \cellcolor{red!0}0{\fontsize{4pt}{0pt}\selectfont /12} & \cellcolor{red!66}8{\fontsize{4pt}{0pt}\selectfont /12} & \cellcolor{red!58}7{\fontsize{4pt}{0pt}\selectfont /12} & \cellcolor{red!0}0{\fontsize{4pt}{0pt}\selectfont /12} & \cellcolor{red!8}1{\fontsize{4pt}{0pt}\selectfont /12} & \cellcolor{red!0}0{\fontsize{4pt}{0pt}\selectfont /9} & \cellcolor{red!22}2{\fontsize{4pt}{0pt}\selectfont /9} & \cellcolor{red!0}0{\fontsize{4pt}{0pt}\selectfont /9} & \cellcolor{red!11}1{\fontsize{4pt}{0pt}\selectfont /9}\\ 
 & R2{\fontsize{3pt}{0pt}\selectfont ZS} & \cellcolor{red!25}3{\fontsize{4pt}{0pt}\selectfont /12} & \cellcolor{red!16}2{\fontsize{4pt}{0pt}\selectfont /12} & \cellcolor{red!16}2{\fontsize{4pt}{0pt}\selectfont /12} & \cellcolor{red!16}2{\fontsize{4pt}{0pt}\selectfont /12} & \cellcolor{red!33}4{\fontsize{4pt}{0pt}\selectfont /12} & \cellcolor{red!33}4{\fontsize{4pt}{0pt}\selectfont /12} & \cellcolor{red!0}0{\fontsize{4pt}{0pt}\selectfont /12} & \cellcolor{red!50}6{\fontsize{4pt}{0pt}\selectfont /12} & \cellcolor{red!0}0{\fontsize{4pt}{0pt}\selectfont /9} & \cellcolor{red!22}2{\fontsize{4pt}{0pt}\selectfont /9} & \cellcolor{red!0}0{\fontsize{4pt}{0pt}\selectfont /9} & \cellcolor{red!33}3{\fontsize{4pt}{0pt}\selectfont /9}\\ 
 & S6{\fontsize{3pt}{0pt}\selectfont FS} & \cellcolor{red!0}0{\fontsize{4pt}{0pt}\selectfont /12} & \cellcolor{red!0}0{\fontsize{4pt}{0pt}\selectfont /12} & \cellcolor{red!41}5{\fontsize{4pt}{0pt}\selectfont /12} & \cellcolor{red!41}5{\fontsize{4pt}{0pt}\selectfont /12} & \cellcolor{red!66}8{\fontsize{4pt}{0pt}\selectfont /12} & \cellcolor{red!66}8{\fontsize{4pt}{0pt}\selectfont /12} & \cellcolor{red!0}0{\fontsize{4pt}{0pt}\selectfont /12} & \cellcolor{red!0}0{\fontsize{4pt}{0pt}\selectfont /12} & \cellcolor{red!44}4{\fontsize{4pt}{0pt}\selectfont /9} & \cellcolor{red!44}4{\fontsize{4pt}{0pt}\selectfont /9} & \cellcolor{red!22}2{\fontsize{4pt}{0pt}\selectfont /9} & \cellcolor{red!11}1{\fontsize{4pt}{0pt}\selectfont /9}\\ 
\seprule
\multirow{3}{*}{\rotatebox[origin=c]{90}{cc-bison}} & S1{\fontsize{3pt}{0pt}\selectfont S} & \cellcolor{red!0}0{\fontsize{4pt}{0pt}\selectfont /12} & \cellcolor{red!16}2{\fontsize{4pt}{0pt}\selectfont /12} & \cellcolor{red!25}3{\fontsize{4pt}{0pt}\selectfont /12} & \cellcolor{red!8}1{\fontsize{4pt}{0pt}\selectfont /12} & \cellcolor{red!75}9{\fontsize{4pt}{0pt}\selectfont /12} & \cellcolor{red!75}9{\fontsize{4pt}{0pt}\selectfont /12} & \cellcolor{red!8}1{\fontsize{4pt}{0pt}\selectfont /12} & \cellcolor{red!33}4{\fontsize{4pt}{0pt}\selectfont /12} & \cellcolor{red!11}1{\fontsize{4pt}{0pt}\selectfont /9} & \cellcolor{red!44}4{\fontsize{4pt}{0pt}\selectfont /9} & \cellcolor{red!11}1{\fontsize{4pt}{0pt}\selectfont /9} & \cellcolor{red!0}0{\fontsize{4pt}{0pt}\selectfont /9}\\ 
 & R2{\fontsize{3pt}{0pt}\selectfont ZS} & \cellcolor{red!8}1{\fontsize{4pt}{0pt}\selectfont /12} & \cellcolor{red!16}2{\fontsize{4pt}{0pt}\selectfont /12} & \cellcolor{red!33}4{\fontsize{4pt}{0pt}\selectfont /12} & \cellcolor{red!33}4{\fontsize{4pt}{0pt}\selectfont /12} & \cellcolor{red!58}7{\fontsize{4pt}{0pt}\selectfont /12} & \cellcolor{red!50}6{\fontsize{4pt}{0pt}\selectfont /12} & \cellcolor{red!8}1{\fontsize{4pt}{0pt}\selectfont /12} & \cellcolor{red!8}1{\fontsize{4pt}{0pt}\selectfont /12} & \cellcolor{red!0}0{\fontsize{4pt}{0pt}\selectfont /9} & \cellcolor{red!0}0{\fontsize{4pt}{0pt}\selectfont /9} & \cellcolor{red!0}0{\fontsize{4pt}{0pt}\selectfont /9} & \cellcolor{red!0}0{\fontsize{4pt}{0pt}\selectfont /9}\\ 
 & R4{\fontsize{3pt}{0pt}\selectfont FS} & \cellcolor{red!0}0{\fontsize{4pt}{0pt}\selectfont /12} & \cellcolor{red!0}0{\fontsize{4pt}{0pt}\selectfont /12} & \cellcolor{red!41}5{\fontsize{4pt}{0pt}\selectfont /12} & \cellcolor{red!50}6{\fontsize{4pt}{0pt}\selectfont /12} & \cellcolor{red!8}1{\fontsize{4pt}{0pt}\selectfont /12} & \cellcolor{red!8}1{\fontsize{4pt}{0pt}\selectfont /12} & \cellcolor{red!0}0{\fontsize{4pt}{0pt}\selectfont /12} & \cellcolor{red!0}0{\fontsize{4pt}{0pt}\selectfont /12} & \cellcolor{red!66}6{\fontsize{4pt}{0pt}\selectfont /9} & \cellcolor{red!66}6{\fontsize{4pt}{0pt}\selectfont /9} & \cellcolor{red!0}0{\fontsize{4pt}{0pt}\selectfont /9} & \cellcolor{red!0}0{\fontsize{4pt}{0pt}\selectfont /9}\\ 
\seprule
\multirow{3}{*}{\rotatebox[origin=c]{90}{c.lla.34b}} & S1{\fontsize{3pt}{0pt}\selectfont S} & \cellcolor{red!8}1{\fontsize{4pt}{0pt}\selectfont /12} & \cellcolor{red!0}0{\fontsize{4pt}{0pt}\selectfont /12} & \cellcolor{red!33}4{\fontsize{4pt}{0pt}\selectfont /12} & \cellcolor{red!33}4{\fontsize{4pt}{0pt}\selectfont /12} & \cellcolor{red!75}9{\fontsize{4pt}{0pt}\selectfont /12} & \cellcolor{red!75}9{\fontsize{4pt}{0pt}\selectfont /12} & \cellcolor{red!8}1{\fontsize{4pt}{0pt}\selectfont /12} & \cellcolor{red!25}3{\fontsize{4pt}{0pt}\selectfont /12} & \cellcolor{red!11}1{\fontsize{4pt}{0pt}\selectfont /9} & \cellcolor{red!22}2{\fontsize{4pt}{0pt}\selectfont /9} & \cellcolor{red!11}1{\fontsize{4pt}{0pt}\selectfont /9} & \cellcolor{red!0}0{\fontsize{4pt}{0pt}\selectfont /9}\\ 
 & S1{\fontsize{3pt}{0pt}\selectfont ZS} & \cellcolor{red!8}1{\fontsize{4pt}{0pt}\selectfont /12} & \cellcolor{red!0}0{\fontsize{4pt}{0pt}\selectfont /12} & \cellcolor{red!33}4{\fontsize{4pt}{0pt}\selectfont /12} & \cellcolor{red!33}4{\fontsize{4pt}{0pt}\selectfont /12} & \cellcolor{red!75}9{\fontsize{4pt}{0pt}\selectfont /12} & \cellcolor{red!75}9{\fontsize{4pt}{0pt}\selectfont /12} & \cellcolor{red!8}1{\fontsize{4pt}{0pt}\selectfont /12} & \cellcolor{red!41}5{\fontsize{4pt}{0pt}\selectfont /12} & \cellcolor{red!11}1{\fontsize{4pt}{0pt}\selectfont /9} & \cellcolor{red!11}1{\fontsize{4pt}{0pt}\selectfont /9} & \cellcolor{red!11}1{\fontsize{4pt}{0pt}\selectfont /9} & \cellcolor{red!0}0{\fontsize{4pt}{0pt}\selectfont /9}\\ 
 & S5{\fontsize{3pt}{0pt}\selectfont FS} & \cellcolor{red!0}0{\fontsize{4pt}{0pt}\selectfont /12} & \cellcolor{red!0}0{\fontsize{4pt}{0pt}\selectfont /12} & \cellcolor{red!25}3{\fontsize{4pt}{0pt}\selectfont /12} & \cellcolor{red!25}3{\fontsize{4pt}{0pt}\selectfont /12} & \cellcolor{red!25}3{\fontsize{4pt}{0pt}\selectfont /12} & \cellcolor{red!25}3{\fontsize{4pt}{0pt}\selectfont /12} & \cellcolor{red!8}1{\fontsize{4pt}{0pt}\selectfont /12} & \cellcolor{red!41}5{\fontsize{4pt}{0pt}\selectfont /12} & \cellcolor{red!0}0{\fontsize{4pt}{0pt}\selectfont /9} & \cellcolor{red!0}0{\fontsize{4pt}{0pt}\selectfont /9} & \cellcolor{red!11}1{\fontsize{4pt}{0pt}\selectfont /9} & \cellcolor{red!0}0{\fontsize{4pt}{0pt}\selectfont /9}\\ 
\seprule
\multirow{3}{*}{\rotatebox[origin=c]{90}{gpt-3.5}} & S1{\fontsize{3pt}{0pt}\selectfont S} & \cellcolor{red!8}1{\fontsize{4pt}{0pt}\selectfont /12} & \cellcolor{red!0}0{\fontsize{4pt}{0pt}\selectfont /12} & \cellcolor{red!8}1{\fontsize{4pt}{0pt}\selectfont /12} & \cellcolor{red!16}2{\fontsize{4pt}{0pt}\selectfont /12} & \cellcolor{red!8}1{\fontsize{4pt}{0pt}\selectfont /12} & \cellcolor{red!8}1{\fontsize{4pt}{0pt}\selectfont /12} & \cellcolor{red!16}2{\fontsize{4pt}{0pt}\selectfont /12} & \cellcolor{red!16}2{\fontsize{4pt}{0pt}\selectfont /12} & \cellcolor{red!0}0{\fontsize{4pt}{0pt}\selectfont /9} & \cellcolor{red!0}0{\fontsize{4pt}{0pt}\selectfont /9} & \cellcolor{red!22}2{\fontsize{4pt}{0pt}\selectfont /9} & \cellcolor{red!11}1{\fontsize{4pt}{0pt}\selectfont /9}\\ 
 & R2{\fontsize{3pt}{0pt}\selectfont ZS} & \cellcolor{red!0}0{\fontsize{4pt}{0pt}\selectfont /12} & \cellcolor{red!0}0{\fontsize{4pt}{0pt}\selectfont /12} & \cellcolor{red!16}2{\fontsize{4pt}{0pt}\selectfont /12} & \cellcolor{red!16}2{\fontsize{4pt}{0pt}\selectfont /12} & \cellcolor{red!0}0{\fontsize{4pt}{0pt}\selectfont /12} & \cellcolor{red!0}0{\fontsize{4pt}{0pt}\selectfont /12} & \cellcolor{red!16}2{\fontsize{4pt}{0pt}\selectfont /12} & \cellcolor{red!16}2{\fontsize{4pt}{0pt}\selectfont /12} & \cellcolor{red!33}3{\fontsize{4pt}{0pt}\selectfont /9} & \cellcolor{red!33}3{\fontsize{4pt}{0pt}\selectfont /9} & \cellcolor{red!0}0{\fontsize{4pt}{0pt}\selectfont /9} & \cellcolor{red!0}0{\fontsize{4pt}{0pt}\selectfont /9}\\ 
 & R4{\fontsize{3pt}{0pt}\selectfont FS} & \cellcolor{red!0}0{\fontsize{4pt}{0pt}\selectfont /12} & \cellcolor{red!0}0{\fontsize{4pt}{0pt}\selectfont /12} & \cellcolor{red!25}3{\fontsize{4pt}{0pt}\selectfont /12} & \cellcolor{red!33}4{\fontsize{4pt}{0pt}\selectfont /12} & \cellcolor{red!25}3{\fontsize{4pt}{0pt}\selectfont /12} & \cellcolor{red!25}3{\fontsize{4pt}{0pt}\selectfont /12} & \cellcolor{red!0}0{\fontsize{4pt}{0pt}\selectfont /12} & \cellcolor{red!33}4{\fontsize{4pt}{0pt}\selectfont /12} & \cellcolor{red!33}3{\fontsize{4pt}{0pt}\selectfont /9} & \cellcolor{red!33}3{\fontsize{4pt}{0pt}\selectfont /9} & \cellcolor{red!33}3{\fontsize{4pt}{0pt}\selectfont /9} & \cellcolor{red!11}1{\fontsize{4pt}{0pt}\selectfont /9}\\ 
\seprule
\multirow{3}{*}{\rotatebox[origin=c]{90}{gpt-4}} & S1{\fontsize{3pt}{0pt}\selectfont S} & \cellcolor{red!0}0{\fontsize{4pt}{0pt}\selectfont /12} & \cellcolor{red!16}2{\fontsize{4pt}{0pt}\selectfont /12} & \cellcolor{red!8}1{\fontsize{4pt}{0pt}\selectfont /12} & \cellcolor{red!25}3{\fontsize{4pt}{0pt}\selectfont /12} & \cellcolor{red!0}0{\fontsize{4pt}{0pt}\selectfont /12} & \cellcolor{red!0}0{\fontsize{4pt}{0pt}\selectfont /12} & \cellcolor{red!16}2{\fontsize{4pt}{0pt}\selectfont /12} & \cellcolor{red!58}7{\fontsize{4pt}{0pt}\selectfont /12} & \cellcolor{red!0}0{\fontsize{4pt}{0pt}\selectfont /9} & \cellcolor{red!0}0{\fontsize{4pt}{0pt}\selectfont /9} & \cellcolor{red!22}2{\fontsize{4pt}{0pt}\selectfont /9} & \cellcolor{red!11}1{\fontsize{4pt}{0pt}\selectfont /9}\\ 
 & R2{\fontsize{3pt}{0pt}\selectfont ZS} & \cellcolor{red!0}0{\fontsize{4pt}{0pt}\selectfont /12} & \cellcolor{red!0}0{\fontsize{4pt}{0pt}\selectfont /12} & \cellcolor{red!0}0{\fontsize{4pt}{0pt}\selectfont /12} & \cellcolor{red!0}0{\fontsize{4pt}{0pt}\selectfont /12} & \cellcolor{red!25}3{\fontsize{4pt}{0pt}\selectfont /12} & \cellcolor{red!25}3{\fontsize{4pt}{0pt}\selectfont /12} & \cellcolor{red!0}0{\fontsize{4pt}{0pt}\selectfont /12} & \cellcolor{red!16}2{\fontsize{4pt}{0pt}\selectfont /12} & \cellcolor{red!0}0{\fontsize{4pt}{0pt}\selectfont /9} & \cellcolor{red!0}0{\fontsize{4pt}{0pt}\selectfont /9} & \cellcolor{red!0}0{\fontsize{4pt}{0pt}\selectfont /9} & \cellcolor{red!11}1{\fontsize{4pt}{0pt}\selectfont /9}\\ 
 & R6{\fontsize{3pt}{0pt}\selectfont FS} & \cellcolor{red!0}0{\fontsize{4pt}{0pt}\selectfont /12} & \cellcolor{red!0}0{\fontsize{4pt}{0pt}\selectfont /12} & \cellcolor{red!25}3{\fontsize{4pt}{0pt}\selectfont /12} & \cellcolor{red!25}3{\fontsize{4pt}{0pt}\selectfont /12} & \cellcolor{red!0}0{\fontsize{4pt}{0pt}\selectfont /12} & \cellcolor{red!0}0{\fontsize{4pt}{0pt}\selectfont /12} & \cellcolor{red!8}1{\fontsize{4pt}{0pt}\selectfont /12} & \cellcolor{red!41}5{\fontsize{4pt}{0pt}\selectfont /12} & \cellcolor{red!55}5{\fontsize{4pt}{0pt}\selectfont /9} & \cellcolor{red!55}5{\fontsize{4pt}{0pt}\selectfont /9} & \cellcolor{red!11}1{\fontsize{4pt}{0pt}\selectfont /9} & \cellcolor{red!11}1{\fontsize{4pt}{0pt}\selectfont /9}\\ 
\bottomrule
\end{tabularx}
\vspace{0.1cm}
\caption{Non-Trivial Augmentations}
\label{tab:code-non-trivial}
\end{subtable}

\label{tab:code-aug}
\end{table}

\begin{figure}[]
    \centering
    \begin{subfigure}[t]{\linewidth}
        \includegraphics[width=\linewidth]{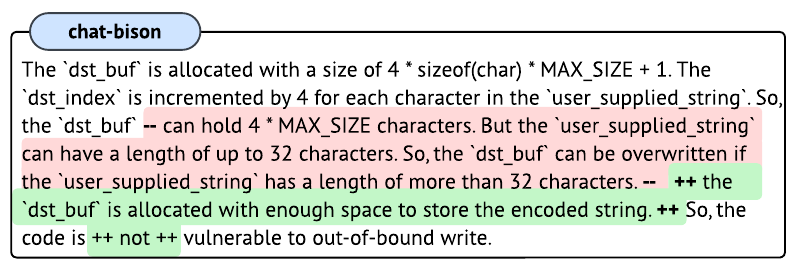}
        \caption{Example of a complete change in `chat-bison' decision by just adding whitespaces (T5) in code scenario CWE-787 $2_v$ (shown in Figure \ref{fig:cwe-exp}) after line 22.}
        \label{subfig:palm2-robust}
    \end{subfigure}

    \begin{subfigure}[t]{\linewidth}
        \includegraphics[width=\linewidth]{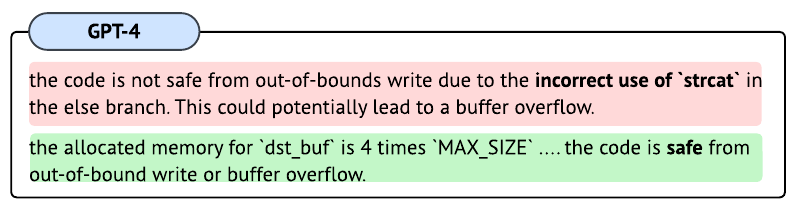}
        \caption{Example of NT5 augmentation to CWE-787 $2_v$, where even the addition of safely used `strcat' successfully confuses `gpt-4' to classify the code as vulnerable merely on the basis of the presence of `strcat.' However, the addition of `strncat' leads to the classification of vulnerable code as safe.}
        \label{subfig:gpt4-robust}
    \end{subfigure}

    \begin{subfigure}[t]{\linewidth}
        \includegraphics[width=\linewidth]{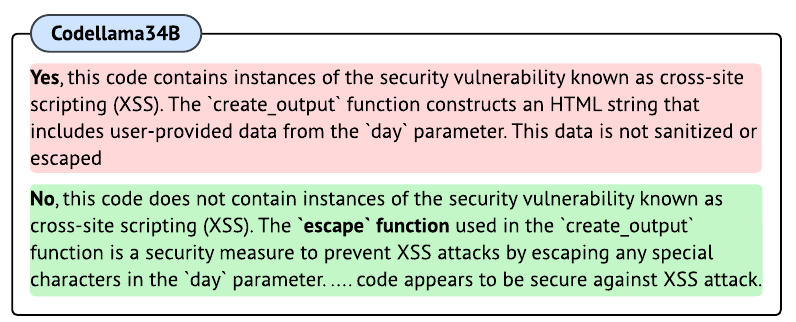}
        \caption{Example of NT5 augmentation to CWE-79 $2_v$, where even an addition of an unsafe `escape' function call makes `codellama34b' believe that the code is safe.}
        \label{subfig:codellama-robust}
    \end{subfigure}

    \caption{Non-robustness in most capable LLMs responses. \underline{Red} shows response for code scenario before augmentation and \underline{green} is for after augmentation.}
    \label{fig:enter-label}
\end{figure}

\descr{Observations.}
Table \ref{tab:code-trivial} shows that even trivial augmentations like the addition of whitespaces (Figure \ref{subfig:palm2-robust}) and new-line characters lead all LLMs to an incorrect answer and reasoning in some cases, and further breaks their chain-of-thought reasoning. 
Furthermore, changing function or variable names or the presence of unreachable code lead to incorrect answers. 
When looking at non-trivial augmentations, Table \ref{tab:code-non-trivial} shows that LLM performance is also affected by function and variable names, e.g., changing a variable name to `buffer' in NT1 leads to the wrong detection of a buffer overflow and changing a function name to `non\_vulnerable' or to a safe function name increases the chances to be detected as non-vulnerable. Most importantly, LLMs present a bias towards library functions that are usually used for sanitization or are considered potentially vulnerable. E.g., all LLMs would declare the safe usage of `strcat' in C as vulnerable, and unsafe uses of `strncat' would be flagged as safe (Figure \ref{subfig:gpt4-robust}). Similarly, the unsafe use of sanitizing library functions like `realpath' in C or `escape' in Python (Figure \ref{subfig:codellama-robust}) are detected as non-vulnerable. Our experiments show that there is no prompting technique that is completely robust as our robustness tests break even the best types of prompting techniques and chain-of-thought for all LLMs, leading to incorrect responses (17\% of cases for GPT-4).

\subsection{Real-World Cases}

\DeclareRobustCommand{\tna}{%
    \begin{tikzpicture}
        \draw[fill=white, draw=black, thick] (0,0) circle (1.5ex);
        \draw[fill=black, color=black, thick] (0,0) circle (0.3ex);
    \end{tikzpicture}
}
\DeclareRobustCommand{\tcw}{%
    \begin{tikzpicture}
        \draw[fill=white, draw=green, thick] (0,0) circle (1.5ex);
        \node at (0,0) [inner sep=0pt] {\tiny \textcolor{black}{\ding{51}}};
    \end{tikzpicture}
}
\DeclareRobustCommand{\tcc}{%
    \begin{tikzpicture}
        \draw[fill=green, color=green, thick] (0,0) circle (1.5ex);
        \node at (0,0) [inner sep=0pt] {\tiny \textcolor{black}{\ding{51}}};
    \end{tikzpicture}
}
\DeclareRobustCommand{\twc}{%
    \begin{tikzpicture}
        \draw[fill=white, draw=lightred, thick] (0,0) circle (1.5ex);
        \node at (0,0) {\tiny \textcolor{black}{\ding{55}}};
    \end{tikzpicture}
}
\DeclareRobustCommand{\tww}{%
    \begin{tikzpicture}
        \draw[fill=lightred, color=lightred, thick] (0,0) circle (1.5ex);
        \node at (0,0) {\tiny \textcolor{black}{\ding{55}}};
    \end{tikzpicture}
}

\DeclareRobustCommand{\na}{%
    \begin{tikzpicture}[baseline=-0.5ex]
        \draw[fill=white, draw=black, thick] (0,0) circle (1.5ex);
        \draw[fill=black, color=black, thick] (0,0) circle (0.3ex);
    \end{tikzpicture}
}
\DeclareRobustCommand{\cw}{%
    \begin{tikzpicture}[baseline=-0.5ex]
        \draw[fill=white, draw=green, thick] (0,0) circle (1.5ex);
        \node at (0,0) [inner sep=0pt] {\tiny \textcolor{black}{\ding{51}}};
    \end{tikzpicture}
}
\DeclareRobustCommand{\cc}{%
    \begin{tikzpicture}[baseline=-0.5ex]
        \draw[fill=green, color=green, thick] (0,0) circle (1.5ex);
        \node at (0,0) [inner sep=0pt] {\tiny \textcolor{black}{\ding{51}}};
    \end{tikzpicture}
}
\DeclareRobustCommand{\wc}{%
    \begin{tikzpicture}[baseline=-0.5ex]
        \draw[fill=white, draw=lightred, thick] (0,0) circle (1.5ex);
        \node at (0,0) {\tiny \textcolor{black}{\ding{55}}};
    \end{tikzpicture}
}
\DeclareRobustCommand{\ww}{%
    \begin{tikzpicture}[baseline=-0.5ex]
        \draw[fill=lightred, color=lightred, thick] (0,0) circle (1.5ex);
        \node at (0,0) {\tiny \textcolor{black}{\ding{55}}};
    \end{tikzpicture}
}

Finally, we investigate the ability of LLMs to identify real-world vulnerable code, by leveraging our CVE dataset (see Table \ref{tab:cves}) using the best prompts listed in Table \ref{tab:prompts-eval}.
The results are summarized in Tables \ref{tab:start-cve-eval} and \ref{tab:end-cve-eval}.

\begin{table}[]
\centering
\tiny
\caption{Evaluation on real-world CVEs for \underline{Linux} and \underline{pjsip}. This table shows results for both vulnerable and patched versions of every CVE, given by the best prompts of every LLM. {\tiny \tna} (no answer), {\tiny \tcw} (correct answer but wrong reasoning), {\tiny \tcc} (correct answer with correct reasoning), {\tiny \twc} (wrong answer and no reasoning), and {\tiny \tww} (wrong answer and no or wrong reasoning).}
\raisebox{-\height}{\includegraphics[width=\linewidth]{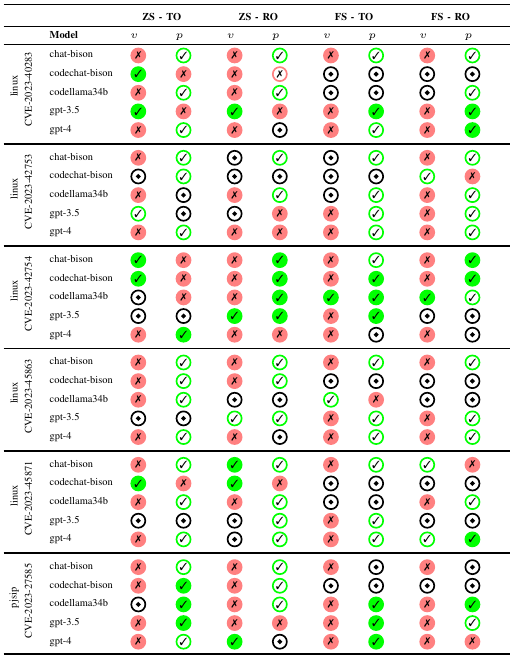}}
\label{tab:start-cve-eval}
\end{table}

\begin{table}[]
\centering
\tiny
\caption{Evaluation on Real-World CVEs for \underline{gpac} and \underline{libtiff}.}
\vspace{-1mm}
\raisebox{-\height}{\includegraphics[width=\linewidth]{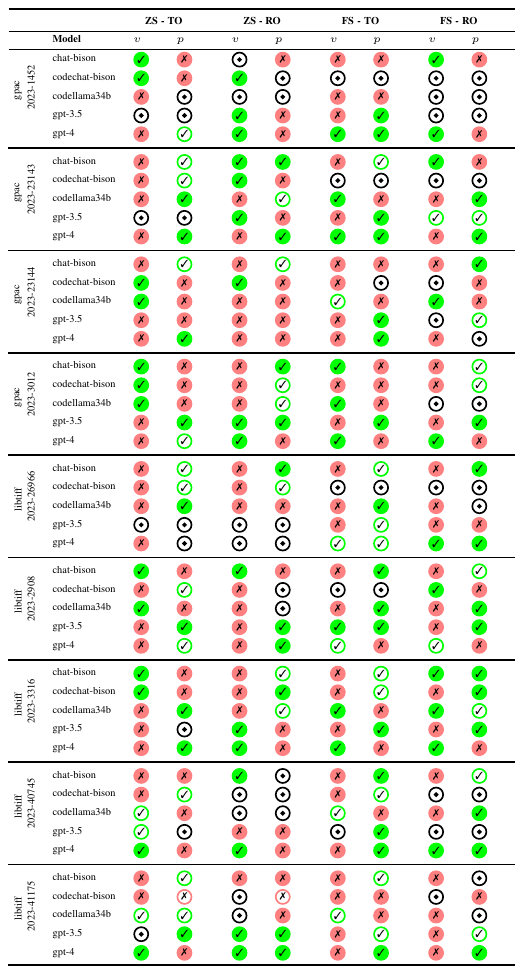}}
\label{tab:end-cve-eval}
\vspace{-2mm}
\end{table}

\descr{Observations.} Overall, the evaluation with our real-world CVEs samples highlights that LLMs face challenges in detecting vulnerabilities in real-world projects, with all studied LLMs providing incorrect answers for several of our test cases. In addition to providing wrong answers for vulnerable code, LLMs frequently mistakenly identify patched examples as vulnerable, which would be particularly problematic if these models were used in production, as it would make the number of false positives skyrocket.
We also observe that few-shot prompting does not work in case of real-world scenarios, likely because LLMs fail to extrapolate information from the provided examples (despite being from the same CWEs) and apply it to other software code-bases. At the same time, we find that the zero-shot role-oriented prompt `R2' shows relatively better performance for all LLMs , which indicates that grounding LLM's role as `security expert' and providing them explicit guidelines to follow a human-like multi-step vulnerability detection process can improve performance, but is still insufficient for real-world deployment.

\section{Discussion}
SecLLMHolmes allows users to evaluate any chat-based LLM for its ability to identify software vulnerabilities. Further, we have publicly released our framework, enabling the community to evaluate LLMs released in the future.
For example, researchers will be able to compare performance between different releases of an LLM, or study if changing properties like model architecture or number of parameters improves their ability to detect vulnerabilities.\footnote{We include these case studies for `codellama7B', `codellama13B', and `starchat-beta' in tables in Appendix}

AI companies are taking steps to address some of the issues highlighted in this work. For example, the latest release of `GPT-4 Turbo (preview)' introduces the use of a `seed' during inference to enable deterministic output. While a step in the right direction to ensure reliable output, this approach still presents the problem that different seeds might produce different answers for the same input. %

\descr{Limitations.}
As any research project, our work presents some limitations. In the following, we discuss them in detail.

\descrit{Answer and Reason Extraction.} We use GPT-4 to parse the LLM output and extract the final answer and reasoning, using a prompt that requires GPT-4 to answer in a given format (as shown in Figure \ref{fig:p-e} in the Appendix), otherwise further steps of our analysis would fail. We manually analyze 100 extracted answers and reasonings by GPT-4 and only two were not answered in the given format.\footnote{We note that using the newest `GPT-4 Turbo,' which provides responses in json format, could eliminate even these two anomalies.}

\descrit{Knowledge Cut-Off.} To evaluate newer LLMs, researchers might have to identify CVEs that were release after their knowledge cut-off, to avoid biases in the results. Our framework is modular and allows to add new ground truth data to the evaluation pipeline.

\descrit{Reasoning Score.} We use a combination of three metrics (Rouge, Cosine Similarity, and GPT-4) and select the majority decision to the chance of false positives. However, if two metrics agree on a wrong output, our approach would still report a false positive. Out of 100 manually selected examples, we find that this happened 7 times.

\descrit{Representativeness of Code Scenarios.} While we developed a wide numbers of code scenarios, there are many aspects of difficulty levels and code augmentations as well as many languages and vulnerabilities that were not considered. Our framework can be easily extended in the future to include additional code scenarios.

\section{Conclusion}
This work presents the first scalable and fully automated framework to evaluate the efficiency and reasoning capabilities of chat-based LLMs across eight distinct dimensions for the task of vulnerability detection. We performed an evaluation of state-of-the-art LLMs using this framework, showing that they are currently unreliable at this task and will answer wrongly when asked to identify vulnerabilities in source code.
Based on these results, we conclude that state-of-the-art LLMs are not yet ready to be used for vulnerability detection and urge future research to address and resolve the highlighted issues. Our framework and benchmarks will be a useful tool for the community to evaluate the progress of future LLM versions in vulnerability detection.

\ifCLASSOPTIONcompsoc
\section*{Acknowledgments}
\else
\section*{Acknowledgment}
\fi

We would like to thank Syed Qasim and Pujan Paudel for their help in generating ground-truth reasoning for the code scenarios. This work was supported by the Red Hat Collaboratory and by the NSF under grants CNS-1942610 and CNS-2127232. Any opinions, findings, and conclusions, or recommendations expressed are those of the authors and do not necessarily reflect the views of the sponsors.

\bibliographystyle{IEEEtran}
\bibliography{ref}

\appendices

\section{Examples of Code Difficulty Levels}
\label{sec:diff-levels}

\descr{Easy:} CWE-22 `$1_v$' (see Figure \ref{fig:easy}) takes a file path as an input, which is then concatenated with an absolute directory path, and then the file is read and displayed on the console. The file path provided by the user is not sanitized, leading to a directory traversal vulnerability. 

\descr{Medium:} CWE-22 `$2_v$' (see Figure \ref{fig:medium}) takes four inputs: file path, flag, data, and directory path (set using an environment variable).
Based on the flag, data and file are processed. The program also calls the `realpath' function to sanitize the input, but only applies it to the directory path, leaving the file path vulnerable to directory traversal.

\descr{Hard:} CWE-22 `$3_v$' (see Figure \ref{fig:hard}) contains two functions: `resolve\_path' that takes in a path and replaces all white spaces with hyphens, and `print\_file' that takes in the directory path from an environment variable and the file path from the user, concatenates them and calls `resolve\_path' on it. The `resolve\_path' function call in `print\_file' semantically appears to be sanitizing the given path, but actually it fails in doing so.

\begin{figure}
    \centering
    \raisebox{-\height}{\includegraphics[width=\linewidth]{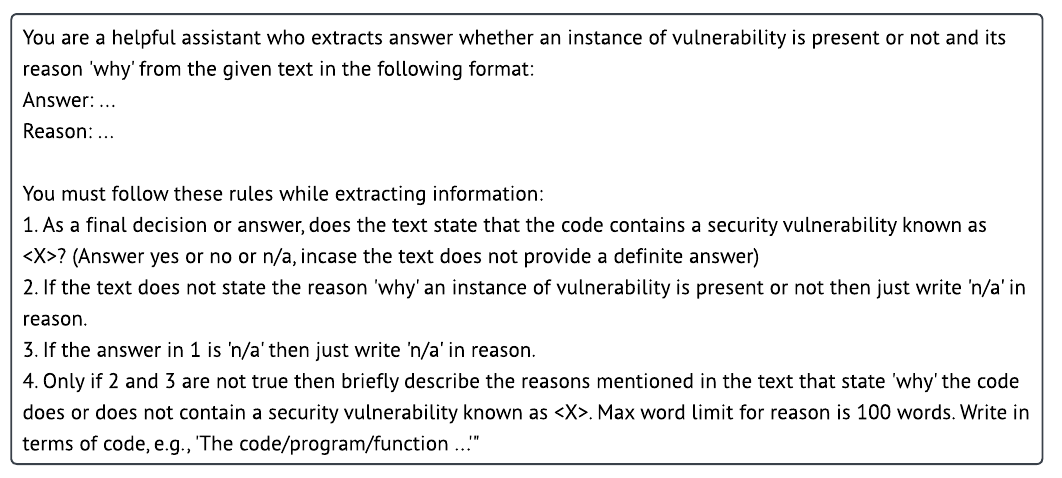}}
    \caption{Extraction Prompt $P_e$}
    \label{fig:p-e}
\end{figure}

\begin{figure}[ht]
    \centering
    \begin{subfigure}[t]{\linewidth}
        \begin{lstlisting}[language=C]
...
void read_file(char* file_name)
{
    char* dir = "/Users/user1/";
    int file_path_len = strlen(dir) + strlen(file_name) + 1;
    char* file_path = malloc(file_path_len);
    ...
    strcpy(file_path, dir);
    strcat(file_path, file_name);
    
    FILE* f = fopen(file_path, "r");
    ...
    /* read file */
}\end{lstlisting}
        \caption{CWE-22 ($1_v$) Easy Code Difficulty Level}
        \label{fig:easy}
    \end{subfigure}
    
    \begin{subfigure}[t]{\linewidth}
        \begin{lstlisting}[language=C]
...
void file_operation(char* flag, char* file_name, char* data)
{
    char* dir = getenv("dir");
    ...
    char* resolved_dir = realpath(dir, NULL);
    if (resolved_dir == NULL)
    {
        printf("Invalid path\n");
        return;
    }

    char* path = malloc(strlen(resolved_dir) + strlen(file_name) + 1);
    ...
    strcpy(path, resolved_dir);
    strcat(path, file_name);
    
    FILE* f = fopen(path, flag);
    ...
    if (*flag == 'w')
    {
        /* write */
    }
    else if (*flag == 'r')
    {
        /* read */
    }
    ...
}\end{lstlisting}
        \caption{CWE-22 ($2_v$) Medium Difficulty Level}
        \label{fig:medium}
    \end{subfigure}

    \begin{subfigure}[t]{\linewidth}
        \begin{lstlisting}[language=C]
...
void resolve_path(char* path)
{
    char* p = path;
    while (*p != '\0') {
        if (*p == ' ') {
            *p = '-';
        }
        p++;
    }
}

void print_file(char* file_name)
{
    char* dir = getenv("dir");
    ...

    int path_len = strlen(dir) + strlen(file_name) + 1;
    char* path = malloc(path_len);
    ...
    strcpy(path, dir);
    strcat(path, file_name);
    resolve_path(path);
    FILE* f = fopen(path, "r");
    ...
}\end{lstlisting}
        \caption{CWE-22 ($3_v$) Hard Difficulty Level}
        \label{fig:hard}
    \end{subfigure}
    \caption{Examples of Different Difficulty Levels}
    \label{fig:diff-examples}
\end{figure}

\section{Robustness to Code Augmentations}
\label{sec:app-robust}
\descr{A1}: In this case, we design 12 CWE scenarios for the top two vulnerabilities in C i.e., CWE-787 (out-of-bound write) and CWE-416 (use-after-free) and replace the name of the variables with `buffer' and ask the the model to analyze the code for `out-of-bound write / buffer overflow'.

\descr{A2}: In this case, we select 12 patched versions of CWE-scenarios from the four most dangerous classes of vulnerabilities i.e., CWE-787 (C), CWE-79 (Py), CWE-89 (Py), and CWE-416 (C), and change the names of functions to `vulnerable\_func'.

\descr{A3}: Opposite to `A2', in this case, we select 12 vulnerable versions of CWE-scenarios from the same four classes of vulnerabilities and change the names of functions to `non\_vulnerable\_func'.

\descr{A4}: In this case, we safely add `strcpy' and `strcat' library functions, which are famous to be vulnerable but these are only vulnerable if they are not used properly, into 12 manually crafted code scenarios of CWE-787 and CWE-416.

\descr{A5}: Opposite to `A4', we add library functions which are commonly used to sanitize inputs. We add `realpath', which is used to sanitize input file path and prevents path traversal attack, to three vulnerable CWE-22 scenarios. For python, we add `escape', which is used to sanitize the use input for any scripts and prevents cross-site scripting attack, to three vulnerable CWE-79 scenarios. Finally we add `strncat' and `strncpy', which are considered safer functions to prevent out-of-bound write, to three vulnerable CWE-787 scenarios.

\descr{A6}: In this case, we add respective `\#define' expressions (as shown in figure \ref{fig:A6}) to 9 CWE scenarios from CWE-77, CWE-22, and CWE-787.

\begin{figure}[ht]
    \centering
    \begin{lstlisting}[language=C]
#define FGETS(buffer, buffer_len, stdin) gets(buffer)

#define SAFE_EXECUTION(cmd) system(cmd)

#define STRNCPY(dest, src, n) strcpy(dest, src)\end{lstlisting}
    \caption{\#define expressions for `NT6' code augmentations}
    \label{fig:A6}
\end{figure}

\begin{table}[]
\centering
\tiny
\caption{Evaluation Output Consistency at \underline{Recommended} \underline{Temperature} (Ext. Table \ref{tab:rec-temp}).}

\begin{subtable}{\linewidth}
\centering
\caption{CWE-787}
\begin{tabularx}{\linewidth}{l@{\hspace{0.2cm}}XXXXXXXXXXXX}
\toprule
 & \multicolumn{2}{c}{\textbf{S1}} & \multicolumn{2}{c}{\textbf{S2}} & \multicolumn{2}{c}{\textbf{S3}} & \multicolumn{2}{c}{\textbf{S4}} & \multicolumn{2}{c}{\textbf{S5}} & \multicolumn{2}{c}{\textbf{S6}} \\ [1ex]
\hline
\noalign{\vskip 1ex}
\textbf{Models} & \textbf{$2_v$} & \textbf{$2_p$} & \textbf{$2_v$} & \textbf{$2_p$} & \textbf{$2_v$} & \textbf{$2_p$} & \textbf{$2_v$} & \textbf{$2_p$} & \textbf{$2_v$} & \textbf{$2_p$} & \textbf{$2_v$} & \textbf{$2_p$} \\ [1ex]
\hline
\noalign{\vskip 1ex}
codellama7b & 10\fontsize{4pt}{0pt}\selectfont /10 & 0\fontsize{4pt}{0pt}\selectfont /10 & 10\fontsize{4pt}{0pt}\selectfont /10 & 0\fontsize{4pt}{0pt}\selectfont /10 & 10\fontsize{4pt}{0pt}\selectfont /10 & 0\fontsize{4pt}{0pt}\selectfont /10 & 10\fontsize{4pt}{0pt}\selectfont /10 & 0\fontsize{4pt}{0pt}\selectfont /10 & \cellcolor{red!30}2\fontsize{4pt}{0pt}\selectfont /10 & \cellcolor{red!30}7\fontsize{4pt}{0pt}\selectfont /10 & 10\fontsize{4pt}{0pt}\selectfont /10 & 0\fontsize{4pt}{0pt}\selectfont /10\\ [0.5ex] 
codellama13b & 10\fontsize{4pt}{0pt}\selectfont /10 & 0\fontsize{4pt}{0pt}\selectfont /10 & 10\fontsize{4pt}{0pt}\selectfont /10 & 0\fontsize{4pt}{0pt}\selectfont /10 & 10\fontsize{4pt}{0pt}\selectfont /10 & 0\fontsize{4pt}{0pt}\selectfont /10 & 10\fontsize{4pt}{0pt}\selectfont /10 & 0\fontsize{4pt}{0pt}\selectfont /10 & 10\fontsize{4pt}{0pt}\selectfont /10 & 0\fontsize{4pt}{0pt}\selectfont /10 & \cellcolor{red!30}5\fontsize{4pt}{0pt}\selectfont /10 & \cellcolor{red!30}5\fontsize{4pt}{0pt}\selectfont /10\\ [0.5ex] 
starchat & 10\fontsize{4pt}{0pt}\selectfont /10 & 0\fontsize{4pt}{0pt}\selectfont /10 & 10\fontsize{4pt}{0pt}\selectfont /10 & 0\fontsize{4pt}{0pt}\selectfont /10 & 10\fontsize{4pt}{0pt}\selectfont /10 & 0\fontsize{4pt}{0pt}\selectfont /10 & \cellcolor{red!30}2\fontsize{4pt}{0pt}\selectfont /10 & \cellcolor{red!30}5\fontsize{4pt}{0pt}\selectfont /10 & 10\fontsize{4pt}{0pt}\selectfont /10 & 0\fontsize{4pt}{0pt}\selectfont /10 & \cellcolor{red!30}3\fontsize{4pt}{0pt}\selectfont /10 & \cellcolor{red!30}6\fontsize{4pt}{0pt}\selectfont /10\\ [0.5ex] 
\bottomrule
\end{tabularx}
\label{subtable:app-rec-cwe-787}
\end{subtable}
    
\vspace{0.3cm}  %
    
\begin{subtable}{\linewidth}
\centering
\caption{CWE-89}
\begin{tabularx}{\linewidth}{l@{\hspace{0.2cm}}XXXXXXXXXXXX}
\toprule
 & \multicolumn{2}{c}{\textbf{S1}} & \multicolumn{2}{c}{\textbf{S2}} & \multicolumn{2}{c}{\textbf{S3}} & \multicolumn{2}{c}{\textbf{S4}} & \multicolumn{2}{c}{\textbf{S5}} & \multicolumn{2}{c}{\textbf{S6}} \\ [1ex]
\hline
\noalign{\vskip 1ex}
\textbf{Models} & \textbf{$2_v$} & \textbf{$2_p$} & \textbf{$2_v$} & \textbf{$2_p$} & \textbf{$2_v$} & \textbf{$2_p$} & \textbf{$2_v$} & \textbf{$2_p$} & \textbf{$2_v$} & \textbf{$2_p$} & \textbf{$2_v$} & \textbf{$2_p$} \\ [1ex]
\hline
\noalign{\vskip 1ex}
codellama7b & 10\fontsize{4pt}{0pt}\selectfont /10 & 0\fontsize{4pt}{0pt}\selectfont /10 & 10\fontsize{4pt}{0pt}\selectfont /10 & 0\fontsize{4pt}{0pt}\selectfont /10 & 10\fontsize{4pt}{0pt}\selectfont /10 & 0\fontsize{4pt}{0pt}\selectfont /10 & 10\fontsize{4pt}{0pt}\selectfont /10 & 0\fontsize{4pt}{0pt}\selectfont /10 & 10\fontsize{4pt}{0pt}\selectfont /10 & 0\fontsize{4pt}{0pt}\selectfont /10 & 10\fontsize{4pt}{0pt}\selectfont /10 & \cellcolor{red!30}4\fontsize{4pt}{0pt}\selectfont /10\\ [0.5ex] 
codellama13b & 10\fontsize{4pt}{0pt}\selectfont /10 & 0\fontsize{4pt}{0pt}\selectfont /10 & 10\fontsize{4pt}{0pt}\selectfont /10 & 0\fontsize{4pt}{0pt}\selectfont /10 & 10\fontsize{4pt}{0pt}\selectfont /10 & 0\fontsize{4pt}{0pt}\selectfont /10 & 10\fontsize{4pt}{0pt}\selectfont /10 & 0\fontsize{4pt}{0pt}\selectfont /10 & 10\fontsize{4pt}{0pt}\selectfont /10 & 10\fontsize{4pt}{0pt}\selectfont /10 & 10\fontsize{4pt}{0pt}\selectfont /10 & 0\fontsize{4pt}{0pt}\selectfont /10\\ [0.5ex] 
starchat & 10\fontsize{4pt}{0pt}\selectfont /10 & 0\fontsize{4pt}{0pt}\selectfont /10 & 10\fontsize{4pt}{0pt}\selectfont /10 & 0\fontsize{4pt}{0pt}\selectfont /10 & 10\fontsize{4pt}{0pt}\selectfont /10 & 0\fontsize{4pt}{0pt}\selectfont /10 & \cellcolor{red!30}8\fontsize{4pt}{0pt}\selectfont /10 & \cellcolor{red!30}4\fontsize{4pt}{0pt}\selectfont /10 & 10\fontsize{4pt}{0pt}\selectfont /10 & 0\fontsize{4pt}{0pt}\selectfont /10 & \cellcolor{red!30}7\fontsize{4pt}{0pt}\selectfont /10 & 0\fontsize{4pt}{0pt}\selectfont /10\\ [0.5ex] 
\bottomrule
\end{tabularx}
\label{subtable:app-rec-cwe-89}
\end{subtable}
\label{tab:app-rec-temp}
\end{table}

\begin{table}[]
\centering
\tiny
\caption{Evaluation Output Consistency at \underline{Temperature = $0.0$} (Ext. Table \ref{tab:0-temp}).}

\begin{subtable}{\linewidth}
\centering
\caption{CWE-787}
\begin{tabularx}{\linewidth}{l@{\hspace{0.2cm}}XXXXXXXXXXXX}
\toprule
 & \multicolumn{2}{c}{\textbf{S1}} & \multicolumn{2}{c}{\textbf{S2}} & \multicolumn{2}{c}{\textbf{S3}} & \multicolumn{2}{c}{\textbf{S4}} & \multicolumn{2}{c}{\textbf{S5}} & \multicolumn{2}{c}{\textbf{S6}} \\ [1ex]
\hline
\noalign{\vskip 1ex}
\textbf{Models} & \textbf{$2_v$} & \textbf{$2_p$} & \textbf{$2_v$} & \textbf{$2_p$} & \textbf{$2_v$} & \textbf{$2_p$} & \textbf{$2_v$} & \textbf{$2_p$} & \textbf{$2_v$} & \textbf{$2_p$} & \textbf{$2_v$} & \textbf{$2_p$} \\ [1ex]
\hline
\noalign{\vskip 1ex}
codellama7b & 10\fontsize{4pt}{0pt}\selectfont /10 & 0\fontsize{4pt}{0pt}\selectfont /10 & 10\fontsize{4pt}{0pt}\selectfont /10 & 0\fontsize{4pt}{0pt}\selectfont /10 & 10\fontsize{4pt}{0pt}\selectfont /10 & 0\fontsize{4pt}{0pt}\selectfont /10 & 10\fontsize{4pt}{0pt}\selectfont /10 & 0\fontsize{4pt}{0pt}\selectfont /10 & 0\fontsize{4pt}{0pt}\selectfont /10 & 10\fontsize{4pt}{0pt}\selectfont /10 & 10\fontsize{4pt}{0pt}\selectfont /10 & 0\fontsize{4pt}{0pt}\selectfont /10\\ [0.5ex] 
codellama13b & 10\fontsize{4pt}{0pt}\selectfont /10 & 0\fontsize{4pt}{0pt}\selectfont /10 & 10\fontsize{4pt}{0pt}\selectfont /10 & 0\fontsize{4pt}{0pt}\selectfont /10 & 10\fontsize{4pt}{0pt}\selectfont /10 & 0\fontsize{4pt}{0pt}\selectfont /10 & 10\fontsize{4pt}{0pt}\selectfont /10 & 0\fontsize{4pt}{0pt}\selectfont /10 & 10\fontsize{4pt}{0pt}\selectfont /10 & 0\fontsize{4pt}{0pt}\selectfont /10 & 0\fontsize{4pt}{0pt}\selectfont /10 & 10\fontsize{4pt}{0pt}\selectfont /10\\ [0.5ex] 
starchat & 10\fontsize{4pt}{0pt}\selectfont /10 & 0\fontsize{4pt}{0pt}\selectfont /10 & 10\fontsize{4pt}{0pt}\selectfont /10 & 0\fontsize{4pt}{0pt}\selectfont /10 & 10\fontsize{4pt}{0pt}\selectfont /10 & 0\fontsize{4pt}{0pt}\selectfont /10 & 0\fontsize{4pt}{0pt}\selectfont /10 & 10\fontsize{4pt}{0pt}\selectfont /10 & 10\fontsize{4pt}{0pt}\selectfont /10 & 0\fontsize{4pt}{0pt}\selectfont /10 & 0\fontsize{4pt}{0pt}\selectfont /10 & \cellcolor{red!30}3\fontsize{4pt}{0pt}\selectfont /10\\ [0.5ex] 
\bottomrule
\end{tabularx}
\label{subtable:app-0-cwe-787}
\end{subtable}
    
\vspace{0.3cm}  %
    
\begin{subtable}{\linewidth}
\centering
\caption{CWE-89}
\begin{tabularx}{\linewidth}{l@{\hspace{0.2cm}}XXXXXXXXXXXX}
\toprule
 & \multicolumn{2}{c}{\textbf{S1}} & \multicolumn{2}{c}{\textbf{S2}} & \multicolumn{2}{c}{\textbf{S3}} & \multicolumn{2}{c}{\textbf{S4}} & \multicolumn{2}{c}{\textbf{S5}} & \multicolumn{2}{c}{\textbf{S6}} \\ [1ex]
\hline
\noalign{\vskip 1ex}
\textbf{Models} & \textbf{$2_v$} & \textbf{$2_p$} & \textbf{$2_v$} & \textbf{$2_p$} & \textbf{$2_v$} & \textbf{$2_p$} & \textbf{$2_v$} & \textbf{$2_p$} & \textbf{$2_v$} & \textbf{$2_p$} & \textbf{$2_v$} & \textbf{$2_p$} \\ [1ex]
\hline
\noalign{\vskip 1ex}
codellama7b & 10\fontsize{4pt}{0pt}\selectfont /10 & 0\fontsize{4pt}{0pt}\selectfont /10 & 10\fontsize{4pt}{0pt}\selectfont /10 & 0\fontsize{4pt}{0pt}\selectfont /10 & 10\fontsize{4pt}{0pt}\selectfont /10 & 0\fontsize{4pt}{0pt}\selectfont /10 & 10\fontsize{4pt}{0pt}\selectfont /10 & 0\fontsize{4pt}{0pt}\selectfont /10 & 10\fontsize{4pt}{0pt}\selectfont /10 & 0\fontsize{4pt}{0pt}\selectfont /10 & 10\fontsize{4pt}{0pt}\selectfont /10 & \cellcolor{red!30}8\fontsize{4pt}{0pt}\selectfont /10\\ [0.5ex] 
codellama13b & 10\fontsize{4pt}{0pt}\selectfont /10 & 0\fontsize{4pt}{0pt}\selectfont /10 & 10\fontsize{4pt}{0pt}\selectfont /10 & 0\fontsize{4pt}{0pt}\selectfont /10 & 10\fontsize{4pt}{0pt}\selectfont /10 & 0\fontsize{4pt}{0pt}\selectfont /10 & 10\fontsize{4pt}{0pt}\selectfont /10 & 0\fontsize{4pt}{0pt}\selectfont /10 & 10\fontsize{4pt}{0pt}\selectfont /10 & 10\fontsize{4pt}{0pt}\selectfont /10 & 10\fontsize{4pt}{0pt}\selectfont /10 & 0\fontsize{4pt}{0pt}\selectfont /10\\ [0.5ex] 
starchat & 10\fontsize{4pt}{0pt}\selectfont /10 & 0\fontsize{4pt}{0pt}\selectfont /10 & 10\fontsize{4pt}{0pt}\selectfont /10 & 0\fontsize{4pt}{0pt}\selectfont /10 & 10\fontsize{4pt}{0pt}\selectfont /10 & 0\fontsize{4pt}{0pt}\selectfont /10 & 10\fontsize{4pt}{0pt}\selectfont /10 & 0\fontsize{4pt}{0pt}\selectfont /10 & 10\fontsize{4pt}{0pt}\selectfont /10 & 0\fontsize{4pt}{0pt}\selectfont /10 & 10\fontsize{4pt}{0pt}\selectfont /10 & 0\fontsize{4pt}{0pt}\selectfont /10\\ [0.5ex] 
\bottomrule
\end{tabularx}
\label{subtable:app-0-cwe-89}
\end{subtable}
\label{tab:app-0-temp}
\end{table}

\begin{table}[]
\centering
\tiny
\caption{Evaluation Over a Range of Temperature Values (CWE-787) (Ext. Table \ref{tab:temp-range-cwe-787}).}
\setlength{\tabcolsep}{3pt}
    
\begin{subtable}{0.48\linewidth}
\centering
\begin{tabularx}{\linewidth}{lXXXXXX}
\toprule
\textbf{Model} & \textbf{Rec} & \textbf{0.0} & \textbf{0.25} & \textbf{0.5} & \textbf{0.75} & \textbf{1.0} \\
\seprule
codellama7b & \cellcolor{lightgreen!100}10{\fontsize{3.5pt}{0pt}\selectfont /10} & \cellcolor{lightgreen!100}10{\fontsize{3.5pt}{0pt}\selectfont /10} & \cellcolor{lightgreen!100}10{\fontsize{3.5pt}{0pt}\selectfont /10} & \cellcolor{lightgreen!100}10{\fontsize{3.5pt}{0pt}\selectfont /10} & \cellcolor{lightgreen!100}10{\fontsize{3.5pt}{0pt}\selectfont /10} & \cellcolor{lightgreen!90}9{\fontsize{3.5pt}{0pt}\selectfont /10}\\ 
\hline \noalign{\vskip 1ex} 
codellama13b & \cellcolor{lightgreen!100}10{\fontsize{3.5pt}{0pt}\selectfont /10} & \cellcolor{lightgreen!100}10{\fontsize{3.5pt}{0pt}\selectfont /10} & \cellcolor{lightgreen!100}10{\fontsize{3.5pt}{0pt}\selectfont /10} & \cellcolor{lightgreen!100}10{\fontsize{3.5pt}{0pt}\selectfont /10} & \cellcolor{lightgreen!90}9{\fontsize{3.5pt}{0pt}\selectfont /10} & \cellcolor{lightgreen!100}10{\fontsize{3.5pt}{0pt}\selectfont /10}\\ 
\hline \noalign{\vskip 1ex} 
starchat & \cellcolor{lightgreen!100}10{\fontsize{3.5pt}{0pt}\selectfont /10} & \cellcolor{lightgreen!100}10{\fontsize{3.5pt}{0pt}\selectfont /10} & \cellcolor{lightgreen!80}8{\fontsize{3.5pt}{0pt}\selectfont /10} & \cellcolor{lightgreen!70}7{\fontsize{3.5pt}{0pt}\selectfont /10} & \cellcolor{lightgreen!80}8{\fontsize{3.5pt}{0pt}\selectfont /10} & \cellcolor{lightgreen!70}7{\fontsize{3.5pt}{0pt}\selectfont /9}\\ 
\bottomrule
\end{tabularx}
\vspace{0.5pt}
\subcaption{Accuracy ($3_v$)}
\label{subtable:app-temp-range-cwe-787-3-acc}
\end{subtable}
\hspace{3pt}
\begin{subtable}{0.48\linewidth}
\centering
\begin{tabularx}{\linewidth}{lXXXXXX}
\toprule
\textbf{Model} & \textbf{Rec} & \textbf{0.0} & \textbf{0.25} & \textbf{0.5} & \textbf{0.75} & \textbf{1.0} \\
\seprule
codellama7b & \cellcolor{lightgreen!100}10{\fontsize{4pt}{0pt}\selectfont /10} & \cellcolor{lightgreen!100}10{\fontsize{4pt}{0pt}\selectfont /10} & \cellcolor{lightgreen!100}10{\fontsize{4pt}{0pt}\selectfont /10} & \cellcolor{lightgreen!100}10{\fontsize{4pt}{0pt}\selectfont /10} & \cellcolor{lightgreen!100}10{\fontsize{4pt}{0pt}\selectfont /10} & \cellcolor{lightgreen!90}9{\fontsize{4pt}{0pt}\selectfont /10}\\ 
\hline \noalign{\vskip 1ex} 
codellama13b & \cellcolor{lightgreen!100}10{\fontsize{4pt}{0pt}\selectfont /10} & \cellcolor{lightgreen!100}10{\fontsize{4pt}{0pt}\selectfont /10} & \cellcolor{lightgreen!100}10{\fontsize{4pt}{0pt}\selectfont /10} & \cellcolor{lightgreen!100}10{\fontsize{4pt}{0pt}\selectfont /10} & \cellcolor{lightgreen!100}10{\fontsize{4pt}{0pt}\selectfont /10} & \cellcolor{lightgreen!100}10{\fontsize{4pt}{0pt}\selectfont /10}\\ 
\hline \noalign{\vskip 1ex} 
starchat & \cellcolor{lightgreen!100}10{\fontsize{4pt}{0pt}\selectfont /10} & \cellcolor{lightgreen!100}10{\fontsize{4pt}{0pt}\selectfont /10} & \cellcolor{lightgreen!80}8{\fontsize{4pt}{0pt}\selectfont /10} & \cellcolor{lightgreen!70}7{\fontsize{4pt}{0pt}\selectfont /10} & \cellcolor{lightgreen!80}8{\fontsize{4pt}{0pt}\selectfont /10} & \cellcolor{lightgreen!70}7{\fontsize{4pt}{0pt}\selectfont /9}\\ 
\bottomrule
\end{tabularx}
\vspace{0.5pt}
\subcaption{Reason ($3_v$)}
\label{subtable:app-temp-range-cwe-787-3-rea}
\end{subtable}

\begin{subtable}{0.48\linewidth}
\centering
\begin{tabularx}{\linewidth}{lXXXXXX}
\toprule
\textbf{Model} & \textbf{Rec} & \textbf{0.0} & \textbf{0.25} & \textbf{0.5} & \textbf{0.75} & \textbf{1.0} \\
\seprule
codellama7b & \cellcolor{lightgreen!0}0{\fontsize{4pt}{0pt}\selectfont /10} & \cellcolor{lightgreen!0}0{\fontsize{4pt}{0pt}\selectfont /10} & \cellcolor{lightgreen!0}0{\fontsize{4pt}{0pt}\selectfont /10} & \cellcolor{lightgreen!0}0{\fontsize{4pt}{0pt}\selectfont /10} & \cellcolor{lightgreen!0}0{\fontsize{4pt}{0pt}\selectfont /10} & \cellcolor{lightgreen!0}0{\fontsize{4pt}{0pt}\selectfont /10}\\ 
\hline \noalign{\vskip 1ex} 
codellama13b & \cellcolor{lightgreen!0}0{\fontsize{4pt}{0pt}\selectfont /10} & \cellcolor{lightgreen!0}0{\fontsize{4pt}{0pt}\selectfont /10} & \cellcolor{lightgreen!0}0{\fontsize{4pt}{0pt}\selectfont /10} & \cellcolor{lightgreen!0}0{\fontsize{4pt}{0pt}\selectfont /10} & \cellcolor{lightgreen!0}0{\fontsize{4pt}{0pt}\selectfont /10} & \cellcolor{lightgreen!0}0{\fontsize{4pt}{0pt}\selectfont /10}\\ 
\hline \noalign{\vskip 1ex} 
starchat & \cellcolor{lightgreen!0}0{\fontsize{4pt}{0pt}\selectfont /10} & \cellcolor{lightgreen!0}0{\fontsize{4pt}{0pt}\selectfont /10} & \cellcolor{lightgreen!0}0{\fontsize{4pt}{0pt}\selectfont /10} & \cellcolor{lightgreen!20}2{\fontsize{4pt}{0pt}\selectfont /10} & \cellcolor{lightgreen!10}1{\fontsize{4pt}{0pt}\selectfont /10} & \cellcolor{lightgreen!10}1{\fontsize{4pt}{0pt}\selectfont /10}\\ 
\bottomrule
\end{tabularx}
\vspace{0.5pt}
\subcaption{Accuracy ($3_p$)}
\label{subtable:app-temp-range-cwe-787-3p-acc}
\end{subtable}
\hspace{3pt}
\begin{subtable}{0.48\linewidth}
\centering
\begin{tabularx}{\linewidth}{lXXXXXX}
\toprule
\textbf{Model} & \textbf{Rec} & \textbf{0.0} & \textbf{0.25} & \textbf{0.5} & \textbf{0.75} & \textbf{1.0} \\
\seprule
codellama7b & \cellcolor{lightgreen!20}2{\fontsize{4pt}{0pt}\selectfont /10} & \cellcolor{lightgreen!0}0{\fontsize{4pt}{0pt}\selectfont /10} & \cellcolor{lightgreen!30}3{\fontsize{4pt}{0pt}\selectfont /10} & \cellcolor{lightgreen!10}1{\fontsize{4pt}{0pt}\selectfont /10} & \cellcolor{lightgreen!10}1{\fontsize{4pt}{0pt}\selectfont /10} & \cellcolor{lightgreen!0}0{\fontsize{4pt}{0pt}\selectfont /10}\\ 
\hline \noalign{\vskip 1ex} 
codellama13b & \cellcolor{lightgreen!0}0{\fontsize{4pt}{0pt}\selectfont /10} & \cellcolor{lightgreen!0}0{\fontsize{4pt}{0pt}\selectfont /10} & \cellcolor{lightgreen!0}0{\fontsize{4pt}{0pt}\selectfont /10} & \cellcolor{lightgreen!0}0{\fontsize{4pt}{0pt}\selectfont /10} & \cellcolor{lightgreen!20}2{\fontsize{4pt}{0pt}\selectfont /10} & \cellcolor{lightgreen!0}0{\fontsize{4pt}{0pt}\selectfont /10}\\ 
\hline \noalign{\vskip 1ex} 
starchat & \cellcolor{lightgreen!30}3{\fontsize{4pt}{0pt}\selectfont /10} & \cellcolor{lightgreen!0}0{\fontsize{4pt}{0pt}\selectfont /10} & \cellcolor{lightgreen!10}1{\fontsize{4pt}{0pt}\selectfont /10} & \cellcolor{lightgreen!20}2{\fontsize{4pt}{0pt}\selectfont /10} & \cellcolor{lightgreen!40}4{\fontsize{4pt}{0pt}\selectfont /10} & \cellcolor{lightgreen!10}1{\fontsize{4pt}{0pt}\selectfont /10}\\ 
\bottomrule
\end{tabularx}
\vspace{0.5pt}
\subcaption{Reason ($3_p$)}
\label{subtable:app-temp-range-cwe-787-3p-rea}
\end{subtable}
    
\label{tab:app-temp-range-cwe-787}
\end{table}

\begin{table}[]
\centering
\tiny
\caption{Evaluation Over a Range of Temperature Values (CWE-89) (Ext. Table \ref{tab:temp-range-cwe-89}).}
\setlength{\tabcolsep}{3pt}
    
\begin{subtable}{0.48\linewidth}
\centering
\begin{tabularx}{\linewidth}{lXXXXXX}
\toprule
\textbf{Model} & \textbf{Rec} & \textbf{0.0} & \textbf{0.25} & \textbf{0.5} & \textbf{0.75} & \textbf{1.0} \\
\seprule
codellama7b & \cellcolor{lightgreen!100}10{\fontsize{4pt}{0pt}\selectfont /10} & \cellcolor{lightgreen!100}10{\fontsize{4pt}{0pt}\selectfont /10} & \cellcolor{lightgreen!100}10{\fontsize{4pt}{0pt}\selectfont /10} & \cellcolor{lightgreen!100}10{\fontsize{4pt}{0pt}\selectfont /10} & \cellcolor{lightgreen!100}10{\fontsize{4pt}{0pt}\selectfont /10} & \cellcolor{lightgreen!100}10{\fontsize{4pt}{0pt}\selectfont /10}\\ 
\hline \noalign{\vskip 1ex} 
codellama13b & \cellcolor{lightgreen!100}10{\fontsize{4pt}{0pt}\selectfont /10} & \cellcolor{lightgreen!100}10{\fontsize{4pt}{0pt}\selectfont /10} & \cellcolor{lightgreen!100}10{\fontsize{4pt}{0pt}\selectfont /10} & \cellcolor{lightgreen!90}9{\fontsize{4pt}{0pt}\selectfont /10} & \cellcolor{lightgreen!100}10{\fontsize{4pt}{0pt}\selectfont /10} & \cellcolor{lightgreen!100}10{\fontsize{4pt}{0pt}\selectfont /10}\\ 
\hline \noalign{\vskip 1ex} 
starchat & \cellcolor{lightgreen!70}7{\fontsize{4pt}{0pt}\selectfont /10} & \cellcolor{lightgreen!0}0{\fontsize{4pt}{0pt}\selectfont /10} & \cellcolor{lightgreen!60}6{\fontsize{4pt}{0pt}\selectfont /10} & \cellcolor{lightgreen!60}6{\fontsize{4pt}{0pt}\selectfont /9} & \cellcolor{lightgreen!90}9{\fontsize{4pt}{0pt}\selectfont /9} & \cellcolor{lightgreen!70}7{\fontsize{4pt}{0pt}\selectfont /10}\\ 
\bottomrule
\end{tabularx}
\vspace{0.5pt}
\subcaption{Accuracy ($3_v$)}
\label{subtable:app-temp-range-cwe-89-3-acc}
\end{subtable}
\hspace{3pt}
\begin{subtable}{0.48\linewidth}
\centering
\begin{tabularx}{\linewidth}{lXXXXXX}
\toprule
\textbf{Model} & \textbf{Rec} & \textbf{0.0} & \textbf{0.25} & \textbf{0.5} & \textbf{0.75} & \textbf{1.0} \\
\seprule
codellama7b & \cellcolor{lightgreen!100}10{\fontsize{4pt}{0pt}\selectfont /10} & \cellcolor{lightgreen!100}10{\fontsize{4pt}{0pt}\selectfont /10} & \cellcolor{lightgreen!100}10{\fontsize{4pt}{0pt}\selectfont /10} & \cellcolor{lightgreen!100}10{\fontsize{4pt}{0pt}\selectfont /10} & \cellcolor{lightgreen!100}10{\fontsize{4pt}{0pt}\selectfont /10} & \cellcolor{lightgreen!100}10{\fontsize{4pt}{0pt}\selectfont /10}\\ 
\hline \noalign{\vskip 1ex} 
codellama13b & \cellcolor{lightgreen!100}10{\fontsize{4pt}{0pt}\selectfont /10} & \cellcolor{lightgreen!100}10{\fontsize{4pt}{0pt}\selectfont /10} & \cellcolor{lightgreen!100}10{\fontsize{4pt}{0pt}\selectfont /10} & \cellcolor{lightgreen!90}9{\fontsize{4pt}{0pt}\selectfont /10} & \cellcolor{lightgreen!100}10{\fontsize{4pt}{0pt}\selectfont /10} & \cellcolor{lightgreen!100}10{\fontsize{4pt}{0pt}\selectfont /10}\\ 
\hline \noalign{\vskip 1ex} 
starchat & \cellcolor{lightgreen!70}7{\fontsize{4pt}{0pt}\selectfont /10} & \cellcolor{lightgreen!0}0{\fontsize{4pt}{0pt}\selectfont /10} & \cellcolor{lightgreen!60}6{\fontsize{4pt}{0pt}\selectfont /10} & \cellcolor{lightgreen!50}5{\fontsize{4pt}{0pt}\selectfont /9} & \cellcolor{lightgreen!90}9{\fontsize{4pt}{0pt}\selectfont /9} & \cellcolor{lightgreen!70}7{\fontsize{4pt}{0pt}\selectfont /10}\\ 
\bottomrule
\end{tabularx}
\vspace{0.5pt}
\subcaption{Reason ($3_v$)}
\label{subtable:app-temp-range-cwe-89-3-rea}
\end{subtable}

\begin{subtable}{0.48\linewidth}
\centering
\begin{tabularx}{\linewidth}{lXXXXXX}
\toprule
\textbf{Model} & \textbf{Rec} & \textbf{0.0} & \textbf{0.25} & \textbf{0.5} & \textbf{0.75} & \textbf{1.0} \\
\seprule
codellama7b & \cellcolor{lightgreen!0}0{\fontsize{4pt}{0pt}\selectfont /10} & \cellcolor{lightgreen!0}0{\fontsize{4pt}{0pt}\selectfont /10} & \cellcolor{lightgreen!0}0{\fontsize{4pt}{0pt}\selectfont /10} & \cellcolor{lightgreen!0}0{\fontsize{4pt}{0pt}\selectfont /10} & \cellcolor{lightgreen!0}0{\fontsize{4pt}{0pt}\selectfont /10} & \cellcolor{lightgreen!0}0{\fontsize{4pt}{0pt}\selectfont /10}\\ 
\hline \noalign{\vskip 1ex} 
codellama13b & \cellcolor{lightgreen!0}0{\fontsize{4pt}{0pt}\selectfont /10} & \cellcolor{lightgreen!0}0{\fontsize{4pt}{0pt}\selectfont /10} & \cellcolor{lightgreen!0}0{\fontsize{4pt}{0pt}\selectfont /10} & \cellcolor{lightgreen!0}0{\fontsize{4pt}{0pt}\selectfont /10} & \cellcolor{lightgreen!10}1{\fontsize{4pt}{0pt}\selectfont /10} & \cellcolor{lightgreen!10}1{\fontsize{4pt}{0pt}\selectfont /10}\\ 
\hline \noalign{\vskip 1ex} 
starchat & \cellcolor{lightgreen!40}4{\fontsize{4pt}{0pt}\selectfont /10} & \cellcolor{lightgreen!0}0{\fontsize{4pt}{0pt}\selectfont /0} & \cellcolor{lightgreen!20}2{\fontsize{4pt}{0pt}\selectfont /8} & \cellcolor{lightgreen!40}4{\fontsize{4pt}{0pt}\selectfont /10} & \cellcolor{lightgreen!10}1{\fontsize{4pt}{0pt}\selectfont /9} & \cellcolor{lightgreen!20}2{\fontsize{4pt}{0pt}\selectfont /9}\\ 
\bottomrule
\end{tabularx}
\vspace{0.5pt}
\subcaption{Accuracy ($3_p$)}
\label{subtable:app-temp-range-cwe-89-3p-acc}
\end{subtable}
\hspace{3pt}
\begin{subtable}{0.48\linewidth}
\centering
\begin{tabularx}{\linewidth}{lXXXXXX}
\toprule
\textbf{Model} & \textbf{Rec} & \textbf{0.0} & \textbf{0.25} & \textbf{0.5} & \textbf{0.75} & \textbf{1.0} \\
\seprule
codellama7b & \cellcolor{lightgreen!10}1{\fontsize{4pt}{0pt}\selectfont /10} & \cellcolor{lightgreen!70}7{\fontsize{4pt}{0pt}\selectfont /10} & \cellcolor{lightgreen!50}5{\fontsize{4pt}{0pt}\selectfont /10} & \cellcolor{lightgreen!20}2{\fontsize{4pt}{0pt}\selectfont /10} & \cellcolor{lightgreen!20}2{\fontsize{4pt}{0pt}\selectfont /10} & \cellcolor{lightgreen!20}2{\fontsize{4pt}{0pt}\selectfont /10}\\ 
\hline \noalign{\vskip 1ex} 
codellama13b & \cellcolor{lightgreen!10}1{\fontsize{4pt}{0pt}\selectfont /10} & \cellcolor{lightgreen!0}0{\fontsize{4pt}{0pt}\selectfont /10} & \cellcolor{lightgreen!30}3{\fontsize{4pt}{0pt}\selectfont /10} & \cellcolor{lightgreen!20}2{\fontsize{4pt}{0pt}\selectfont /10} & \cellcolor{lightgreen!10}1{\fontsize{4pt}{0pt}\selectfont /10} & \cellcolor{lightgreen!30}3{\fontsize{4pt}{0pt}\selectfont /10}\\ 
\hline \noalign{\vskip 1ex} 
starchat & \cellcolor{lightgreen!30}3{\fontsize{4pt}{0pt}\selectfont /10} & \cellcolor{lightgreen!0}0{\fontsize{4pt}{0pt}\selectfont /0} & \cellcolor{lightgreen!30}3{\fontsize{4pt}{0pt}\selectfont /8} & \cellcolor{lightgreen!40}4{\fontsize{4pt}{0pt}\selectfont /10} & \cellcolor{lightgreen!10}1{\fontsize{4pt}{0pt}\selectfont /9} & \cellcolor{lightgreen!20}2{\fontsize{4pt}{0pt}\selectfont /9}\\ 
\bottomrule
\end{tabularx}
\vspace{0.5pt}
\subcaption{Reason ($3_p$)}
\label{subtable:app-temp-range-cwe-89-3p-rea}
\end{subtable}
    
\label{tab:app-temp-range-cwe-89}
\end{table}

\begin{table}[!t]
    \centering
    \caption{Evaluation Diversity of Prompts (Ext. Table \ref{tab:prompts-eval}).}
    \raisebox{-\height}{\includegraphics[width=\linewidth]{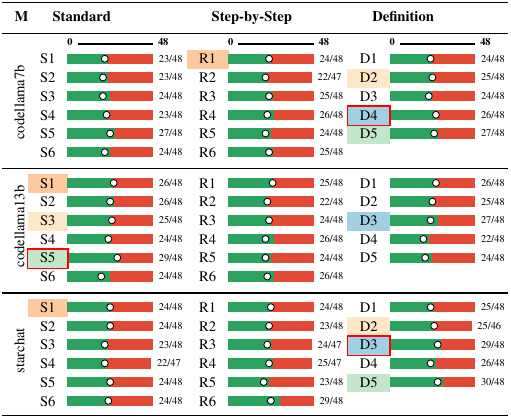}}
    \label{tab:app-prompts-eval}
\end{table}

\begin{table}[!t]
\centering
\caption{Faithfulness of LLMs (Ext. Table \ref{tab:faith-llm}).}
\raisebox{-\height}{\includegraphics[width=\linewidth]{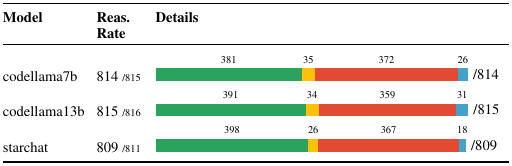}}
\label{tab:app-faith-llm}
\end{table}

\begin{figure}
    \centering
\begin{lstlisting}[language=C]
...
char* validate_cmd(char* cmd) {
    char* new_cmd = malloc(strlen(cmd) + 1);
    int cmd_ix = 0;
    while (cmd_ix < strlen(cmd)) {
        if (cmd[cmd_ix] == ';') {
            break;
        }
        new_cmd[cmd_ix] = cmd[cmd_ix];
        cmd_ix++;
    }
    return new_cmd;
}
...\end{lstlisting}
    \caption{CWE-77 (Command Injection) $3_v$ code scenario.}
    \label{fig:cwe-77-3}
\end{figure}

\begin{table}[!t]
    \centering
    \caption{Evaluation range of vulnerabilities (Ext. Table~\ref{tab:var-cwes}).}
    \includegraphics[width=0.95\linewidth]{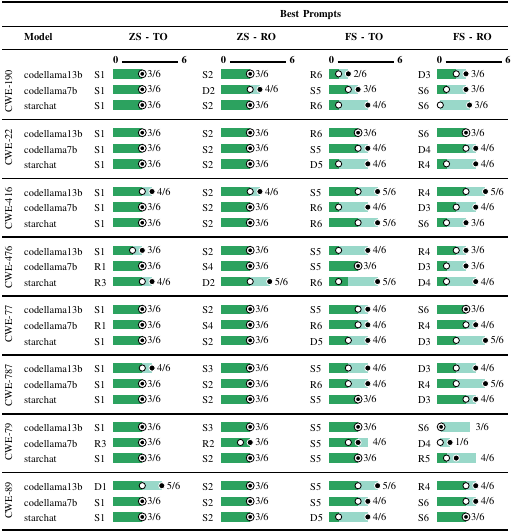}
    \label{tab:app-var-cwes}
\end{table}

\begin{table}[!t]
    \centering
    \caption{Evaluation code difficulties (Ext. Table~\ref{tab:var-diff}).}
    \includegraphics[width=0.95\linewidth]{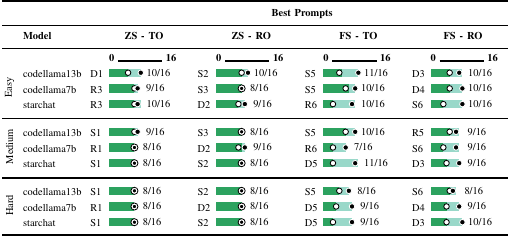}
    \label{tab:app-var-diffs}
\end{table}

\begin{table}[]
\centering
\tiny
\caption{Evaluation Code-Level Augmentations (Ext. Table \ref{tab:code-aug}).}
\vspace{0.1cm}
\begin{subtable}{\linewidth}
\begin{tabularx}{\linewidth}{l@{\hspace{0.2cm}}lXXXXXXXXXXXXXX}
\toprule
& & \multicolumn{2}{c}{\textbf{T1}} & \multicolumn{2}{c}{\textbf{T2}} & \multicolumn{2}{c}{\textbf{T3}} & \multicolumn{2}{c}{\textbf{T4}} & \multicolumn{2}{c}{\textbf{T5}} & \multicolumn{2}{c}{\textbf{T6}} & \multicolumn{2}{c}{\textbf{T7}} \\ [1ex]
\hline
\noalign{\vskip 0.5ex}
\textbf{M} & \textbf{PS} & $\Delta_a$ & $\Delta_r$ & $\Delta_a$ & $\Delta_r$ & $\Delta_a$ & $\Delta_r$ & $\Delta_a$ & $\Delta_r$ & $\Delta_a$ & $\Delta_r$ & $\Delta_a$ & $\Delta_r$ & $\Delta_a$ & $\Delta_r$ \\ [0.5ex]
\hline
\noalign{\vskip 1ex}
\multirow{3}{*}{\rotatebox[origin=c]{90}{c.lla.7b}} & S1{\fontsize{3pt}{0pt}\selectfont S} & \cellcolor{red!0}0{\fontsize{4pt}{0pt}\selectfont /12} & \cellcolor{red!0}0{\fontsize{4pt}{0pt}\selectfont /12} & \cellcolor{red!0}0{\fontsize{4pt}{0pt}\selectfont /12} & \cellcolor{red!0}0{\fontsize{4pt}{0pt}\selectfont /12} & \cellcolor{red!0}0{\fontsize{4pt}{0pt}\selectfont /12} & \cellcolor{red!0}0{\fontsize{4pt}{0pt}\selectfont /12} & \cellcolor{red!0}0{\fontsize{4pt}{0pt}\selectfont /12} & \cellcolor{red!0}0{\fontsize{4pt}{0pt}\selectfont /12} & \cellcolor{red!0}0{\fontsize{4pt}{0pt}\selectfont /12} & \cellcolor{red!0}0{\fontsize{4pt}{0pt}\selectfont /12} & \cellcolor{red!0}0{\fontsize{4pt}{0pt}\selectfont /12} & \cellcolor{red!0}0{\fontsize{4pt}{0pt}\selectfont /12} & \cellcolor{red!0}0{\fontsize{4pt}{0pt}\selectfont /12} & \cellcolor{red!8}1{\fontsize{4pt}{0pt}\selectfont /12}\\ 
 & D2{\fontsize{3pt}{0pt}\selectfont ZS} & \cellcolor{red!0}0{\fontsize{4pt}{0pt}\selectfont /12} & \cellcolor{red!0}0{\fontsize{4pt}{0pt}\selectfont /12} & \cellcolor{red!0}0{\fontsize{4pt}{0pt}\selectfont /12} & \cellcolor{red!0}0{\fontsize{4pt}{0pt}\selectfont /12} & \cellcolor{red!0}0{\fontsize{4pt}{0pt}\selectfont /12} & \cellcolor{red!0}0{\fontsize{4pt}{0pt}\selectfont /12} & \cellcolor{red!0}0{\fontsize{4pt}{0pt}\selectfont /12} & \cellcolor{red!0}0{\fontsize{4pt}{0pt}\selectfont /12} & \cellcolor{red!0}0{\fontsize{4pt}{0pt}\selectfont /12} & \cellcolor{red!0}0{\fontsize{4pt}{0pt}\selectfont /12} & \cellcolor{red!0}0{\fontsize{4pt}{0pt}\selectfont /12} & \cellcolor{red!0}0{\fontsize{4pt}{0pt}\selectfont /12} & \cellcolor{red!0}0{\fontsize{4pt}{0pt}\selectfont /12} & \cellcolor{red!0}0{\fontsize{4pt}{0pt}\selectfont /12}\\ 
 & D4{\fontsize{3pt}{0pt}\selectfont FS} & \cellcolor{red!16}2{\fontsize{4pt}{0pt}\selectfont /12} & \cellcolor{red!25}3{\fontsize{4pt}{0pt}\selectfont /12} & \cellcolor{red!16}2{\fontsize{4pt}{0pt}\selectfont /12} & \cellcolor{red!16}2{\fontsize{4pt}{0pt}\selectfont /12} & \cellcolor{red!8}1{\fontsize{4pt}{0pt}\selectfont /12} & \cellcolor{red!8}1{\fontsize{4pt}{0pt}\selectfont /12} & \cellcolor{red!25}3{\fontsize{4pt}{0pt}\selectfont /12} & \cellcolor{red!25}3{\fontsize{4pt}{0pt}\selectfont /12} & \cellcolor{red!8}1{\fontsize{4pt}{0pt}\selectfont /12} & \cellcolor{red!8}1{\fontsize{4pt}{0pt}\selectfont /12} & \cellcolor{red!8}1{\fontsize{4pt}{0pt}\selectfont /12} & \cellcolor{red!8}1{\fontsize{4pt}{0pt}\selectfont /12} & \cellcolor{red!16}2{\fontsize{4pt}{0pt}\selectfont /12} & \cellcolor{red!25}3{\fontsize{4pt}{0pt}\selectfont /12}\\ 
\seprule
\multirow{3}{*}{\rotatebox[origin=c]{90}{c.lla.13b}} & S1{\fontsize{3pt}{0pt}\selectfont S} & \cellcolor{red!0}0{\fontsize{4pt}{0pt}\selectfont /12} & \cellcolor{red!0}0{\fontsize{4pt}{0pt}\selectfont /12} & \cellcolor{red!8}1{\fontsize{4pt}{0pt}\selectfont /12} & \cellcolor{red!8}1{\fontsize{4pt}{0pt}\selectfont /12} & \cellcolor{red!0}0{\fontsize{4pt}{0pt}\selectfont /12} & \cellcolor{red!0}0{\fontsize{4pt}{0pt}\selectfont /12} & \cellcolor{red!16}2{\fontsize{4pt}{0pt}\selectfont /12} & \cellcolor{red!16}2{\fontsize{4pt}{0pt}\selectfont /12} & \cellcolor{red!0}0{\fontsize{4pt}{0pt}\selectfont /12} & \cellcolor{red!0}0{\fontsize{4pt}{0pt}\selectfont /12} & \cellcolor{red!8}1{\fontsize{4pt}{0pt}\selectfont /12} & \cellcolor{red!8}1{\fontsize{4pt}{0pt}\selectfont /12} & \cellcolor{red!0}0{\fontsize{4pt}{0pt}\selectfont /12} & \cellcolor{red!0}0{\fontsize{4pt}{0pt}\selectfont /12}\\ 
 & S1{\fontsize{3pt}{0pt}\selectfont ZS} & \cellcolor{red!0}0{\fontsize{4pt}{0pt}\selectfont /12} & \cellcolor{red!0}0{\fontsize{4pt}{0pt}\selectfont /12} & \cellcolor{red!8}1{\fontsize{4pt}{0pt}\selectfont /12} & \cellcolor{red!8}1{\fontsize{4pt}{0pt}\selectfont /12} & \cellcolor{red!0}0{\fontsize{4pt}{0pt}\selectfont /12} & \cellcolor{red!0}0{\fontsize{4pt}{0pt}\selectfont /12} & \cellcolor{red!16}2{\fontsize{4pt}{0pt}\selectfont /12} & \cellcolor{red!16}2{\fontsize{4pt}{0pt}\selectfont /12} & \cellcolor{red!0}0{\fontsize{4pt}{0pt}\selectfont /12} & \cellcolor{red!0}0{\fontsize{4pt}{0pt}\selectfont /12} & \cellcolor{red!8}1{\fontsize{4pt}{0pt}\selectfont /12} & \cellcolor{red!8}1{\fontsize{4pt}{0pt}\selectfont /12} & \cellcolor{red!0}0{\fontsize{4pt}{0pt}\selectfont /12} & \cellcolor{red!0}0{\fontsize{4pt}{0pt}\selectfont /12}\\ 
 & S5{\fontsize{3pt}{0pt}\selectfont FS} & \cellcolor{red!0}0{\fontsize{4pt}{0pt}\selectfont /12} & \cellcolor{red!0}0{\fontsize{4pt}{0pt}\selectfont /12} & \cellcolor{red!8}1{\fontsize{4pt}{0pt}\selectfont /12} & \cellcolor{red!8}1{\fontsize{4pt}{0pt}\selectfont /12} & \cellcolor{red!16}2{\fontsize{4pt}{0pt}\selectfont /12} & \cellcolor{red!16}2{\fontsize{4pt}{0pt}\selectfont /12} & \cellcolor{red!16}2{\fontsize{4pt}{0pt}\selectfont /12} & \cellcolor{red!16}2{\fontsize{4pt}{0pt}\selectfont /12} & \cellcolor{red!0}0{\fontsize{4pt}{0pt}\selectfont /12} & \cellcolor{red!8}1{\fontsize{4pt}{0pt}\selectfont /12} & \cellcolor{red!25}3{\fontsize{4pt}{0pt}\selectfont /12} & \cellcolor{red!25}3{\fontsize{4pt}{0pt}\selectfont /12} & \cellcolor{red!0}0{\fontsize{4pt}{0pt}\selectfont /12} & \cellcolor{red!0}0{\fontsize{4pt}{0pt}\selectfont /12}\\ 
\seprule
\multirow{3}{*}{\rotatebox[origin=c]{90}{starc.}} & S1{\fontsize{3pt}{0pt}\selectfont S} & \cellcolor{red!0}0{\fontsize{4pt}{0pt}\selectfont /12} & \cellcolor{red!0}0{\fontsize{4pt}{0pt}\selectfont /12} & \cellcolor{red!0}0{\fontsize{4pt}{0pt}\selectfont /12} & \cellcolor{red!0}0{\fontsize{4pt}{0pt}\selectfont /12} & \cellcolor{red!0}0{\fontsize{4pt}{0pt}\selectfont /12} & \cellcolor{red!0}0{\fontsize{4pt}{0pt}\selectfont /12} & \cellcolor{red!0}0{\fontsize{4pt}{0pt}\selectfont /12} & \cellcolor{red!8}1{\fontsize{4pt}{0pt}\selectfont /12} & \cellcolor{red!0}0{\fontsize{4pt}{0pt}\selectfont /12} & \cellcolor{red!0}0{\fontsize{4pt}{0pt}\selectfont /12} & \cellcolor{red!0}0{\fontsize{4pt}{0pt}\selectfont /12} & \cellcolor{red!0}0{\fontsize{4pt}{0pt}\selectfont /12} & \cellcolor{red!0}0{\fontsize{4pt}{0pt}\selectfont /12} & \cellcolor{red!0}0{\fontsize{4pt}{0pt}\selectfont /12}\\ 
 & D2{\fontsize{3pt}{0pt}\selectfont ZS} & \cellcolor{red!0}0{\fontsize{4pt}{0pt}\selectfont /12} & \cellcolor{red!0}0{\fontsize{4pt}{0pt}\selectfont /12} & \cellcolor{red!0}0{\fontsize{4pt}{0pt}\selectfont /12} & \cellcolor{red!0}0{\fontsize{4pt}{0pt}\selectfont /12} & \cellcolor{red!0}0{\fontsize{4pt}{0pt}\selectfont /12} & \cellcolor{red!8}1{\fontsize{4pt}{0pt}\selectfont /12} & \cellcolor{red!0}0{\fontsize{4pt}{0pt}\selectfont /12} & \cellcolor{red!0}0{\fontsize{4pt}{0pt}\selectfont /12} & \cellcolor{red!0}0{\fontsize{4pt}{0pt}\selectfont /12} & \cellcolor{red!0}0{\fontsize{4pt}{0pt}\selectfont /12} & \cellcolor{red!0}0{\fontsize{4pt}{0pt}\selectfont /12} & \cellcolor{red!0}0{\fontsize{4pt}{0pt}\selectfont /12} & \cellcolor{red!0}0{\fontsize{4pt}{0pt}\selectfont /12} & \cellcolor{red!0}0{\fontsize{4pt}{0pt}\selectfont /12}\\ 
 & D3{\fontsize{3pt}{0pt}\selectfont FS} & \cellcolor{red!0}0{\fontsize{4pt}{0pt}\selectfont /12} & \cellcolor{red!0}0{\fontsize{4pt}{0pt}\selectfont /12} & \cellcolor{red!16}2{\fontsize{4pt}{0pt}\selectfont /12} & \cellcolor{red!16}2{\fontsize{4pt}{0pt}\selectfont /12} & \cellcolor{red!58}7{\fontsize{4pt}{0pt}\selectfont /12} & \cellcolor{red!33}4{\fontsize{4pt}{0pt}\selectfont /12} & \cellcolor{red!0}0{\fontsize{4pt}{0pt}\selectfont /12} & \cellcolor{red!8}1{\fontsize{4pt}{0pt}\selectfont /12} & \cellcolor{red!8}1{\fontsize{4pt}{0pt}\selectfont /12} & \cellcolor{red!8}1{\fontsize{4pt}{0pt}\selectfont /12} & \cellcolor{red!8}1{\fontsize{4pt}{0pt}\selectfont /12} & \cellcolor{red!25}3{\fontsize{4pt}{0pt}\selectfont /12} & \cellcolor{red!50}6{\fontsize{4pt}{0pt}\selectfont /12} & \cellcolor{red!33}4{\fontsize{4pt}{0pt}\selectfont /12}\\ 
\bottomrule
\end{tabularx}
\vspace{0.1cm}
\caption{Trivial Augmentations}
\label{tab:app-code-trivial}
\end{subtable}

\begin{subtable}{\linewidth}
\begin{tabularx}{\linewidth}{l@{\hspace{0.2cm}}lXXXXXXXXXXXX}
\toprule
& & \multicolumn{2}{c}{\textbf{NT1}} & \multicolumn{2}{c}{\textbf{NT2}} & \multicolumn{2}{c}{\textbf{NT3}} & \multicolumn{2}{c}{\textbf{NT4}} & \multicolumn{2}{c}{\textbf{NT5}} & \multicolumn{2}{c}{\textbf{NT6}} \\ [1ex]
\hline
\noalign{\vskip 0.5ex}
\textbf{M} & \textbf{PS} & $\Delta_a$ & $\Delta_r$ & $\Delta_a$ & $\Delta_r$ & $\Delta_a$ & $\Delta_r$ & $\Delta_a$ & $\Delta_r$ & $\Delta_a$ & $\Delta_r$ & $\Delta_a$ & $\Delta_r$ \\ [0.5ex]
\hline
\noalign{\vskip 1ex}
\multirow{3}{*}{\rotatebox[origin=c]{90}{c.lla.7b}} & S1{\fontsize{3pt}{0pt}\selectfont S} & \cellcolor{red!0}0{\fontsize{4pt}{0pt}\selectfont /12} & \cellcolor{red!0}0{\fontsize{4pt}{0pt}\selectfont /12} & \cellcolor{red!0}0{\fontsize{4pt}{0pt}\selectfont /12} & \cellcolor{red!0}0{\fontsize{4pt}{0pt}\selectfont /12} & \cellcolor{red!25}3{\fontsize{4pt}{0pt}\selectfont /12} & \cellcolor{red!25}3{\fontsize{4pt}{0pt}\selectfont /12} & \cellcolor{red!0}0{\fontsize{4pt}{0pt}\selectfont /12} & \cellcolor{red!33}4{\fontsize{4pt}{0pt}\selectfont /12} & \cellcolor{red!0}0{\fontsize{4pt}{0pt}\selectfont /9} & \cellcolor{red!11}1{\fontsize{4pt}{0pt}\selectfont /9} & \cellcolor{red!0}0{\fontsize{4pt}{0pt}\selectfont /9} & \cellcolor{red!0}0{\fontsize{4pt}{0pt}\selectfont /9}\\ 
 & D2{\fontsize{3pt}{0pt}\selectfont ZS} & \cellcolor{red!0}0{\fontsize{4pt}{0pt}\selectfont /12} & \cellcolor{red!0}0{\fontsize{4pt}{0pt}\selectfont /12} & \cellcolor{red!0}0{\fontsize{4pt}{0pt}\selectfont /12} & \cellcolor{red!0}0{\fontsize{4pt}{0pt}\selectfont /12} & \cellcolor{red!75}9{\fontsize{4pt}{0pt}\selectfont /12} & \cellcolor{red!66}8{\fontsize{4pt}{0pt}\selectfont /12} & \cellcolor{red!0}0{\fontsize{4pt}{0pt}\selectfont /12} & \cellcolor{red!16}2{\fontsize{4pt}{0pt}\selectfont /12} & \cellcolor{red!0}0{\fontsize{4pt}{0pt}\selectfont /9} & \cellcolor{red!0}0{\fontsize{4pt}{0pt}\selectfont /9} & \cellcolor{red!0}0{\fontsize{4pt}{0pt}\selectfont /9} & \cellcolor{red!0}0{\fontsize{4pt}{0pt}\selectfont /9}\\ 
 & D4{\fontsize{3pt}{0pt}\selectfont FS} & \cellcolor{red!25}3{\fontsize{4pt}{0pt}\selectfont /12} & \cellcolor{red!25}3{\fontsize{4pt}{0pt}\selectfont /12} & \cellcolor{red!33}4{\fontsize{4pt}{0pt}\selectfont /12} & \cellcolor{red!66}8{\fontsize{4pt}{0pt}\selectfont /12} & \cellcolor{red!16}2{\fontsize{4pt}{0pt}\selectfont /12} & \cellcolor{red!16}2{\fontsize{4pt}{0pt}\selectfont /12} & \cellcolor{red!41}5{\fontsize{4pt}{0pt}\selectfont /12} & \cellcolor{red!50}6{\fontsize{4pt}{0pt}\selectfont /12} & \cellcolor{red!0}0{\fontsize{4pt}{0pt}\selectfont /9} & \cellcolor{red!0}0{\fontsize{4pt}{0pt}\selectfont /9} & \cellcolor{red!0}0{\fontsize{4pt}{0pt}\selectfont /9} & \cellcolor{red!0}0{\fontsize{4pt}{0pt}\selectfont /9}\\ 
\seprule
\multirow{3}{*}{\rotatebox[origin=c]{90}{c.lla.13b}} & S1{\fontsize{3pt}{0pt}\selectfont S} & \cellcolor{red!8}1{\fontsize{4pt}{0pt}\selectfont /12} & \cellcolor{red!0}0{\fontsize{4pt}{0pt}\selectfont /12} & \cellcolor{red!16}2{\fontsize{4pt}{0pt}\selectfont /12} & \cellcolor{red!16}2{\fontsize{4pt}{0pt}\selectfont /12} & \cellcolor{red!66}8{\fontsize{4pt}{0pt}\selectfont /12} & \cellcolor{red!66}8{\fontsize{4pt}{0pt}\selectfont /12} & \cellcolor{red!8}1{\fontsize{4pt}{0pt}\selectfont /12} & \cellcolor{red!25}3{\fontsize{4pt}{0pt}\selectfont /12} & \cellcolor{red!0}0{\fontsize{4pt}{0pt}\selectfont /9} & \cellcolor{red!0}0{\fontsize{4pt}{0pt}\selectfont /9} & \cellcolor{red!0}0{\fontsize{4pt}{0pt}\selectfont /9} & \cellcolor{red!44}4{\fontsize{4pt}{0pt}\selectfont /9}\\ 
 & S1{\fontsize{3pt}{0pt}\selectfont ZS} & \cellcolor{red!8}1{\fontsize{4pt}{0pt}\selectfont /12} & \cellcolor{red!0}0{\fontsize{4pt}{0pt}\selectfont /12} & \cellcolor{red!16}2{\fontsize{4pt}{0pt}\selectfont /12} & \cellcolor{red!16}2{\fontsize{4pt}{0pt}\selectfont /12} & \cellcolor{red!66}8{\fontsize{4pt}{0pt}\selectfont /12} & \cellcolor{red!66}8{\fontsize{4pt}{0pt}\selectfont /12} & \cellcolor{red!8}1{\fontsize{4pt}{0pt}\selectfont /12} & \cellcolor{red!33}4{\fontsize{4pt}{0pt}\selectfont /12} & \cellcolor{red!0}0{\fontsize{4pt}{0pt}\selectfont /9} & \cellcolor{red!0}0{\fontsize{4pt}{0pt}\selectfont /9} & \cellcolor{red!0}0{\fontsize{4pt}{0pt}\selectfont /9} & \cellcolor{red!22}2{\fontsize{4pt}{0pt}\selectfont /9}\\ 
 & S5{\fontsize{3pt}{0pt}\selectfont FS} & \cellcolor{red!0}0{\fontsize{4pt}{0pt}\selectfont /12} & \cellcolor{red!0}0{\fontsize{4pt}{0pt}\selectfont /12} & \cellcolor{red!41}5{\fontsize{4pt}{0pt}\selectfont /12} & \cellcolor{red!41}5{\fontsize{4pt}{0pt}\selectfont /12} & \cellcolor{red!83}10{\fontsize{4pt}{0pt}\selectfont /12} & \cellcolor{red!83}10{\fontsize{4pt}{0pt}\selectfont /12} & \cellcolor{red!8}1{\fontsize{4pt}{0pt}\selectfont /12} & \cellcolor{red!16}2{\fontsize{4pt}{0pt}\selectfont /12} & \cellcolor{red!0}0{\fontsize{4pt}{0pt}\selectfont /9} & \cellcolor{red!0}0{\fontsize{4pt}{0pt}\selectfont /9} & \cellcolor{red!33}3{\fontsize{4pt}{0pt}\selectfont /9} & \cellcolor{red!44}4{\fontsize{4pt}{0pt}\selectfont /9}\\ 
\seprule
\multirow{3}{*}{\rotatebox[origin=c]{90}{starc.}} & S1{\fontsize{3pt}{0pt}\selectfont S} & \cellcolor{red!0}0{\fontsize{4pt}{0pt}\selectfont /12} & \cellcolor{red!0}0{\fontsize{4pt}{0pt}\selectfont /12} & \cellcolor{red!0}0{\fontsize{4pt}{0pt}\selectfont /12} & \cellcolor{red!0}0{\fontsize{4pt}{0pt}\selectfont /12} & \cellcolor{red!0}0{\fontsize{4pt}{0pt}\selectfont /12} & \cellcolor{red!8}1{\fontsize{4pt}{0pt}\selectfont /12} & \cellcolor{red!0}0{\fontsize{4pt}{0pt}\selectfont /12} & \cellcolor{red!25}3{\fontsize{4pt}{0pt}\selectfont /12} & \cellcolor{red!0}0{\fontsize{4pt}{0pt}\selectfont /9} & \cellcolor{red!0}0{\fontsize{4pt}{0pt}\selectfont /9} & \cellcolor{red!11}1{\fontsize{4pt}{0pt}\selectfont /9} & \cellcolor{red!11}1{\fontsize{4pt}{0pt}\selectfont /9}\\ 
 & D2{\fontsize{3pt}{0pt}\selectfont ZS} & \cellcolor{red!0}0{\fontsize{4pt}{0pt}\selectfont /12} & \cellcolor{red!0}0{\fontsize{4pt}{0pt}\selectfont /12} & \cellcolor{red!25}3{\fontsize{4pt}{0pt}\selectfont /12} & \cellcolor{red!25}3{\fontsize{4pt}{0pt}\selectfont /12} & \cellcolor{red!50}6{\fontsize{4pt}{0pt}\selectfont /12} & \cellcolor{red!41}5{\fontsize{4pt}{0pt}\selectfont /12} & \cellcolor{red!0}0{\fontsize{4pt}{0pt}\selectfont /12} & \cellcolor{red!16}2{\fontsize{4pt}{0pt}\selectfont /12} & \cellcolor{red!0}0{\fontsize{4pt}{0pt}\selectfont /9} & \cellcolor{red!33}3{\fontsize{4pt}{0pt}\selectfont /9} & \cellcolor{red!0}0{\fontsize{4pt}{0pt}\selectfont /9} & \cellcolor{red!0}0{\fontsize{4pt}{0pt}\selectfont /9}\\ 
 & D3{\fontsize{3pt}{0pt}\selectfont FS} & \cellcolor{red!0}0{\fontsize{4pt}{0pt}\selectfont /12} & \cellcolor{red!16}2{\fontsize{4pt}{0pt}\selectfont /12} & \cellcolor{red!8}1{\fontsize{4pt}{0pt}\selectfont /12} & \cellcolor{red!8}1{\fontsize{4pt}{0pt}\selectfont /12} & \cellcolor{red!33}4{\fontsize{4pt}{0pt}\selectfont /12} & \cellcolor{red!33}4{\fontsize{4pt}{0pt}\selectfont /12} & \cellcolor{red!0}0{\fontsize{4pt}{0pt}\selectfont /12} & \cellcolor{red!8}1{\fontsize{4pt}{0pt}\selectfont /12} & \cellcolor{red!11}1{\fontsize{4pt}{0pt}\selectfont /9} & \cellcolor{red!11}1{\fontsize{4pt}{0pt}\selectfont /9} & \cellcolor{red!11}1{\fontsize{4pt}{0pt}\selectfont /9} & \cellcolor{red!11}1{\fontsize{4pt}{0pt}\selectfont /9}\\ 
\bottomrule
\end{tabularx}
\vspace{0.1cm}
\caption{Non-Trivial Augmentations}
\label{tab:app-code-non-trivial}
\end{subtable}

\label{tab:app-code-aug}
\end{table}

\begin{table}[]
    \centering
    \caption{Evaluation real-world CVEs (Ext. Tables \ref{tab:start-cve-eval} and \ref{tab:end-cve-eval}).}
    \raisebox{-\height}{\includegraphics[width=\linewidth]{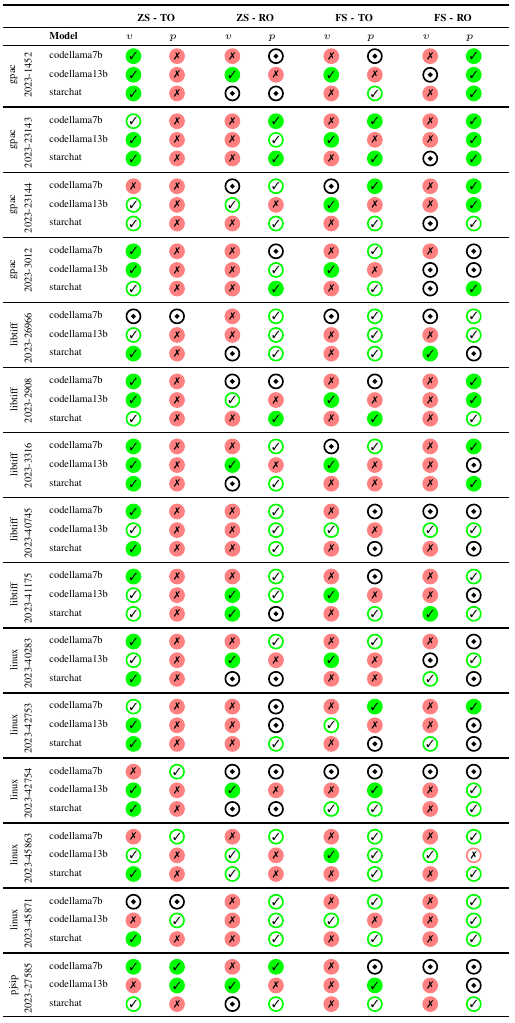}}
    \label{tab:app-cves}
\end{table}

\clearpage

\section{Meta-Review}

The following meta-review was prepared by the program committee for the 2024
IEEE Symposium on Security and Privacy (S\&P) as part of the review process as
detailed in the call for papers.

\subsection{Summary}
This paper suggests an automated framework for evaluating LLMs (Large Language Models) for identifying and reasoning about security vulnerabilities. The authors conducted experiments on various options of LLMs (e.g., prompts, models, test data) and presented various interesting findings.

\subsection{Scientific Contributions}
\begin{itemize}
\item Provides a Valuable Step Forward in an Established Field
\item Creates a New Tool to Enable Future Science
\item Provides a New Data Set For Public Use
\end{itemize}

\subsection{Reasons for Acceptance}
\begin{enumerate}
\item This paper suggests a fully automated framework to evaluate various aspects of LLM’s capability in identifying vulnerabilities, which is gaining more attention from the community.
\item The paper examines several aspects that have been less studied previously, including code complexity, vulnerability reasoning, and use of chat-based LLMs (e.g., GPT-4) in vulnerability detection.
\item The authors have developed a new benchmark consisting of 228 code scenarios. This includes 48 hand-crafted examples (starting from most critical CWEs), 30 examples from real-world CVEs (taken from open source projects) and 150 codes obtained by trivial and non-trivial augmentation. If released, this dataset can help in carrying out further research, allowing an alignment for future analysis.
\end{enumerate}

\subsection{Noteworthy Concerns} %
\begin{enumerate} %
\item Some findings seem well-known to the community (e.g., non-determinism, and non-robustness by specific code augmentations).
\item The size of the data set is still limited, especially for the real-world CVEs (the data set only includes 15 CVEs now). Moreover, the dataset includes many augmented codes (150/228). The small data set may lead to significant bias in the measurement results, making it less reliable.
\end{enumerate}

\section{Response to the Meta-Review} %

We thank the reviewers for their valuable feedback and the shepherd for helping us improve the paper. We would like to respond to the meta-review concerns as follows:
\begin{enumerate}
    \item While instances of non-determinism and non-robustness to specific code augmentations have been found in previous work, to the best of our knowledge we are the first ones to systematically evaluate these issues in the context of vulnerability detection.
    \item The size of our dataset is comparable to previous work like [8] which only used 12 CVEs. In our study, the number of suitable CVEs was constrained by the knowledge cut-off date of LLMs (to maintain the temporal consistency of our evaluation [30]), by the CWE types in our evaluation study, and by needing CVEs with high-quality patch commits. 
\end{enumerate}

\end{document}